\documentclass[10pt]{article}

\usepackage{amsmath}
\usepackage{amssymb}

\usepackage{graphicx}

\usepackage{cite}

\usepackage[labelfont=bf,labelsep=period,justification=raggedright]{caption}

\makeatletter
\renewcommand{\@biblabel}[1]{\quad#1.}
\makeatother

\date{}

\begin{document}

\begin{flushleft}
{\Large \textbf{Emergence of scale-free close-knit friendship
structure in online social networks}
}
\\
Ai-Xiang Cui$^{1}$,
Zi-Ke Zhang$^{2,1}$,
Ming Tang$^{1,\ast}$,
Pak Ming Hui$^{3}$,
Yan Fu$^{1}$
\\
\bf{1} Web Sciences Center, University of Electronic Science and Technology of China, Chengdu 610054, People's Republic of China
\\
\bf{2} Institute for Information Economy, Hangzhou Normal University, Hangzhou 310036, People's Republic of China
\\
\bf{3} Department of Physics, The Chinese University of Hong Kong, Shatin, Hong Kong, People's Republic of China
\\
$\ast$ E-mail: tangminghuang521@hotmail.com
\end{flushleft}

\section*{Abstract}
Despite the structural properties of online social networks have attracted
much attention, the properties of the close-knit
friendship structures remain an important question. Here, we
mainly focus on how these mesoscale structures are
affected by the local and global structural properties. Analyzing the data of
four large-scale online social networks reveals several common structural properties. It is found that
not only the local structures given by the indegree, outdegree,
and reciprocal degree distributions follow a similar scaling behavior, the mesoscale structures represented by the distributions of close-knit friendship structures also exhibit a similar scaling law.  The degree correlation is very weak over a wide range of the degrees.
We propose a simple directed network model that captures the observed properties.
The model incorporates two mechanisms: reciprocation and
preferential attachment.
Through rate equation analysis of our model, the local-scale and mesoscale
structural properties are
derived. In the local-scale, the same scaling behavior of indegree and outdegree distributions stems from indegree and outdegree of nodes both growing as the same function of the introduction time, and the reciprocal degree distribution also shows the same power-law due to the linear relationship between the reciprocal degree and in/outdegree of nodes. In the mesoscale, the distributions of four closed triples representing close-knit friendship structures are found to exhibit identical power-laws, a behavior attributed to the negligible degree correlations.
Intriguingly, all the power-law exponents of the distributions in the local-scale and mesoscale depend only on one global parameter
-- the mean in/outdegree, while both the mean in/outdegree
and the reciprocity together determine the ratio of the reciprocal
degree of a node to its in/outdegree.
Structural properties of numerical simulated networks are analyzed and compared with each of the four real
networks. This work helps understand the
interplay between structures on different scales in online
social networks.


\section*{Introduction}
In recent years, an increasing number of online social
systems~(\textit{e.g.}, \textit{YouTube} and
\textit{Facebook})~have been attracting wide attention from
different
fields\cite{corten2011composition,gjoka2010walking,ahn2007analysis}.
Online social networks provide a platform
for web surfers to make acquaintance with congenial
friends\cite{hu2009evolution}, exchange photos and personal
news\cite{mislove2008growth}, share
videos\cite{benevenuto2008understanding}, establish
communities or forums on
focused issues\cite{gomez2008statistical}, etc. These online
interactive behaviors, which partly reflect
real-life social relationships among people, provide an
unprecedented opportunity to study and
understand the dazzling characteristics of real-life social
systems~\cite{scott1988social, scott2000social}.

Complex network theory has been proven to be a
powerful framework to understand
the structure and dynamics of complex systems
\cite{albert2002statistical,
dorogovtsev2002evolution, newman2003structure,
boccaletti2006complex,
costa2007characterization,dorogovtsev2008critical,barthelemy2011spatial}.
Online social systems have been treated as undirected
networks~\cite{castellano2009statistical,centola2010spread},
which have been applied successfully in exploring various
systems~\cite{albert2002statistical}. This
simplification, however cannot
describe the asymmetric interactions among users.
Taking~\textit{Flickr} as an example,
if a user $A$ designates another user $B$ as a friend, user $A$
can see the photos of user $B$, but not
the other way round unless user $B$ also
designates user $A$ as his friend. Technically, an asymmetric
interaction represents one directed link, and many online social
systems are thus directed networks in nature. The directionality of links is
important in characterizing the functioning of many systems, \textit{e.g.},
leadership structure of social reputation
\cite{l¨¹2011leaders,zhou2011emergence}, reciprocal behavior in
evolutionary games \cite{nowak2005evolution}, information
hierarchy of the World Wide Web~\cite{brin1998anatomy,
kleinberg1999authoritative}, citation relationship of scientific
publications \cite{newman2001structure,leicht2007large}, etc.
Much effort has been devoted to understanding the
structural properties of these directed networks, including the
indegree and outdegree
distributions~\cite{rodgers2001properties}, average shortest
distance \cite{rodgers2001properties}, degree correlation
\cite{foster2010edge}, and community structure
\cite{palla2007directed,leicht2008community,kim2010finding}.
Correspondingly, there are many models proposed
for the underlying mechanisms of the statistical
properties.  Dorogovtsev~\textit{et
al.}~\cite{dorogovtsev2000structure} generalized the
Barab{\'a}si-Albert(BA)
model~\cite{barabasi1999emergence} and obtained the exact form of
the indegree distribution of growing networks in the
thermodynamic limit.  Krapivsky~\textit{et
al.}~\cite{krapivsky2001degree} introduced a directed network
model that generates correlated indegree and outdegree
distributions. Zhou~\textit{et al.}~\cite{zhou2011emergence}
argued that the ``good get richer" mechanism would facilitate the
emergence of scale-free leadership structure in online social
networks.

Up to now, most of the work on
complex networks can be classified into studies
on three scales: the local scale based on the
single node properties (through statistical distributions), the
macro-scale based on the global properties of
networks (with global parameters), and the mesoscale
based on properties due to a group of nodes (via modular
properties)~\cite{almendral2010announcement,almendral2011introduction,reichardt2011interplay}.
However, a majority of studies
focused on the first two scales. In view of the significant role
of modularity in the functionality of real networks, it has
become increasing important to study the mesoscale structures.
Communities and motifs are two key mesoscale structures of real
complex networks. Community structures at mesoscale level are
ubiquitous in a variety of real complex
systems~\cite{fortunato2010community,lancichinetti2011finding},
such as \textit{Facebook}, \textit{YouTube}, and
\textit{Xiaonei}. There are more connections among members of the
same community than among members in different communities.
Lancichinetti \textit{et al.} analyzed the statistical properties
of communities in five categories of real complex networks, and
found that communities detected in networks of the same category
display similar structural
characteristics~\cite{lancichinetti2010characterizing}. Motifs,
which are defined as subgraphs that occur much more often than
expected in a random network, play a significant role in our
understanding of the interplay between the structures and
dynamics of real complex
networks~\cite{Watts1998,milo2002network,milo2004superfamilies,huang2007bridge,fagiolo2007clustering,
ahnert2008clustering}.

In spite of the structural features revealed at
the three scales, understanding the interplay between the
different scales has remained a major
challenge~\cite{almendral2010announcement,almendral2011introduction,reichardt2011interplay}.
In the present work, we study how the close-knit friendship
structures of online social networks at the mesoscale level and
the structural properties at the two other scales are affecting
each other. In social networks, the close-knit friendship
structure describes the closest unit, which is
usually represented by the closed triples. In a directed network, there are $13$
different possible three-node subgraphs~\cite{milo2002network}.
For situations without
reciprocal links, a focal node has three possible unclosed
triples.  Each unclosed triple can be closed by adding a directed
link between the two unconnected nodes, giving rise to four types
of closed triples as shown in
Figure~\ref{four-triangles}~\cite{fagiolo2007clustering,
ahnert2008clustering}. The four closed triples
fall into two groups: one is a feedback ($FB$) loop and the three
others are feedforward (\textit{i.e.}, $FF_{a}$, $FF_{b}$, and
$FF_{c}$) loops.  Structurally, the roles of three nodes in the
$FB$ loop are equivalent, but it is not the case in the $FF$
loops. Any $FF_{a}$ loop (from the perspective of the focal node)
becomes a $FF_b$ loop for another node and a $FF_{c}$ loop for the
third node, and thus the numbers of three feedforward loops are equal
in directed networks. Compared to the unclosed triples,
the closed triples play a more important role in
dynamical processes on online social
networks~\cite{mangan2003structure,mangan2006incoherent}, such as
opinion formation~\cite{sousa2005consensus}, game
dynamics~\cite{ghoneim2008characterizing}, and cooperation
evolution~\cite{hales2008motifs}.

In online social networks, the closed triples are a good
indicator of close-knit friendships among people.  To understand
the mesoscale structural properties of online social
networks, we analyze data of popular online
social networks, establish the empirical facts, and introduce a
directed network model.  We analyze four
large-scale online social networks, namely \textit{Epinions},
\textit{Slashdot}, \textit{Flickr}, and \textit{Youtube}, and
establish that the distributions in each scale follow
a similar power law. We propose a simple directed network model
incorporating two processes: external reciprocation and internal evolution.
Theoretical analysis shows that the distributions of four closed triples
display almost identical scaling laws due to the negligible
degree correlations, and the distribution exponents depend only on one global
parameter - the mean in/outdegree. Simulation results based on the
model are basically consistent with both the
empirical results and theoretical analysis.

\section*{Results}

\subsection*{Empirical Results}

We first analyze four representative directed
online social networks and establish the
empirical features.  As listed in
Table~\ref{tab:basiccharacteristic}, these four datasets are: (i)
\textit{Epinions} Social
Network (ESN, http://snap.stanford.edu\\/data/soc-Epinions1.html)~\cite{leskovec2010signed}: a who-trust-whom online social
network of a general consumer review site \textit{Epinions.com}
in which members can decide whether to ``trust" each other or
not, and subsequently all the trusted relationships
form a so-called social trust
network. (ii) \textit{Slashdot} Social
Network (SSN, http://snap.stanford.edu/data/soc-Slashdot0902.html)~\cite{leskovec2010signed}: a friendship network of a
technology-related news website \textit{Slashdot.com}. Nodes are
the users and links represent the
friendships among the users. (iii) \textit{Flickr}
Social Network (FSN, http://socialnetworks.mpi-sws.org/data-imc2007.html)~\cite{mislove2007measurement}: a friendship network of a
photo-sharing site \textit{Flickr.com} that allows users to
designate others as ``contacts" or ``friends" and track their
activities in real time. This network contains all the friendship
links among the users of Flickr. (iv) \textit{YouTube} Social
Netowrk (YSN, http://socialnetworks.mpi-sws.org/data-imc2007.html)~\cite{mislove2007measurement}: a friendship network of a
popular video-sharing website \textit{YouTube.com} on which users
can upload, share and view videos.  The nodes in the network are
the users of YouTube, and a directed link is established from a
user $A$ to a user $B$ when user $A$ declares user $B$ as a
friend.  Table~\ref{tab:basiccharacteristic} summarizes the basic
global features of the four online social networks.  These networks
all show a large reciprocity $r$, defined by
$r=E_r/(E-E_r)$~\cite{garlaschelli2004patterns} with $E_r$ and
$E$ being the numbers of reciprocal links and
single directed links,
respectively. Note that a reciprocal link
contributes two single directed links.  For example,
$r\approx0.25$ for ESN, $r\approx0.73$ for SSN, $r\approx0.45$
for FSN, and $r\approx0.65$ for YSN.

We also studied the local-scale structural properties of these
social networks via statistical distributions.
The results of ESN are presented as an example. Figure~\ref{kin-and-kout-of-Epinions}
shows the indegree and outdegree distributions
(black squares) on a log-log plot.
The data span more than two decades.  The
distributions follow a power law with
approximately the same exponent, \textit{i.e.},
~$P(k_{in})\sim k_{in}^{-\gamma_{in}}$ and $P(k_{out})\sim
k_{out}^{-\gamma_{out}}$, with
$\gamma_{in}\approx1.73$ and $\gamma_{out}\approx1.71$ obtained
by the maximum likelihood
estimation\cite{clauset2009power,stumpf2012critical}.  More
details about the power-law fits are given
in Table S1 of Appendix SI.
Figure~\ref{kintokout-of-Epinions} shows that the
indegree~$k_{in}$ of each node is nearly proportional to its
outdegree~$k_{out}$ (also see Figures S4-S6 of Appendix SI), which is consistent with the
similar scaling law of their
distributions. In growing networks, the fat-tail
power-law behavior in the degree distribution  suggests that
directed links are not drawn toward and from existing users
uniformly. Mislove \textit{et al.} showed that there is a
positive correlation between the number of links a user has and
its probability of creating or receiving new links in online
social networks~\cite{mislove2008growth}.  This phenomenon is
called ``preferential
attachment''~\cite{barabasi1999emergence,krapivsky2001degree,mislove2008growth,capocci2006preferential}.
The behavior $k_{in}\approx k_{out}$ for any node
implies that a node with large $k_{in}$ has a
strong ability to attract links from other nodes and also
a strong tendency to link to other nodes. This is
reminiscences of the product $k_{out}^{i}k_{in}^{j}$ used in the
prediction of a link between the nodes $i$ and
$j$~\cite{l¨¹2011link}, \textit{i.e.}, a larger product gives a larger
probability of having a directed link from $i$ to $j$.  These
results lead us to incorporate a preferential attachment
mechanism related to $k_{out}^{i}k_{in}^{j}$ into the mechanism
of how the links grow in a network.

The reciprocal degree is the number of reciprocal
links that a node possesses.
Figure~\ref{kr-of-Epinions} shows that the reciprocal degree
distribution also follows a power law $P(k_{r}) \sim
k_{r}^{\gamma_{r}}$ with an exponent
$\gamma_{r}\approx1.69$ as examined by the maximum likelihood
estimation\cite{clauset2009power,stumpf2012critical}, similar to
that of the indegree and outdegree distributions.
Figure~\ref{kinavebiedge-koutavebiedge-Epinions}
shows that the mean reciprocal degree of the
nodes with the same indegree  $\langle k_{r}(k_{in})\rangle$ is
approximately linearly proportional to the indegree $k_{in}$
(also see Figures S10-S12 of Appendix SI), \textit{i.e.}, $\langle
k_{r}(k_{in})\rangle\sim k_{in}$, and in a
similar fashion $\langle k_{r}(k_{out})\rangle\sim k_{out}$,
implying that the probability that a randomly chosen directed
link happens to be a reciprocal link is roughly a constant.
All these features are consistent with the
observation that the indegree, outdegree, and
reciprocal degree distributions all follow a similar exponent.

For mesoscale structures, we focus on the
four closed triples \textit{i.e.}, $FB$, $FF_{a}$, $FF_{b}$
and $FF_{c}$. As the numbers of three feedforward loops are equal, \textit{i.e.},
$N_{FF_{a}}=N_{FF_{b}}= N_{FF_{c}}$, we only look at the total
numbers of $FB$ and $FF_{a}$ closed triples.  For ESN,
$N_{FB}=740,310$ and $N_{FF_{a}}=3,586,403$ as shown in
Table~\ref{tab:basiccharacteristic}.
Considering the feedforward loops as the same up
to the permutation of the focal node, it is interesting to see
that $N_{FB}:N_{FF_{a}}\!\approx\!1:5$.  This implies the
existence of some underlying mechanism. Since the indegree and outdegree
distributions are heterogeneous, we study the numbers of
the four closed triples~(\textit{i.e.},
$n_{FB}, n_{FF_{a}}, n_{FF_{b}}$, and $n_{FF_{c}}$)~at different
nodes and their distributions.
Figure~\ref{four-triangle-of-Epinions} shows that, although
the numbers of feedback and feedforward loops are
different, their distributions follow similar
scaling laws, \textit{i.e.}, $P(n_{FB})\sim
{n_{FB}}^{-{\gamma}_{FB}}$ and $P(n_{FF})\sim
{n_{FF}}^{-{\gamma}_{FF}}$, with $\gamma_{FB}\approx1.37$,
$\gamma_{FF_{a}}\approx1.39$, $\gamma_{FF_{b}}\approx1.35$ and
$\gamma_{FF_{c}}\approx1.38$ as determined by the maximum
likelihood estimation\cite{clauset2009power,stumpf2012critical}.
More details on the exponents are given
in Table S1 of Appendix SI. Moreover,
although the numbers of three feedforward loops are equal, their
distributions look slightly different in detail. This is a
phenomenon worthy of further research.

To understand this phenomenon, we consider the three unclosed
triples in Figure~\ref{four-triangles}. For a node with indegree
$k_{in}$ and outdegree $k_{out}$, there are $C_{k_{out}}^{2}$
unclosed triples $A$, $C_{k_{in}}^{1}C_{k_{out}}^{1}$ unclosed
triples $B$, and $C_{k_{in}}^{2}$ unclosed triples $C$
when reciprocal links are forbidden, where
$C_{n}^{m}=n!/[m!(n-m)!]$ denotes the binomial coefficient.
These unclosed triples
would generate closed triples in the ratio
$n'_{FB}:n'_{FF_a}=(C_{k_{in}}^{1}C_{k_{out}}^{1}/2):(C_{k_{out}}^{2}+C_{k_{in}}^{2}+C_{k_{in}}^{1}C_{k_{out}}^{1}/2)$.
Accounting for all the nodes, we can obtain the
total number of optional closed
triples~$N'_{FB}=\sum_{i}^{N}(C_{k_{in}^{i}}^{1}C_{k_{out}^{i}}^{1}/2)$ and
$N'_{FF_a}\!=\!\sum_{i=1}^{N}(C_{k_{out}^{i}}^{2}+C_{k_{in}^{i}}^{2}+C_{k_{in}^{i}}^{1}C_{k_{out}^{i}}^{1}/2)$, respectively.
Assuming there is no degree correlation and making use of
$k_{in}\approx k_{out}$, we have $N_{FB}:N_{FF_a}\approx1:5$,
which is basically consistent with the ratio found in ESN.
The assumption of no degree distribution is supported by
the results in Figure~\ref{knn-of-Epinions}(a), in which the network shows a very weak degree correlation over two decades that can be treated almost as no degree correlation (further quantitative evidence is given by the Pearson correlation coefficient in Table S2 of Appendix SI)\cite{newman2002assortative}. In this case, the number of closed
triples at a node depends only on its indegree
$k_{in}$ and outdegree $k_{out}$, \textit{i.e.}, $n_{FB}\!\sim\!
k_{in}^{2}$ and $n_{FF_a}\!\sim\! k_{in}^{2}$ for large $k_{in}$,
$n_{FB}\!\sim\!k_{out}^{2}$ and $n_{FF_a}\!\sim\!k_{out}^{2}$ for
large $k_{out}$.  This behavior is confirmed in
Figure~\ref{kinAveTriangle-of-Epinions} and
Figure~\ref{koutAveTriangle-of-Epinions} (also see Figures S19-S24 of Appendix SI). This
also gives the reason why the distributions of four closed
triples follow similar scaling laws.
Results of analyzing the other three
networks (\textit{i.e.}, \textit{Slashdot}, \textit{Flickr} and
\textit{YouTube}) also exhibit similar
phenomena (see Figures S1-S24 of Appendix SI).

\subsection*{Directed Network Model}
We propose a growing
network model with node and link creation processes incorporating
link directionality that reproduces the empirical
features.  In the model, we consider two evolutionary
ingredients: reciprocation and preferential attachment. On one
hand, many empirical results
show that the reciprocity $r$ of online social networks is much
greater than in sparse random directed networks with
$r\!\rightarrow\!0$
~\cite{garlaschelli2004patterns,mislove2008growth}.
Our results of $r\approx0.45$ of FSN and
$r\approx0.65$ of YSN provide further evidence.
The high reciprocity implies that there is a good chance that the creation
of a directed link prompts the establishment of a reversed link.
For example, users of Flickr
often respond to an incoming link by quickly establishing a
reversed link as a matter of
courtesy~\cite{mislove2008growth}. Thus, reciprocation is
believed to be an independent growth mechanism in large-scale
online social networks. On the other hand, preferential
attachment has been proven to be an important and basic growing
mechanism in online social
networks\cite{barabasi1999emergence,krapivsky2001degree,mislove2008growth,capocci2006preferential}.
Users with large indegrees and outdegrees are more likely to receive
incoming links and create outgoing links, respectively.  This
motivated us to incorporate a preferential attachment mechanism
depending on the product~$k_{out}^{i}k_{in}^{j}$ in creating new links.

The model starts with an initial seed consisting
of $m_{0}$ nodes. At each time step, a new node is added
and $2+m+mp$ new directed
links are introduced according to two processes:
external reciprocation and internal evolution.

(1) External reciprocation. The new
node in every time step establishes a new
directed link with an existing nodes $i$ in
the network with a probability
\begin{equation}
p_{i}=\frac{k_{in}^{i}}{\sum_{j}k_{in}^{j}}
\end{equation}
proportional to the indegree
$k_{in}^{i}$ of node $i$.  To incorporate the reciprocation
mechanism, the node $i$ that receives the link
creates a reversed link to the new node. Consequently, a reciprocal link is created between
these two nodes. This mechanism is reasonable in
that a strong motivation of a new user joining a social network
is to get connected to and interact with someone already in the
network. As we shall see, this process can be treated
conveniently in the mathematical analysis of the model.


(2) Internal evolution. In each time step,
$m$ new directed links, representing the
activity of the network, are created among the existing
nodes according to the preferential attachment mechanism. Consider
two unconnected nodes $i$ and $j$ up to that time
step, a new directed link from node $i$ to node $j$ is created with the
probability
\begin{equation}\label{pij}
p_{ij}=\frac{k_{out}^{i}k_{in}^{j}}{\sum_{x,y,x\notin\Gamma_{in}(y)}k_{out}^{x}k_{in}^{y}},
\end{equation}
where $k_{out}^{i}$ and $k_{in}^{j}$ are the outdegree of node
$i$ and the indegree of the target node $j$, respectively, and
$\Gamma_{in}(y)$ in the normalization factor is the set of
incoming neighbors of node $y$ at that time step.  This attachment
probability is proportional to the
product~$k_{out}^{i}k_{in}^{j}$. The larger the product is, the
greater probability a new directed link is created between them.
For each of the new directed links created, a
reversed link will be established with the reciprocation
probability $p$.  Therefore, $m+mp$ directed links are introduced
into the network through internal evolution in each time step. It
should be noted that multiple links between two
nodes and self-connections are prohibited in the model.

\section*{Materials and Methods}

\subsection*{Rate Equation Analysis}
We first analyze the indegree and
outdegree distributions of the model. After $t$ steps, the
growing directed network has $N\!=\!m_{0}\!+\!t$ nodes and
$(2\!+\!m\!+\!mp)t$ directed links, where the
tiny number of initial links in the seed are ignored. Meanwhile, the
sum of indegree and the sum of outdegree are equal, \textit{i.e.},
$\sum_{j}k_{in}^{j}\!=\!\sum_{j}k_{out}^{j}\!=\!(2\!+\!m\!+\!mp)t$.
For a sparse network with mean indegree~$\langle
k\rangle\!=\!2\!+\!m\!+\!mp\!<<\!N$, we have
$\sum_{x,y,x\notin\Gamma_{in}(y)}k_{out}^{x}k_{in}^{y}\approx\sum
k_{out}^{x}\times\sum k_{in}^{y}=[(2+m+mp)t]^2$ so that
Eq.~\eqref{pij} can be approximated by
\begin{equation}
p_{ij}\approx\frac{k_{out}^{i}k_{in}^{j}}{[(2+m+mp)t]^{2}}.
\end{equation}
Consider the creation of one new
directed link via the internal evolution at
step~$t$.  The probability $p_{k_{in}^{i}}^{+}$ that the indegree
$k_{in}^{i}$ of node $i$ increases by one due to
the creation of one link is
\begin{equation}
p_{k_{in}^{i}}^{+}=\sum_{j\notin\Gamma
_{in}(i)}\frac{k_{out}^{j}k_{in}^{i}}{[(2+m+mp)t]^{2}}
+p\sum_{j\notin\Gamma_{out}(i)}\frac{k_{out}^{i}k_{in}^{j}}{[(2+m+mp)t]^{2}},
\end{equation}
where the first term gives the probability that the node $i$
receives a new incoming link from one of the other nodes and the
second term gives the probability that a reversed
link is created back to node $i$ when a new directed link was
created from node $i$ to some node $j$. According to
$\sum_{j\notin\Gamma_{in}(i)}k_{out}^{j}\simeq
\sum_{j\notin\Gamma_{out}(i)}k_{in}^{j}\approx(2+m+mp)t$,
$p_{k_{in}^{i}}^{+}$ is approximately given by
\begin{equation}\label{add indegree}
p_{k_{in}^{i}}^{+}\approx\frac{k_{in}^{i}+pk_{out}^{i}}{(2+m+mp)t}.
\end{equation}
Similarly, the probability $p_{k_{out}^{i}}^{+}$ that
the outdegree $k_{out}^{i}$ of
node $i$ increases by one due to the creation of one link is
\begin{equation}\label{add outdegree}
p_{k_{out}^{i}}^{+}\approx\frac{k_{out}^{i}+
pk_{in}^{i}}{(2+m+mp)t}.
\end{equation}
Equations for the rate of change of the expected indegree~$k_{in}^{i}$~and
outdegree~$k_{out}^{i}$ can then be written down.
Taking $k_{in}^{i}$~and~$k_{out}^{i}$ as continuous variables, the
dynamical equations are
\begin{eqnarray}\label{Continuum theory1}
\frac{dk_{in}^{i}(t)}{dt}&=&\frac{k_{in}^{i}}{\sum_{j}k_{in}^{j}}+mp_{k_{in}}^{+},\nonumber \\
\frac{dk_{out}^{i}(t)}{dt}&=&\frac{k_{in}^{i}}{\sum_{j}k_{in}^{j}}+mp_{k_{out}}^{+},
\end{eqnarray}
where the first term in the equations comes from
the newly added node in a time step. The
difference of the two equations gives
\begin{equation}\label{Continuum theory2}
\frac{d[k_{in}^{i}(t)-k_{out}^{i}(t)]}{dt}=
mp_{k_{in}}^{+}-mp_{k_{out}}^{+}
=\frac{m(1-p)(k_{in}^{i}-k_{out}^{i})}{(2+m+mp)t},
\end{equation}
where Eqs.~\eqref{add indegree} and~\eqref{add
outdegree} have been used.
Let $t_{i}$ be the time that the node $i$ is
introduced, \textit{i. e.,} $k_{in}^{i}(t_{i}) =
k_{out}^{i}(t_{i}) = 1$. It
follows from Eq.~\eqref{Continuum theory2} that
~$k_{in}^{i}(t)=k_{out}^{i}(t)$~at any time $t$. Although the expected
value of the difference between indegrees and outdegrees of a
node does not grow over time mathematically, the difference does
exist in a particular realization of the model in simulations.
Eq.~\eqref{Continuum theory1}
and the initial condition
$k_{in}^{i}(t_i)\!=\!k_{out}^{i}(t_i)\!=\!1$
gives
\begin{equation}\label{solutions}
k_{in}^{i}(t)=k_{out}^{i}(t)=(\frac{t}{t_i})^{\beta},
\end{equation}
where~$\beta=(1+m+mp)/(2+m+mp)$. The indegree and
outdegree of the nodes both grow over time in the same functional
form, with older nodes having higher indegrees and outdegrees.

Let $N_{k_{in}}(t)$ and $N_{k_{out}}(t)$ be the number of
nodes with expected indegree $k_{in}$ and
outdegree $k_{out}$ at the time step $t$, respectively.
The rate equation of $N_{k_{in}}(t)$
is then given by
\begin{equation}\label{dkin/dt}
\frac{dN_{k_{in}}(t)}{dt}=\frac{k_{in}-1}{\sum_{j}k_{in}^{j}}N_{k_{in}-1}-\frac{k_{in}}{\sum_{j}k_{in}^{j}}N_{k_{in}}
+mp_{k_{in}-1}^{+}N_{k_{in}-1}-mp_{k_{in}}^{+}N_{k_{in}}+\delta_{k_{in},1}.
\end{equation}
The first and third terms on the right-hand side account for the
increase of $N_{k_{in}}(t)$ due to the external reciprocation and
internal evolution, respectively; and the second and fourth terms
account for the decrease due to the processes.  The last term
accounts for the introduction of a new node with indegree
$k_{in}\!=\!1$ at time $t$. Eq.~\eqref{dkin/dt} is
valid for all $k_{in}\!\ge\!1$.

After many steps $t$, there are $N = m_{0}+t
\!\approx\!t$ nodes in the network. In the asymptotic limit, we
substitute $N_{k_{in}}(t)=tP(k_{in})$, where $P(k_{in})$ is the
indegree distribution~\cite{krapivsky2000connectivity}, and
$\sum_{j}k_{in}^{j}\!=\!(2\!+\!m\!+\!mp)t$ into
Eq.~\eqref{dkin/dt} to obtain the simple recursive relation
\begin{equation}\label{kinrelation}
[2+m+mp+(1+m+mp)k_{in}]P(k_{in})
=(1+m+mp)(k_{in}-1)P(k_{in}-1)+(2+m+mp)\delta_{k_{in},1}.
\end{equation}
Using the initial condition that $k_{in}=1$ at the time that a
node was introduced, the solution of
Eq.\eqref{kinrelation} is
\begin{equation}\label{kinsolutioneuqal}
P(k_{in})=A\frac{\Gamma(k_{in})}{\Gamma(k_{in}+2+\frac{1}{1+m+mp})},
\end{equation}
where $A=\frac{2+m+mp}{3+2m+2mp}\Gamma(\frac{1}{1+m+mp}+3)$
and $\Gamma$ is the Euler gamma function. Using the
asymptotic form $\Gamma(x+\lambda)\rightarrow x^\lambda$ as
$x\rightarrow\infty$, we can extract the scaling form
\begin{equation}\label{kinsolution}
P(k_{in})\approx Ak_{in}^{-(2+\frac{1}{1+m+mp})}.
\end{equation}

Similarly, the rate equation of $N_{k_{out}}(t)$ is given by
\begin{equation}\label{dkout/dt}
\frac{dN_{k_{out}}(t)}{dt}
=\frac{k_{in}-1}{\sum_{j}k_{in}^{j}}N_{k_{out}-1}-
\frac{k_{in}}{\sum_{j}k_{in}^{j}}N_{k_{out}}
+mp_{k_{out}-1}^{+}N_{k_{out}-1}-mp_{k_{out}}^{+}N_{k_{out}}
+\delta_{k_{out},1}.
\end{equation}
The first (second) and third (fourth) terms on the right-hand
side account for the increase (decrease) in
$N_{k_{out}}$ due to the external reciprocation and internal
evolution, respectively; and the last term accounts for the introduction of a
new node with $k_{out}=1$ at time $t$.  Substituting
$N_{k_{out}}(t)=tP(k_{out})$, where $P(k_{out})$ is the outdegree
distribution, and $\sum_{j}k_{in}^{j}=(2+m+mp)t$ into
Eq.~\eqref{dkout/dt}, the recursive relation for
$P(k_{out})$ is
\begin{equation}\label{koutrelation}
[2+m+mp+(1+m+mp)k_{out}+1]
P(k_{out})=(1+m+mp)(k_{out}-1)P(k_{out}-1)+(2+m+mp)\delta_{k_{out}1},
\end{equation}
which is identical to
Eq.~\eqref{kinrelation} for $P(k_{in})$.  It follows that
\begin{equation}\label{koutsolution}
P(k_{out})\approx Ak_{out}^{-(2+\frac{1}{1+m+mp})}.
\end{equation}
The results show that the expected indegree and
outdegree grow over time following the same functional form of
Eq.~\eqref{solutions}, and the indegree and outdegree
distributions follow the same scaling law with an exponent
\begin{equation}\label{gamma}
\gamma=2+\frac{1}{1+m+mp}.
\end{equation}

Next, we consider the
reciprocal degree distribution $P(k_r)$. For a
node $i$ with~$k_{in}^{i}=k_{out}^{i}$,
$k_r^{i}$ satisfies the dynamical equation
\begin{equation}\label{kr}
\frac{dk_{r}^{i}(t)}{dt}= \frac{k_{in}^{i}(t)}{(2+m+mp)t}
+\frac{mpk_{in}^{i}(t)+mpk_{out}^{i}(t)}{(2+m+mp)t}.
\end{equation}
Substituting Eq.~\eqref{solutions} into Eq.~\eqref{kr} and using
the initial condition that $k_{r}^{i}(t_i)=1$ at
the time that node $i$ was introduced, the solution to Eq.~\eqref{kr} is
\begin{equation}\label{kr-solution}
k_{r}^{i}(t)=\frac{2mp+1}{1+m+mp}[k_{in}^{i}(t)-1]+1.
\end{equation}
For large~$k_{in}^{i}$, we have
\begin{equation}\label{kr-scaling}
k_{r}^{i}\sim\frac{2mp+1}{1+m+mp}k_{in}^{i}.
\end{equation}
Using $P(k_{r})dk_{r}=P(k_{in})dk_{in}$, the distribution
$P(k_{r})$ follows
\begin{equation}\label{r-exponent}
P(k_{r})\sim {k_{r}}^{-\gamma},
\end{equation}
where $\gamma$ is given by Eq.~\eqref{gamma}
as for the indegree and outdegree distributions.

Furthermore, we analyze the degree correlations between connected
nodes by the rate equation approach.
Let $N^{l_{out}}_{k_{in}}$ be the number of
links that originate from a
node with an expected outdegree $l_{out}$
to a node with an expected indegree
$k_{in}$~\cite{krapivsky2001organization}.  Generally,
$P^{l_{out}}_{k_{in}}$ is defined for $k_{in}\geq 1$ and
$l_{out}\geq2$. The quantity $N^{l_{out}}_{k_{in}}(t)$ evolves
according to
\begin{equation}\label{degreecorrelation}
\begin{aligned}
\frac{dN^{l_{out}}_{k_{in}}(t)}{dt}
  &= \frac{(k_{in}-1)N^{l_{out}}_{k_{in}-1}-k_{in}N^{l_{out}}_{k_{in}}}{\sum{k_{in}N_{k_{in}}}}
   +\frac{(l_{out}-1)N^{l_{out}-1}_{k_{in}}-l_{out}N^{l_{out}}_{k_{in}}}{\sum{k_{in}N_{k_{in}}}}
   +\frac{(l_{out}-1)N_{l_{out}-1}}{\sum{k_{in}N_{k_{in}}}}\delta_{1,k_{in}}\\
  &+ mp_{k_{in}-1}^{+}N^{l_{out}}_{k_{in}-1}-mp_{k_{in}}^{+}N^{l_{out}}_{k_{in}}
   + mp_{l_{out}-1}^{+}N^{l_{out}-1}_{k_{in}}-mp_{l_{out}}^{+}N^{l_{out}}_{k_{in}}\\
  &+(m+mp)\frac{(l_{out}-1)(k_{in}-1)}{\sum_{x,y,x\notin\Gamma_{in}(y)}k_{out}^{x}k_{in}^{y}}N_{l_{out}-1}N_{k_{in}-1}
   -(m+mp)\frac{l_{out}k_{in}}{\sum_{x,y,x\notin\Gamma_{in}(y)}k_{out}^{x}k_{in}^{y}}N_{l_{out}}N_{k_{in}},
\end{aligned}
\end{equation}
where the first two terms on the
right-hand side account for the changes due to the
introduction of a new node,
including the gains when the new node is
connected to a node with indegree $(k_{in}-1)$ (outdegree
$(l_{out}-1)$) which is already connected to a node with
outdegree $l_{out}$ (indegree $k_{in}$), and the losses when the
new node is connected to either end of a link that connects a
node with outdegree $l_{out}$ and another node with indegree
$k_{in}$. The third term accounts for the gain in
$N_{1}^{l_{out}}$ due to the addition of the new node. The
remaining terms take into account the changes
due to the internal evolution process with the
introduction of $m+mp$ directed links.

Asymptotically,
$N^{l_{out}}_{k_{in}}\!\to\!{(2+m+mp)tP^{l_{out}}_{k_{in}}}$,
$N_{k_{in}}\!\to\!{tP(k_{in})}$ and
$N_{l_{out}}\!\to\!{tP(l_{out})}$.  Considering
$\sum{k_{in}N_{k_{in}}}=\sum_{j}{k_{in}^{j}}=(2\!+\!m\!+\!mp)t$
and
$\sum_{x,y,x\notin\Gamma_{in}(y)}k_{out}^{x}k_{in}^{y}\approx\sum
k_{out}^{x}\times\sum k_{in}^{y}=[(2+m+mp)t]^2$,
Eq.~\eqref{degreecorrelation} gives a recursive relation
\begin{equation}\label{correlationrelation}
\begin{aligned}
{[2+m+mp+(1+m+mp)(k_{in}\!+\!l_{out})]}P^{l_{out}}_{k_{in}}
   &=(1\!+\!m\!+\!mp)[(k_{in}\!-\!1)P^{l_{out}}_{k_{in}\!-\!1}\!+(l_{out}\!-\!1)P^{l_{out}-1}_{k_{in}}]\\
   &+\frac{1}{2+m+mp}(l_{out}\!-\!1)P(l_{out}\!-\!1)\delta_{k_{in},1}\\
   &\!+\!\frac{m+mp}{(2\!+\!m\!+\!mp)^2}[(l_{out}\!-\!1)(k_{in}\!-\!1)P(l_{out}\!-\!1)P(k_{in}\!-\!1)\!\\
   &-\!l_{out}k_{in}P(l_{out})P(k_{in})].
\end{aligned}
\end{equation}
Solving Eq.~\eqref{correlationrelation} directly for
$P^{l_{out}}_{k_{in}}$ is difficult, however, it is observed that
decomposing $P^{l_{out}}_{k_{in}}$ into
\begin{equation}\label{scaledrelation}
P^{l_{out}}_{k_{in}}\sim l_{out}P(l_{out})k_{in}P(k_{in}),
\end{equation}
with $P(l_{out})$ given by Eq.~\eqref{koutsolution} and
$P(k_{in})$ given by Eq.~\eqref{kinsolution} satisfies
Eq.~\eqref{correlationrelation} in the scaling regime, as one can
readily show by substituting Eq.~\eqref{scaledrelation} into
Eq.~\eqref{correlationrelation} and taking the limits of
$l_{out}\!\to\!\infty$ and $k_{in}\!\to\!\infty$.
Eq.~\eqref{scaledrelation} implies that there is no degree
correlation, a feature that is supported by the empirical results in
Figure~\ref{knn-of-Epinions} for ESN over a wide range of degrees (also see Figures S16-S18 of Appendix SI). It also follows from
$k^{i}_{in}= k^{i}_{out}$ and Eq.~\eqref{scaledrelation} that
$P^{l_{out}}_{k_{in}}=P^{l_{out}}_{k_{out}}=P^{l_{in}}_{k_{in}}=P^{l_{in}}_{k_{out}}$.
Interpreting $P_{k_{in}}^{l_{out}}$ as a joint probability, the lack of degree correlation as expressed in Eq.~(\ref{scaledrelation}) implies that the conditional probability
\begin{equation}\label{conditional probability}
P(k_{in}|l_{out})
\sim k_{in}P(k_{in}),
\end{equation}
which is independent of $l_{out}$.
For a node $i$ with large $k_{in}^{i}=k_{out}^{i}$, the average
nearest neighbor function can be calculated as
\begin{equation}\label{knn}
k_{in}^{nn}(k_{out})=\sum_{k'_{in}}k'_{in}P(k'_{in}|k_{out}) \sim
\sum_{k'_{in}} {k'}_{in}^{2} P(k'_{in}),
\end{equation}
which is also independent of $k_{out}$.  This is consistent with
the behavior of $k_{in}^{nn}(k_{out})$ in ESN, as shown in
Figure~\ref{knn-of-Epinions}.

The number of $FB$ loops can be formally written
as~\cite{catanzaro2005generation}
\begin{equation}\label{nFB}
n_{FB}=\frac{C_{k_{in}}^{1}C_{k_{out}}^{1}}{2}\sum_{k'_{in},k''_{in}}P(k'_{in}|k_{out})P(k''_{out}|k_{in})P_{k'_{in}}^{k''_{out}},
\end{equation}
where $P_{k'_{in}}^{k''_{out}}$ is the probability that a link
connects a node with outdegree $k''_{out}$ to a node with
indegree $k'_{in}$. The lack of degree correlations
makes the summations independent of $k_{in}$ and $k_{out}$, and thus $n_{FB}$ scales as
\begin{equation}\label{nFB-scaling}
n_{FB}\sim{k_{in}}^{2}.
\end{equation}
Similarly, the numbers of four closed triples
$n_{\Delta}$ at a node with large indegree and outdegree follow
the scaling behavior $n_{\Delta}\sim k_{in}^{2}$ or
$n_{\Delta}\sim k_{out}^{2}$. Combining $n_{\Delta}\sim
k_{in}^{2}$ with Eq.~\eqref{kinsolution} ($P(k_{in})\sim
k_{in}^{\gamma_{in}}$), the distributions of four closed triples
have the same scaling behavior as follows:
\begin{equation}\label{P(n_Delta)}
P(n_{\Delta})\sim n_{\Delta}^{\gamma_{\Delta}},
\end{equation}
where the exponent $\gamma_{\Delta}$ can be readily found by
using $P(n_{\Delta})dn_{\Delta}=P(k_{in})dk_{in}$ to be
\begin{equation}\label{gamma-Delta}
\gamma_{\Delta}=\frac{3}{2}+\frac{1}{2(1+m+mp)}.
\end{equation}
The exponent $\gamma_{\Delta}$ is determined by the parameters
$m$ and $p$ and it falls into the range $(1.5,\,2]$.

\subsection*{Simulation Results}
We also carried out numerical simulations to
study the structural properties of the model and compared results
with data of real online social networks. The activity $m$ and reciprocation
probability $p$ are two important parameters of the model.  They
determine the reciprocity $r\!=\!(1\!+\!mp)/(1\!+\!m)$ and mean
indegree $\langle k\rangle=2\!+\!m\!+\!mp$ of simulated
networks.  In order to compare results with real online social
networks, we take three parameters from real data, namely the
number of nodes $N$, the reciprocity $r$ and the mean indegree
(outdegree) $\langle k \rangle$, and determine the parameter $m$
and $p$ in the model through
\begin{eqnarray}\label{m-p}
m&=&\frac{\langle k\rangle}{1+r}\!-\!1;\nonumber \\
p&=&\frac{\langle k\rangle r\!-\!r\!-\!1}{\langle k\rangle \!-\!r\!-\!1}.
\end{eqnarray}

Taking ESN as an example, we have $\langle k\rangle\approx6.7$,
$r\approx0.25$, and $N=75879$.  The model parameters are then
fixed at $m\approx4.34$ and $p\approx0.08$
according to Eq.\eqref{m-p}. With the values of
$m$ and $p$, a network of $N=75879$ nodes is simulated.  For a
non-integer value of $m$, it is implemented in a probabilistic
way.  For ESN with $m=4.34$, for example, the initiation of the
fifth new directed link through the internal evolution process is
implemented with a probability $0.34$ after establishing four new
directed links in every time step. The structural properties
of the simulated network are analyzed for each of the quantities
studied for the real data. Results are shown in
Figures~\ref{kin-and-kout-of-Epinions}-\ref{koutAveTriangle-of-Epinions}
as red circles for comparison (also see Figures S1-S24 of Appendix SI).  The model basically reproduces
the key properties of ESN.

For the indegree and outdegree distributions (see
Figure~\ref{kin-and-kout-of-Epinions}) and the reciprocal degree
distribution (see Figure~\ref{kr-of-Epinions}), the simulation
results also show similar scaling law, with the
exponents $\gamma_{in}\approx1.95$, $\gamma_{out}\approx1.96$ and
$\gamma_{r}\approx2.1$ determined by the maximum likelihood
estimation\cite{clauset2009power,stumpf2012critical} (see
Table S1 of Appendix SI for more detail). These values are slightly larger
than the corresponding values of the exponents in ESN.
According to
Eqs.~\eqref{gamma},~\eqref{r-exponent} and~\eqref{m-p}, these
exponents are equal and the theoretical value is
~$\gamma=2+1/(\langle k\rangle-1)\approx2.17$.
Note that the rate equation analysis assumes an
infinite system. The difference between the simulated results
and the theoretical value comes from the finite size of simulated
network, as well as the approximations made in getting at the
values of the exponent. The indegree and outdegree
distributions of simulated network are in reasonable agreement with
the empirical results of ESN. The model, however, gives a
reciprocal degree distribution smaller than the ESN empirical
results over a wide range of $k_{r}$. This discrepancy implies
that there are some network growing mechanisms in ESN that are
not included in the model, \textit{e.g.}, different reciprocation
probabilities for different nodes \cite{gallos2012people}.
This, together with a possibly very weak degree correlation in Figure~\ref{knn-of-Epinions} that we ignored, may be the reason for the simulation results in
Figures~\ref{kintokout-of-Epinions} and~\ref{kinavebiedge-koutavebiedge-Epinions} to be bigger than the empirical values for large in/outdegrees,
and for the small differences in the tails in Figures~\ref{kin-and-kout-of-Epinions} and \ref{kr-of-Epinions}\cite{wu2010evidence,mossa2002truncation}.

For the distributions of the four closed triples,
the distributions from simulations follow a
power-law behavior with almost the same exponent (see
Figure~\ref{four-triangle-of-Epinions}), where
$\gamma_{FB}\approx1.47$, $\gamma_{FF_{a}}\approx1.46$,
$\gamma_{FF_{b}}\approx1.46$ and $\gamma_{FF_{c}}\approx1.46$ as
determined by the maximum likelihood estimation.  These values
are slightly larger than the exponents found in ESN.
Theoretically, $\gamma_\Delta=3/2+1/[2(\langle
k\rangle-1)]\approx1.58$ according to Eqs.~\eqref{gamma-Delta}
and~\eqref{m-p}. We note
that the theoretical values of both $\gamma$ and
$\gamma_\triangle$ depend only on the mean indegree $\langle
k\rangle$, which in turn is determined by the two model
parameters $m$ and $p$. Figure~\ref{exponents} shows
the values of all the $\gamma$-exponents of the distributions for
the four online social networks and the corresponding simulated
networks, which are determined by the maximum likelihood estimation.

The two parameters $m$ and $p$ affect the
reciprocal degree of nodes $k_r$ through Eq.\eqref{kr-scaling}.
Substituting Eq.~\eqref{m-p} into Eq.~\eqref{kr-scaling},
we have $k_r\sim(2\langle k\rangle
r-r-1)k_{in}/[(1+r)(\langle k\rangle-1)] \approx0.3k_{in}$ for
ESN.  The reciprocal degree $k_{r}$ of a node and its $k_{in}$
are related by a factor depending on the two
global parameters $\langle k\rangle$~and $r$.
This linear relationship between $k_{r}$ and
$k_{in}$ ($k_{out}$) with a slope $0.3$ is observed in simulation
results, as shown in
Figure~\ref{kinavebiedge-koutavebiedge-Epinions}, but the ESN
data show a faster increase of $k_{r}$ with $k_{in}$ and
$k_{out}$. When the network has a larger
reciprocity, such as $r\approx0.73$ for \textit{Slashdot},
$r\approx0.45$ for \textit{Flicker}, and $r\approx0.65$ for
\textit{YouTube}, a better agreement is observed (see
Figures S10-S12 of Appendix
SI). Despite some small differences in the tail
in Figures~\ref{kinAveTriangle-of-Epinions} and
\ref{koutAveTriangle-of-Epinions}, which may be caused by local
proximity bias in link creation~\cite{mislove2008growth},
simulation results for the dependence of the number of closed
triples with $k_{in}$ and $k_{out}$ are basically in accordance
with empirical results.

More comparison of results between the model and
large-scale online social networks are given in Appendix SI (see Figures S1-S24).  The
results further support the notions that the two mechanisms
incorporated in our model provide a potential explanation of the local and
mesoscale structures in these online social networks.

\section*{Discussion}
With the advancement in information technology, online social
systems become an increasingly
important part of modern life.  It is, therefore, of great
significance to study the structures and dynamics of these
systems.  In this study, we focused on the local scale, mesoscale
and macroscale structural properties of online
social networks, especially the influence of
properties on the local scale and macroscale on the
mesoscale structures.  We analyzed the data and extracted the
local scale and macroscale structural
properties of four large-scale online social
networks.  It was found that the indegree and
outdegree distributions follow a similar
scaling law, which follows from the fact that $k_{in}\approx
k_{out}$ for most of the nodes.  It implies that there is a
preferential attachment mechanism in which the
product $k_{out}^{i}k_{in}^{j}$ is
important in the establishment of links during the evolution of
online social networks.  In addition, the very large reciprocity
$r$ observed in these networks suggests the existence of a
reciprocation mechanism in online social networks.  The
reciprocal degree distribution also shows a
similar exponent as that of the
indegree distribution due to
the roughly linear relationship between the
reciprocal degree $k_{r}$ and the indegree $k_{in}$ of
nodes~(\textit{i.e.}, $k_r\sim k_{in}$), which
in turn implies a fixed probability of reciprocal
links between connected nodes. In the mesoscale, the close-knit
friendship structures are determined by both local scale
(\textit{i.e.}, indegree and outdegree~$k_{in}\approx k_{out}$)
and macroscale (\textit{i.e.}, mean in/outdegree~$\langle k\rangle$)
structural properties. For a
node with large~$k_{in}\approx k_{out}$, the numbers of the
four closed triples show the same
scaling behavior: $n_{FB}\sim k_{in}^{2}$ and
$n_{FF}\sim k_{in}^{2}$, as a result of the negligible degree
correlations in these networks.  For all nodes, the distributions
of these closed triples also follow a similar
scaling law. Despite the numbers of the three feedforward
loops are equal, their distributions look somewhat different in
detail.

To reproduce the empirical features,
we proposed and studied a simple directed network
model incorporating an external reciprocation process and an
internal evolution process.  The two parameters in the model are
the activity $m$ and the reciprocation probability $p$.
They can be inferred from the reciprocity $r$ and
mean indegree $\langle k\rangle$ of real online social networks
according to Eq.\eqref{m-p}, so as to ensure that the simulated
network and the real network have the same reciprocity and mean
indegree. Analytically, we derived the structural properties in the
local-scale and mesoscale. The results
show that the exponents characterizing the distributions of
indegree, outdegree, reciprocal degree and four
closed triples depend only on the mean indegree
$\langle k\rangle$, \textit{i.e.},
$\gamma\!=\!2\!+\!1/(\langle k\rangle\!-\!1)$ and
$\gamma_\triangle\!=\!3/2\!+\!1/[2(\langle k\rangle\!-\!1)]$.
In addition, the mean indegree $\langle k\rangle$~and the reciprocity $r$
together determine the ratio of the reciprocal
degree to the directed in/outdegree, \textit{i.e.},
$k_r\!\sim\!(2\langle k\rangle r -r-1)k_{in}/[(\langle
k\rangle-1)(1+r)]$. The expected indegree and
outdegree of nodes in the model grow as the same function of the
time that the nodes are introduced, with very old nodes having
very high indegrees and outdegrees. This phenomenon, coupled with
an essentially fixed rate of reciprocation, reproduces almost all
the properties of the online social networks studied here.

The mesoscale structural properties reported in our work help us
understand the interplay between structural properties on
different scales in online social networks.  More specifically,
the mesoscale structures in these online social networks are
determined by global parameters as well as by local
distributions.  This provides a useful perspective of future
studies in social network analysis.
Our work also provides a better
understanding of the evolution of online social networks,
especially the emergence of close-knit friendship structures with
a scaling behavior in their distributions. The two
processes (reciprocation and preferential attachment) provide a
possible explanation of the mechanisms underlying the local scale
and mesoscale structural properties of online social networks.
The former reflects that users often respond to a
new incoming link by quickly establishing a reversed link.  The
latter means that a well-known user with a large $k_{in}$ is more
likely to attract new connections and an active user with a large
$k_{out}$ is more likely to create new connections. Our
model may also be applied to other growing directed networks in
which the indegree and outdgree distributions show a similar
scaling behavior and the reciprocation mechanism is valid.
However, the model is not applicable to the symmetric online social networks that lack
the power-law degree distributions\cite{corten2011composition,gjoka2010walking,ahn2007analysis}
(\textit{e.g.}, Facebook), and to the
WWW~\cite{krapivsky2001degree} and
Wikipedia~\cite{capocci2006preferential} as the indegree and
outdegree distributions in these systems carry different exponents
and the reciprocation mechanism is absent.  Similarly, it does
not apply to the citation network as a paper can only cite
published papers, but not vice versa.

Although simulated results of our model basically
reproduced the structural properties of the online social
networks at different scales, the differences in the exponents
characterizing the distributions
and in the tails of the distributions
in real online social networks
(\textit{e.g.}, Figures~\ref{kr-of-Epinions},
\ref{kinavebiedge-koutavebiedge-Epinions},
\ref{kinAveTriangle-of-Epinions},
\ref{koutAveTriangle-of-Epinions}) imply that there exist other
factors, such as individual users of different reciprocation
probabilities and local proximity bias, that are ignored in the
model.  These factors are good ingredients for future work.  It
is also important to study the emergence of communities in online
social networks.  The present work also forms the basis for the
understanding of the impact of mesoscale structural properties on
dynamical processes on online social networks, such as
information diffusion, opinions formation, and cooperation
evolution.  An interesting problem for future work
is to investigate whether the model can be applied to offline real
social networks.  Such a work would help reveal the difference
between online and offline social networks.

\section*{Supporting information}
\textbf{Appendix SI} Appendix to the manuscript.\\
(PDF)

\setlength{\parindent}{0pt}
\textbf{Table S1 The exponents of various distributions obtained by
power-law fits of real online social networks and the simulated
network based on the model using the maximum likelihood
estimation.} $x_{min}$ is the lower bound of the range for
fitting a power-law distribution, $\gamma$ is the corresponding
exponent and $KS$ is the goodness-of-fit value based on the
Kolmogorov-Smirnov statistic.\\
(PDF)

\textbf{Table S2 Pearson correlation coefficient.} $r(in,in)$
quantifies the tendency of nodes with a high indegree to be
connected to another node with a high indegree.  The other
quantities carry a similar interpretation.\\
(PDF)

\textbf{Figure S1 Indegree (a) and outdegree (b)
distributions of the Slashdot social network (black squares) and
simulation results (red circles) based on the model.}
The dashed lines in both panels have a
slope $-2.1$ as the analytic results in Eqs. (17) and (31)
suggested. The simulated network is generated by the model with
the parameters $N=82168$, $m\approx5.14$ and $p\approx0.67$ as
determined by the mean degree $\langle k \rangle$ and reciprocity
of the Slashdot social network. Data points are averages over the logarithmic bins of the indegree $k_{in}$ and outdegree $k_{out}$, respectively.\\
(PDF)

\textbf{Figure S2 Indegree (a) and outdegree (b)
distributions of the Flickr social network (black squares) and
simulation results (red circles) based on the model.}
The dashed lines in both panels have a
slope $-2.08$ as the analytic results in Eqs. (17) and (31)
suggested. The simulated network is generated by the model with
the parameters $N=100000$, $m\approx8.07$ and $p\approx0.39$ as
determined by the mean degree $\langle k \rangle$ and reciprocity
of the Flickr social network. Data points are averages over the logarithmic bins of the indegree $k_{in}$ and outdegree $k_{out}$, respectively.\\
(PDF)

\textbf{Figure S3 Indegree (a) and outdegree (b)
distributions of the YouTube social network (black squares) and
simulation results (red circles) based on the model.}
The dashed lines in both panels have a
slope $-2.3$ as the analytic results in Eqs. (17) and (31)
suggested. The simulated network is generated by the model with
the parameters $N=100000$, $m\approx4.34$ and $p\approx0.08$ as
determined by the mean degree $\langle k \rangle$ and reciprocity
of the YouTube social network. Data points are averages over the logarithmic bins of the indegree $k_{in}$ and outdegree $k_{out}$, respectively.\\
(PDF)

\textbf{Figure S4 Relationship between the
indegree and the outdegree of nodes in the Slashdot social
network and the model.}
Results of the Slashdot social network (black
squares) and simulation results (red circles) based on the model
are shown. The blue dash line represents the relation function
$k_{in}=k_{out}$. Data points are averages over the logarithmic bins of the indegree $k_{in}$.\\
(PDF)

\textbf{Figure S5 Relationship between the
indegree and the outdegree of nodes in the Flickr social network
and the model.}
Results of the Flickr social network (black squares) and simulation
results (red circles) based on the model are shown. The blue dash
line represents the relation function $k_{in}=k_{out}$. Data points are averages over the logarithmic bins of the indegree $k_{in}$.\\
(PDF)

\textbf{Figure S6 Relationship between the
indegree and the outdegree of nodes in the YouTube social network
and the model.}
Results of the YouTube social network (black
squares) and simulation results (red circles) based on the model
are shown. The blue dash line represents the relation function
$k_{in}=k_{out}$. Data points are averages over the logarithmic bins of the indegree $k_{in}$.\\
(PDF)

\textbf{Figure S7 Reciprocal degree distributions of
the Slashdot social network and the model.}
Results of the Slashdot social network (black
squares) and simulation results (red circles) based on the model
are shown. Analytic treatment (see Eqs. (17) and (31)) suggests a
scaling behavior with an exponent $-2.1$, as shown by the dash
line. Data points are averages over the logarithmic bins of the reciprocal degree $k_{r}$.\\
(PDF)

\textbf{Figure S8 Reciprocal degree distributions of
the Flickr social network and the model.}
Results of the Flickr social network (black
squares) and simulation results (red circles) based on the model
are shown. Analytic treatment (see Eqs. (17) and (31)) suggests a
scaling behavior with an exponent $-2.08$, as shown by the dash
line. Data points are averages over the logarithmic bins of the reciprocal degree $k_{r}$.\\
(PDF)

\textbf{Figure S9 Reciprocal degree distributions of
the YouTube social network and the model.}
Results of the YouTube social network (black
squares) and simulation results (red circles) based on the model
are shown. Analytic treatment (see Eqs. (17) and (31)) suggests a
scaling behavior with an exponent $-2.3$, as shown by the dash
line. Data points are averages over the logarithmic bins of the reciprocal degree $k_{r}$.\\
(PDF)

\textbf{Figure S10 Mean reciprocal degree of nodes
with (a) the same indegree and (b) the same outdegree in the
Slashdot social network and in the model.}
Results of the Slashdot social network (black
squares) and simulation results (red circles) based on the model
are shown in a log-log scale in the main panels. Analytic
treatment suggests that $\langle k_{r}\rangle$ is linearly
dependent on $k_{in}$ and $k_{out}$, and the blue dash lines of
slope $1$ show its dependence. The inset in each panel shows the
results in a linear scale and the dash line has a slope of
$0.82$, as given by Eqs. $(20)$ and $(31)$.
Data points are averages over the logarithmic bins of the indegree $k_{in}$ and outdegree $k_{out}$, respectively.\\
(PDF)

\textbf{Figure S11 Mean reciprocal degree of nodes
with (a) the same indegree and (b) the same outdegree in the
Flickr social network and in the model.}
Results of the Flickr social network (black
squares) and simulation results (red circles) based on the model
are shown in a log-log scale in the main panels. Analytic
treatment suggests that $\langle k_{r}\rangle$ is linearly
dependent on $k_{in}$ and $k_{out}$, and the blue dash lines of
slope $1$ show its dependence. The inset in each panel shows the
results in a linear scale and the dash line has a slope of
$0.59$, as given by Eqs. $(20)$ and $(31)$.
Data points are averages over the logarithmic bins of the indegree $k_{in}$ and outdegree $k_{out}$, respectively.\\
(PDF)

\textbf{Figure S12 Mean reciprocal degree of nodes
with (a) the same indegree and (b) the same outdegree in the
YouTube social network and in the model.}
Results of the YouTube social network (black
squares) and simulation results (red circles) based on the model
are shown in a log-log scale in the main panels. Analytic
treatment suggests that $\langle k_{r}\rangle$ is linearly
dependent on $k_{in}$ and $k_{out}$, and the blue dash lines of
slope $1$ show its dependence. The inset in each panel shows the
results in a linear scale and the dash line has a slope of
$0.73$, as given by Eqs. $(20)$ and $(31)$.
Data points are averages over the logarithmic bins of the indegree $k_{in}$ and outdegree $k_{out}$, respectively.\\
(PDF)

\textbf{Figure S13 Distributions of four basic
closed triples in the slashdot social network and the model.}
Distributions of closed triples
corresponding to (a) $FB$, (b) $FF_{a}$, (c) $FF_{b}$, and (d)
$FF_{c}$ loops in the Slashdot social network (black squares) and
in the simulated network based on the model (red circles).
Analytic treatment (see Eqs. $(30)$ and $(31)$) suggests a
scaling behavior with an exponent $-1.55$, as shown by the dash
lines. Data points are averages over the logarithmic bins of the $n_{FB}$, $n_{FFa}$, $n_{FFb}$ and $n_{FFc}$, respectively.\\
(PDF)

\textbf{Figure S14 Distributions of four basic
closed triples in the Flickr social network and the model.}
Distributions of closed triples
corresponding to (a) $FB$, (b) $FF_{a}$, (c) $FF_{b}$, and (d)
$FF_{c}$ loops in the Flickr social network (black squares) and
in the simulated network based on the model (red circles).
Analytic treatment (see Eqs. $(30)$ and $(31)$) suggests a
scaling behavior with an exponent $-1.54$, as shown by the dash
lines. Data points are averages over the logarithmic bins of the $n_{FB}$, $n_{FFa}$, $n_{FFb}$ and $n_{FFc}$, respectively.\\
(PDF)

\textbf{Figure S15 Distributions of four basic
closed triples in the YouTube social network and the model.}
Distributions of closed triples
corresponding to (a) $FB$, (b) $FF_{a}$, (c) $FF_{b}$, and (d)
$FF_{c}$ loops in the YouTube social network (black squares) and
in the simulated network based on the model (red circles).
Analytic treatment (see Eqs. $(30)$ and $(31)$) suggests a
scaling behavior with an exponent $-1.65$, as shown by the dash
lines. Data points are averages over the logarithmic bins of the $n_{FB}$, $n_{FFa}$, $n_{FFb}$ and $n_{FFc}$, respectively.\\
(PDF)

\textbf{Figure S16 Degree correlations in the
Slashdot social network and the model.} Results
of degree correlations as measured by four quantities
corresponding to the average nearest neighbor degree
$<k_{in}^{nn}(k_{in})>$ (squares), $<k_{out}^{nn}(k_{in})>$
(circles), $<k_{out}^{nn}(k_{out})>$ (triangles), and
$<k_{in}^{nn}(k_{out})>$ (inverted triangles) for (a) Slashdot
social network and (b) simulated network based on the model.
Data points are averages over the logarithmic bins of the indegree $k_{in}$ or outdegree $k_{out}$.\\
(PDF)

\textbf{Figure S17 Degree correlations in the
Flickr social network and the model.} Results of
degree correlations as measured by four quantities corresponding
to the average nearest neighbor degree $<k_{in}^{nn}(k_{in})>$
(squares), $<k_{out}^{nn}(k_{in})>$ (circles),
$<k_{out}^{nn}(k_{out})>$ (triangles), and
$<k_{in}^{nn}(k_{out})>$ (inverted triangles) for (a) Flickr
social network and (b) simulated network based on the model.
Data points are averages over the logarithmic bins of the indegree $k_{in}$ or outdegree $k_{out}$.\\
(PDF)

\textbf{Figure S18 Degree correlations in the
YouTube social netowrk and the model.} Results
of degree correlations as measured by four quantities
corresponding to the average nearest neighbor degree
$<k_{in}^{nn}(k_{in})>$ (squares), $<k_{out}^{nn}(k_{in})>$
(circles), $<k_{out}^{nn}(k_{out})>$ (triangles), and
$<k_{in}^{nn}(k_{out})>$ (inverted triangles) for (a) YouTube
social network and (b) simulated network based on the model.
Data points are averages over the logarithmic bins of the indegree $k_{in}$ or outdegree $k_{out}$.\\
(PDF)

\textbf{Figure S19 Mean number of the four closed
triples for nodes with the same indegree in the Slashdot social
network and the model.} Results for the mean number of closed triples corresponding to (a) $FB$, (b)
$FF_{a}$, (c) $FF_{b}$, and (d) $FF_{c}$ loops for nodes with the
same indegree are shown for the Slashdot social network (black
squares) and simulated network (red circles) based on the model.
Analytic treatment (see Eq. $(28)$) gives a scaling behavior with
an exponent $2$, as indicated by the dash line.
Data points are averages over the logarithmic bins of the indegree $k_{in}$.\\
(PDF)

\textbf{Figure S20 Mean number of the four closed
triples for nodes with the same indegree in the Flickr social
network and the model.} Results for the mean
number of closed triples corresponding to (a) $FB$, (b) $FF_{a}$,
(c) $FF_{b}$, and (d) $FF_{c}$ loops for nodes with the same
indegree are shown for the Flickr social network (black squares)
and simulated network (red circles) based on the model. Analytic
treatment (see Eq. $(28)$) gives a scaling behavior with an
exponent $2$, as indicated by the dash line.
Data points are averages over the logarithmic bins of the indegree $k_{in}$.\\
(PDF)

\textbf{Figure S21 Mean number of the four closed
triples for nodes with the same indegree in the YouTube social
network and the model.} Results for the
mean number of closed triples corresponding to (a) $FB$, (b)
$FF_{a}$, (c) $FF_{b}$, and (d) $FF_{c}$ loops for nodes with the
same indegree are shown for the YouTube social network (black
squares) and simulated network (red circles) based on the model.
Analytic treatment (see Eq. $(28)$) gives a scaling behavior with
an exponent $2$, as indicated by the dash line.
Data points are averages over the logarithmic bins of the indegree $k_{in}$.\\
(PDF)

\textbf{Figure S22 Mean number of the four closed
triples for nodes with the same outdegree in the Slashdot social
network and the model.} Results for the mean
number of closed triples corresponding to (a) $FB$, (b) $FF_{a}$,
(c) $FF_{b}$, and (d) $FF_{c}$ loops for nodes with the same
outdegree are shown for the Slashdot social network (black
squares) and simulated network (red circles) based on the model.
Analytic treatment (see Eq. $(28)$) gives a scaling behavior with
an exponent $2$, as indicated by the dash line.
Data points are averages over the logarithmic bins of the outdegree $k_{out}$.\\
(PDF)

\textbf{Figure S23 Mean number of the four closed
triples for nodes with the same outdegree in the Flickr social
network and the model.} Results for the mean
number of closed triples corresponding to (a) $FB$, (b) $FF_{a}$,
(c) $FF_{b}$, and (d) $FF_{c}$ loops for nodes with the same
outdegree are shown for the Flickr social network (black squares)
and simulated network (red circles) based on the model. Analytic
treatment (see Eq. $(28)$) gives a scaling behavior with an
exponent $2$, as indicated by the dash line.
Data points are averages over the logarithmic bins of the outdegree $k_{out}$.\\
(PDF)

\textbf{Figure S24 Mean number of the four closed
triples for nodes with the same outdegree in the YouTube social
network and the model.} Results for the mean
number of closed triples corresponding to (a) $FB$, (b) $FF_{a}$,
(c) $FF_{b}$, and (d) $FF_{c}$ loops for nodes with the same
outdegree are shown for the YouTube social network (black
squares) and simulated network (red circles) based on the model.
Analytic treatment (see Eq. $(28)$) gives a scaling behavior with
an exponent $2$, as indicated by the dash line.
Data points are averages over the logarithmic bins of the outdegree $k_{out}$.\\
(PDF)

\section*{Acknowledgments}
We acknowledge Tao Zhou and Ying Liu for valuable discussions.


\bibliography{reference}

\begin{thebibliography}{10}
\providecommand{\url}[1]{\texttt{#1}}
\providecommand{\urlprefix}{URL }
\expandafter\ifx\csname urlstyle\endcsname\relax
  \providecommand{\doi}[1]{doi:\discretionary{}{}{}#1}\else
  \providecommand{\doi}{doi:\discretionary{}{}{}\begingroup
  \urlstyle{rm}\Url}\fi
\providecommand{\bibAnnoteFile}[1]{%
  \IfFileExists{#1}{\begin{quotation}\noindent\textsc{Key:} #1\\
  \textsc{Annotation:}\ \input{#1}\end{quotation}}{}}
\providecommand{\bibAnnote}[2]{%
  \begin{quotation}\noindent\textsc{Key:} #1\\
  \textsc{Annotation:}\ #2\end{quotation}}
\providecommand{\eprint}[2][]{\url{#2}}

\bibitem{corten2011composition}
Corten R (2012) Composition and structure of a large online social network in
  the netherlands.
\newblock PLoS ONE 7: e34760.
\bibAnnoteFile{corten2011composition}

\bibitem{gjoka2010walking}
Gjoka~M BCTMA Kurant~M (2010) Walking in facebook: A case study of unbiased
  sampling of osns.
\newblock In: Proceedings of IEEE INFOCOM. ACM.
\bibAnnoteFile{gjoka2010walking}

\bibitem{ahn2007analysis}
Ahn YY, Han S, Kwak H, Moon S, Jeong H (2007) Analysis of topological
  characteristics of huge online social networking services.
\newblock In: Proceedings of the 16th international conference on World Wide
  Web. ACM, pp. 835--844.
\bibAnnoteFile{ahn2007analysis}

\bibitem{hu2009evolution}
Hu HB, Wang XF (2009) Evolution of a large online social network.
\newblock Phys Lett A 373: 1105--1110.
\bibAnnoteFile{hu2009evolution}

\bibitem{mislove2008growth}
Mislove A, Koppula HS, Gummadi KP, Druschel P, Bhattacharjee B (2008) Growth of
  the flickr social network.
\newblock In: Proceedings of the first workshop on Online social networks. ACM,
  pp. 25--30.
\bibAnnoteFile{mislove2008growth}

\bibitem{benevenuto2008understanding}
Benevenuto F, Duarte F, Rodrigues T, Almeida VAF, Almeida JM, et~al. (2008)
  Understanding video interactions in youtube.
\newblock In: Proceeding of the 16th ACM international conference on
  Multimedia. ACM, pp. 761--764.
\bibAnnoteFile{benevenuto2008understanding}

\bibitem{gomez2008statistical}
G{\'o}mez V, Kaltenbrunner A, L{\'o}pez V (2008) Statistical analysis of the
  social network and discussion threads in slashdot.
\newblock In: Proceeding of the 17th international conference on World Wide
  Web. ACM, pp. 645--654.
\bibAnnoteFile{gomez2008statistical}

\bibitem{scott1988social}
Scott J (1988) Social network analysis.
\newblock Sociology 22: 109-127.
\bibAnnoteFile{scott1988social}

\bibitem{scott2000social}
Scott J (2000) Social network analysis: a handbook.
\newblock Sage.
\bibAnnoteFile{scott2000social}

\bibitem{albert2002statistical}
Albert R, Barab{\'a}si AL (2002) Statistical mechanics of complex networks.
\newblock Rev Mod Phys 74: 47-97.
\bibAnnoteFile{albert2002statistical}

\bibitem{dorogovtsev2002evolution}
Dorogovtsev SN, Mendes JFF (2002) Evolution of networks.
\newblock Adv Phys 51: 1079--1187.
\bibAnnoteFile{dorogovtsev2002evolution}

\bibitem{newman2003structure}
Newman MEJ (2003) The structure and function of complex networks.
\newblock SIAM Rev 45: 167--256.
\bibAnnoteFile{newman2003structure}

\bibitem{boccaletti2006complex}
Boccaletti S, Latora V, Moreno Y, Chavez M, Hwang DU (2006) Complex networks:
  Structure and dynamics.
\newblock Phys Rep 424: 175--308.
\bibAnnoteFile{boccaletti2006complex}

\bibitem{costa2007characterization}
Costa LF, Rodrigues FA, Travieso G, Boas PRV (2007) Characterization of complex
  networks: A survey of measurements.
\newblock Adv Phys 56: 167--242.
\bibAnnoteFile{costa2007characterization}

\bibitem{dorogovtsev2008critical}
Dorogovtsev SN, Goltsev AV, Mendes JFF (2008) Critical phenomena in complex
  networks.
\newblock Rev Mod Phys 80: 1275-1335.
\bibAnnoteFile{dorogovtsev2008critical}

\bibitem{barthelemy2011spatial}
Barth{\'e}lemy M (2011) Spatial networks.
\newblock Phys Rep 499: 1--101.
\bibAnnoteFile{barthelemy2011spatial}

\bibitem{castellano2009statistical}
Castellano C, Fortunato S, Loreto V (2009) Statistical physics of social
  dynamics.
\newblock Rev Mod Phys 81: 591-646.
\bibAnnoteFile{castellano2009statistical}

\bibitem{centola2010spread}
Centola D (2010) The spread of behavior in an online social network experiment.
\newblock Science 329: 1194-1197.
\bibAnnoteFile{centola2010spread}

\bibitem{l¨¹2011leaders}
L{\"u} L, Zhang YC, Yeung CH, Zhou T (2011) Leaders in social networks, the
  delicious case.
\newblock PloS ONE 6: e21202.
\bibAnnoteFile{l¨¹2011leaders}

\bibitem{zhou2011emergence}
Zhou T, Medo M, Cimini G, Zhang ZK, Zhang YC (2011) Emergence of scale-free
  leadership structure in social recommender systems.
\newblock PloS ONE 6: e20648.
\bibAnnoteFile{zhou2011emergence}

\bibitem{nowak2005evolution}
Nowak MA, Sigmund K (2005) Evolution of indirect reciprocity.
\newblock Nature 437: 1291--1298.
\bibAnnoteFile{nowak2005evolution}

\bibitem{brin1998anatomy}
Brin S, Page L (1999) The anatomy of a large-scale hypertextual web search
  engine.
\newblock Computer Networks and ISDN Systems 30: 107--117.
\bibAnnoteFile{brin1998anatomy}

\bibitem{kleinberg1999authoritative}
Kleinberg JM (1999) Authoritative sources in a hyperlinked environment.
\newblock JACM 46: 604--632.
\bibAnnoteFile{kleinberg1999authoritative}

\bibitem{newman2001structure}
Newman MEJ (2001) The structure of scientific collaboration networks.
\newblock Proc Natl Acad Sci USA 98: 404-409.
\bibAnnoteFile{newman2001structure}

\bibitem{leicht2007large}
Leicht EA, Clarkson G, Shedden K, Newman MEJ (2007) Large-scale structure of
  time evolving citation networks.
\newblock Eur Phys J B 59: 75--83.
\bibAnnoteFile{leicht2007large}

\bibitem{rodgers2001properties}
Rodgers GJ, Darby-Dowman K (2001) Properties of a growing random directed
  network.
\newblock Eur Phys J B 23: 267--271.
\bibAnnoteFile{rodgers2001properties}

\bibitem{foster2010edge}
Foster JG, Foster DV, Grassberger P, Paczuski M (2010) Edge direction and the
  structure of networks.
\newblock Proc Natl Acad Sci USA 107: 10815-10820.
\bibAnnoteFile{foster2010edge}

\bibitem{palla2007directed}
Palla G, Farkas IJ, Pollner P, Derenyi I, Vicsek T (2007) Directed network
  modules.
\newblock New J Phys 9: 186-206.
\bibAnnoteFile{palla2007directed}

\bibitem{leicht2008community}
Leicht EA, Newman MEJ (2008) Community structure in directed networks.
\newblock Phys Rev Lett 100: 118703.
\bibAnnoteFile{leicht2008community}

\bibitem{kim2010finding}
Kim Y, Son SW, Jeong H (2010) Finding communities in directed networks.
\newblock Phys Rev E 81: 016103.
\bibAnnoteFile{kim2010finding}

\bibitem{dorogovtsev2000structure}
Dorogovtsev SN, Mendes JFF, Samukhin AN (2000) Structure of growing networks
  with preferential linking.
\newblock Phys Rev Lett 85: 4633--4636.
\bibAnnoteFile{dorogovtsev2000structure}

\bibitem{barabasi1999emergence}
Barab{\'a}si AL, Albert R (1999) Emergence of scaling in random networks.
\newblock Science 286: 509-512.
\bibAnnoteFile{barabasi1999emergence}

\bibitem{krapivsky2001degree}
Krapivsky PL, Rodgers GJ, Redner S (2001) Degree distributions of growing
  networks.
\newblock Phys Rev Lett 86: 5401--5404.
\bibAnnoteFile{krapivsky2001degree}

\bibitem{almendral2010announcement}
Almendral JA, Criado R, Leyva I, Buld{\'u} JM, Sendi\~{n}a Nadal I (2010)
  Announcement: Focus issue on ``mesoscales in complex networks''.
\newblock CHAOS 20: 010202.
\bibAnnoteFile{almendral2010announcement}

\bibitem{almendral2011introduction}
Almendral JA, Criado R, Leyva I, Buld{\'u} JM, Sendi\~{n}a Nadal I (2011)
  Introduction to focus issue: Mesoscales in complex networks.
\newblock CHAOS 21: 016101.
\bibAnnoteFile{almendral2011introduction}

\bibitem{reichardt2011interplay}
Reichardt J, Alamino R, Saad D (2011) The interplay between microscopic and
  mesoscopic structures in complex networks.
\newblock PloS ONE 6: e21282.
\bibAnnoteFile{reichardt2011interplay}

\bibitem{fortunato2010community}
Fortunato S (2010) Community detection in graphs.
\newblock Phys Rep 486: 75--174.
\bibAnnoteFile{fortunato2010community}

\bibitem{lancichinetti2011finding}
Lancichinetti A, Radicchi F, Ramasco JJ, Fortunato S (2011) Finding
  statistically significant communities in networks.
\newblock PloS ONE 6: e18961.
\bibAnnoteFile{lancichinetti2011finding}

\bibitem{lancichinetti2010characterizing}
Lancichinetti A, Kivel{\"a} M, Saram{\"a}ki J, Fortunato S (2010)
  Characterizing the community structure of complex networks.
\newblock PloS ONE 5: e11976.
\bibAnnoteFile{lancichinetti2010characterizing}

\bibitem{Watts1998}
Watts DJ, Strogatz SH (1998) {Collective dynamics of `small-world' networks}.
\newblock Nature 393: 440-442.
\bibAnnoteFile{Watts1998}

\bibitem{milo2002network}
Milo R, Shen-Orr S, Itzkovitz S, Kashtan N, Chklovskii D, et~al. (2002) Network
  motifs: simple building blocks of complex networks.
\newblock Science 298: 824-827.
\bibAnnoteFile{milo2002network}

\bibitem{milo2004superfamilies}
Milo R, Itzkovitz S, Kashtan N, Levitt R, Shen-Orr S, et~al. (2004)
  Superfamilies of evolved and designed networks.
\newblock Science 303: 1538-1542.
\bibAnnoteFile{milo2004superfamilies}

\bibitem{huang2007bridge}
Huang CY, Sun CT, Cheng CY, Hsieh JL (2007) Bridge and brick motifs in complex
  networks.
\newblock Physica A 377: 340--350.
\bibAnnoteFile{huang2007bridge}

\bibitem{fagiolo2007clustering}
Fagiolo G (2007) Clustering in complex directed networks.
\newblock Phys Rev E 76: 026107.
\bibAnnoteFile{fagiolo2007clustering}

\bibitem{ahnert2008clustering}
Ahnert SE, Fink TMA (2008) Clustering signatures classify directed networks.
\newblock Phys Rev E 78: 036112.
\bibAnnoteFile{ahnert2008clustering}

\bibitem{mangan2003structure}
Mangan S, Alon U (2003) Structure and function of the feed-forward loop network
  motif.
\newblock Proc Natl Acad of Sci USA 100: 11980-11985.
\bibAnnoteFile{mangan2003structure}

\bibitem{mangan2006incoherent}
Mangan S, Itzkovitz S, Zaslaver A, Alon U (2006) The incoherent feed-forward
  loop accelerates the response-time of the gal system of escherichia coli.
\newblock J Mol Biol 356: 1073--1081.
\bibAnnoteFile{mangan2006incoherent}

\bibitem{sousa2005consensus}
Sousa AO (2005) Consensus formation on a triad scale-free network.
\newblock Physica A 348: 701--710.
\bibAnnoteFile{sousa2005consensus}

\bibitem{ghoneim2008characterizing}
Ghoneim A, Abbass H, Barlow M (2008) Characterizing game dynamics in two-player
  strategy games using network motifs.
\newblock IEEE Trans Syst Man Cybern Part B: Cybern 38: 682--690.
\bibAnnoteFile{ghoneim2008characterizing}

\bibitem{hales2008motifs}
Hales D, Arteconi S (2008) Motifs in evolving cooperative networks look like
  protein structure networks.
\newblock Networks and Heterogeneous Media 3: 239-249.
\bibAnnoteFile{hales2008motifs}

\bibitem{leskovec2010signed}
Leskovec J, Huttenlocher D, Kleinberg J (2010) Signed networks in social media.
\newblock In: Proceedings of the 28th international conference on Human factors
  in computing systems. ACM, pp. 1361--1370.
\bibAnnoteFile{leskovec2010signed}

\bibitem{mislove2007measurement}
Mislove A, Marcon M, Gummadi KP, Druschel P, Bhattacharjee B (2007) Measurement
  and analysis of online social networks.
\newblock In: Proceedings of the 7th ACM SIGCOMM conference on Internet
  measurement. ACM, pp. 29--42.
\bibAnnoteFile{mislove2007measurement}

\bibitem{garlaschelli2004patterns}
Garlaschelli D, Loffredo MI (2004) Patterns of link reciprocity in directed
  networks.
\newblock Phys Rev Lett 93: 268701.
\bibAnnoteFile{garlaschelli2004patterns}

\bibitem{clauset2009power}
Clauset A, Shalizi CR, Newman MEJ (2009) Power-law distributions in empirical
  data.
\newblock SIAM Rev 51: 661-703.
\bibAnnoteFile{clauset2009power}

\bibitem{stumpf2012critical}
Stumpf MPH, Porter MA (2012) Critical truths about power laws.
\newblock Science 335: 665--666.
\bibAnnoteFile{stumpf2012critical}

\bibitem{capocci2006preferential}
Capocci A, Servedio VDP, Colaiori F, Buriol LS, Donato D, et~al. (2006)
  Preferential attachment in the growth of social networks: The internet
  encyclopedia wikipedia.
\newblock Phys Rev E 74: 036116.
\bibAnnoteFile{capocci2006preferential}

\bibitem{l¨¹2011link}
L{\"u} L, Zhou T (2011) Link prediction in complex networks: A survey.
\newblock Physica A 390: 1150--1170.
\bibAnnoteFile{l¨¹2011link}

\bibitem{newman2002assortative}
Newman MEJ (2002) Assortative mixing in networks.
\newblock Phys Rev Lett 89: 208701.
\bibAnnoteFile{newman2002assortative}

\bibitem{krapivsky2000connectivity}
Krapivsky PL, Redner S, Leyvraz F (2000) Connectivity of growing random
  networks.
\newblock Phys Rev Lett 85: 4629--4632.
\bibAnnoteFile{krapivsky2000connectivity}

\bibitem{krapivsky2001organization}
Krapivsky PL, Redner S (2001) Organization of growing random networks.
\newblock Phys Rev E 63: 066123.
\bibAnnoteFile{krapivsky2001organization}

\bibitem{catanzaro2005generation}
Catanzaro M, Bogu{\~n}{\'a} M, Pastor-Satorras R (2005) Generation of
  uncorrelated random scale-free networks.
\newblock Phys Rev E 71: 027103.
\bibAnnoteFile{catanzaro2005generation}

\bibitem{gallos2012people}
Gallos LK, Rybski D, Liljeros F, Havlin S, Makse HA (2012) How people interact
  in evolving online affiliation networks.
\newblock Phys Rev X 2: 031014.
\bibAnnoteFile{gallos2012people}

\bibitem{wu2010evidence}
Wu Y, Zhou C, Xiao J, Kurths J, Schellnhuber HJ (2010) Evidence for a bimodal
  distribution in human communication.
\newblock Proc Natl Acad Sci USA 107: 18803--18808.
\bibAnnoteFile{wu2010evidence}

\bibitem{mossa2002truncation}
Mossa S, Barthelemy M, Eugene~Stanley H, Nunes~Amaral LA (2002) Truncation of
  power law behavior in ¡°scale-free¡± network models due to information
  filtering.
\newblock Phys Rev Lett 88: 138701.
\bibAnnoteFile{mossa2002truncation}

\end{thebibliography}

\newpage


\begin{table}[!ht]
\caption{ \bf{Basic statistics of the four online
social network datasets.} Properties of each network: number of
users $N$, number of directed links $E$, reciprocity $r$, number
of feedback ($FB$) loop $N_{FB}$, number of feedforward loops
$N_{FF_a}$. The numbers of the three feedforward
loops ($FF_{a}$, $FF_{b}$, $FF_{c}$) are equal, because every
$FF_{a}$ loop from the perspective of the focal node constitutes a $FF_b$
loop and a $FF_c$ loop from the perspective of the another two
nodes.} \centering
\begin{tabular}{lrrrrr}
\hline
Data sets & Epinions & Slashdot & Flickr &  YouTube\\
\hline
$N$  &75,879 &82,168 &1,715,255 &1,138,499\\
$E$  &508,825 &870,161 &22,613,980 &4,945,382\\
$k_{in}^{max}$ &3035 &2552 &16255 &25519\\
$k_{out}^{max}$ &1801 &2510 &26185 &28644\\
$r$  &0.25 &0.73 &0.45 &0.65\\
$N_{FB}$  &740,310 &899,316  &435,829,822 &5,320,127\\
$N_{FF_a}$  &3,586,403 &2,881,727 &1,667,179,686  &16,287,794\\
\hline
\end{tabular}
\label{tab:basiccharacteristic}
\end{table}

\section*{Figure Legends}
\begin{figure}[!ht]
\begin{center}
\includegraphics[width=4in,height=3in]{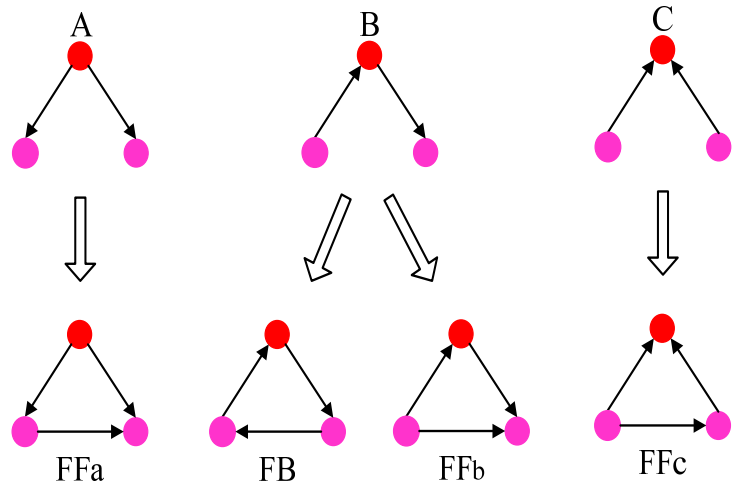}
\end{center}
\caption{\textbf{Three possible unclosed triples and four basic
closed triples for a focal node
(red).} The basic closed triples correspond to
one feedback ($FB$) loop and three
feedforward ($FF$) loops. The three feedforward
loops differ in the indegree $k_{in}$ of the focal node:
$k_{in}=0$ for $FF_{a}$, $k_{in}=1$ for $FF_{b}$ and $k_{in}=2$
for $FF_{c}$. The numbers of the three feedforward loops are
equal because every $FF_{a}$ loop from the perspective of the
focal node constitutes a $FF_b$ loop and a $FF_c$ loop from the
perspective of the another two nodes, but the loops may arise
from different growth histories.} \label{four-triangles}
\end{figure}

\begin{figure}[!ht]
\begin{center}
\includegraphics[width=4in]{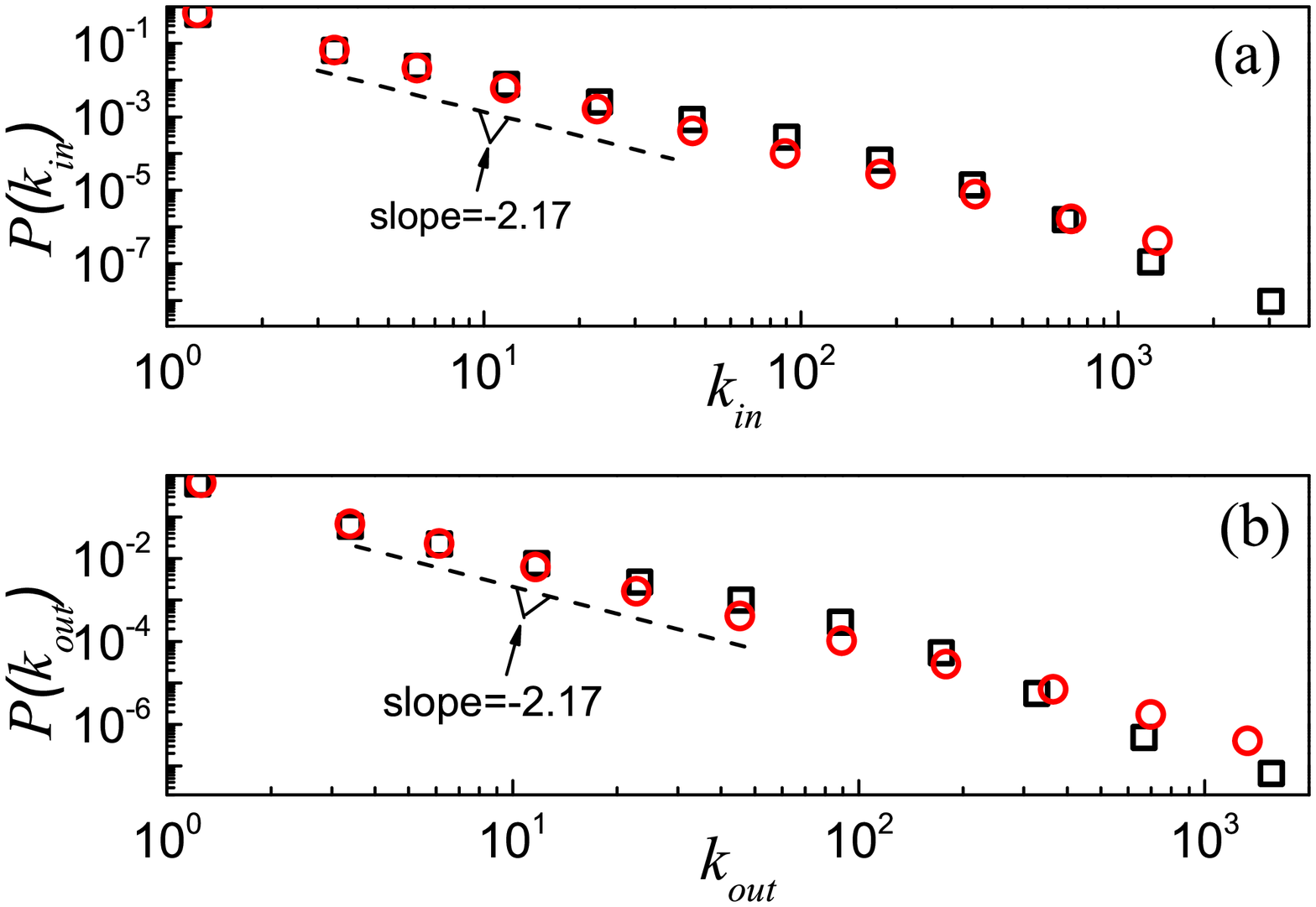}
\end{center}
\caption{\textbf{Indegree (a) and outdegree (b) distributions of
the Epinions social network (black squares) and simulation results (red circles)
based on the model.}  The dashed lines in both
panels have a slope $-2.17$ as the analytic
results in Eqs.~\eqref{gamma} and~\eqref{m-p} suggested. The
simulated network is generated by the model with the parameters
$N=75879$, $m \approx 4.34$ and $p \approx 0.08$, as determined by
the mean degree $\langle k \rangle$ and reciprocity $r$ of the
Epinions social network. Data points are averages over the logarithmic bins of the indegree $k_{in}$ and outdegree $k_{out}$, respectively.}
\label{kin-and-kout-of-Epinions}
\end{figure}

\begin{figure}[!ht]
\begin{center}
\includegraphics[width=3in]{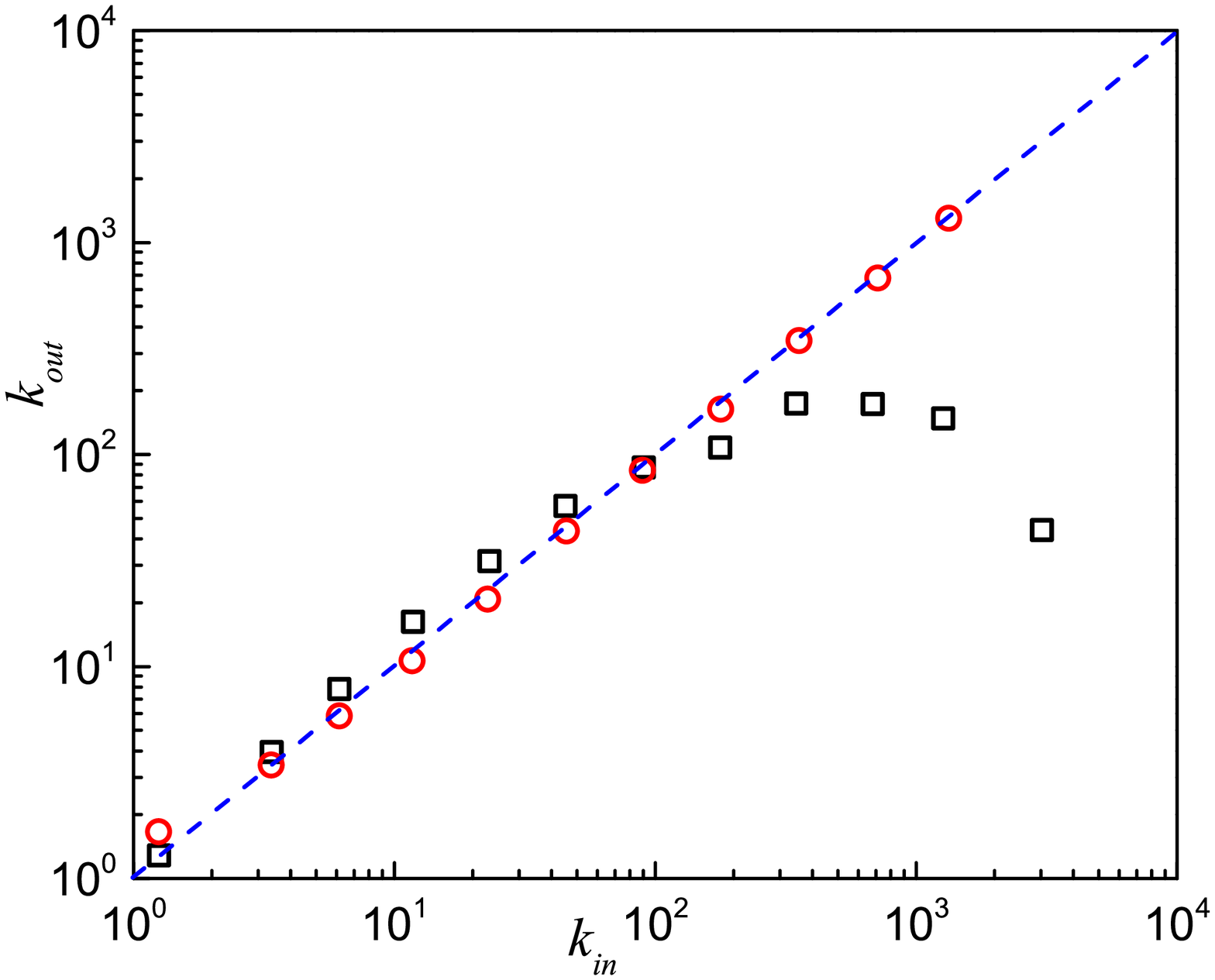}
\end{center}
\caption{\textbf{Relationship between the indegree and the
outdegree of nodes in the Epinions social network and the model.} Results of the Epinions social network
(black squares) and simulation results (red circles) based on the
model are shown.  The blue dash line represents the relation
function $k_{in}=k_{out}$. Data points are averages over the logarithmic bins of the indegree $k_{in}$.}
\label{kintokout-of-Epinions}
\end{figure}

\begin{figure}[!ht]
\begin{center}
\includegraphics[width=4in]{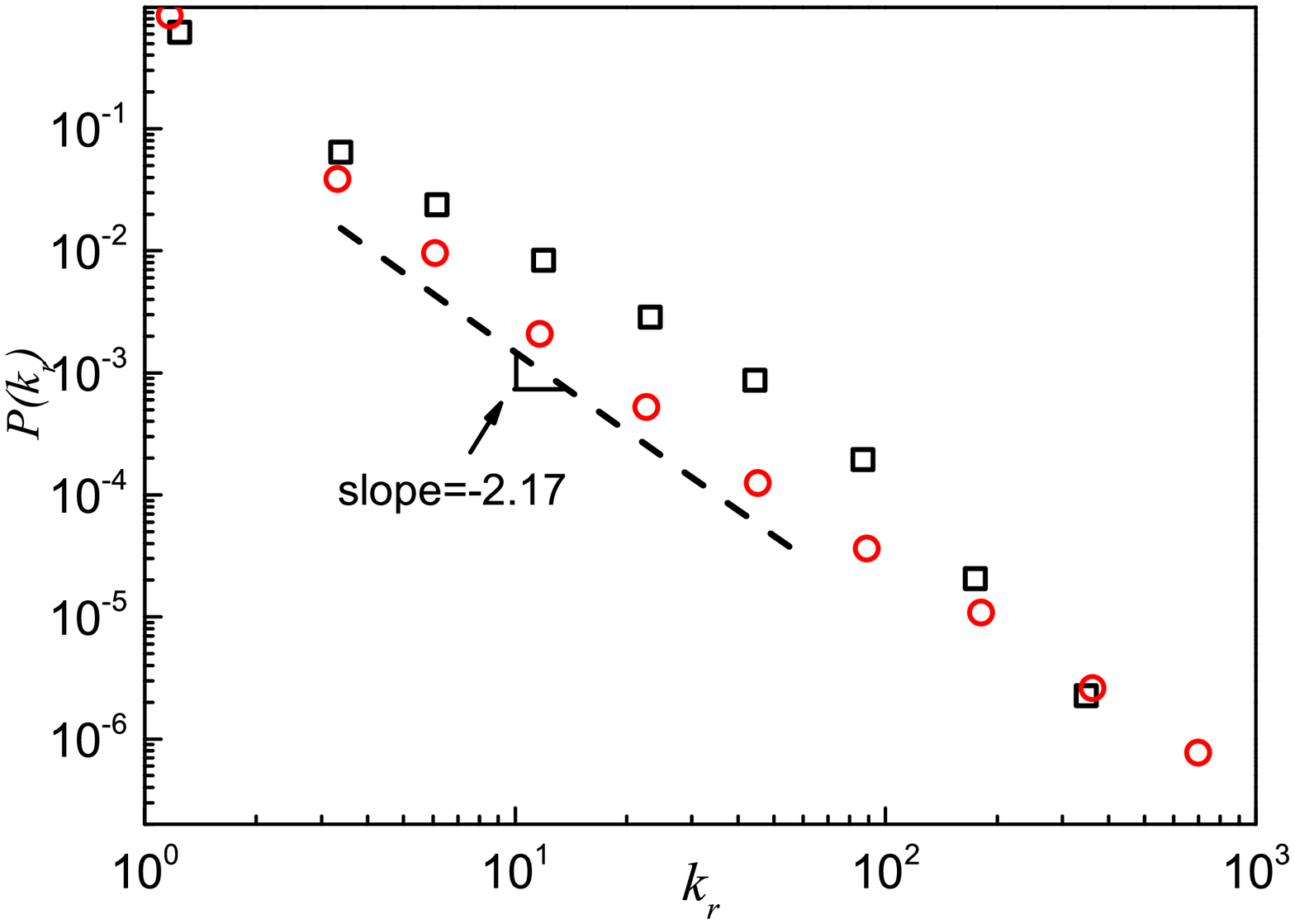}
\end{center}
\caption{\textbf{Reciprocal degree distributions of the Epinions
social network and the model.}  Results of the Epinions social
network (black squares) and simulation results (red circles) based
on the model are shown.  Analytic treatment (see
Eqs.~\eqref{gamma} and~\eqref{m-p}) suggests a scaling behavior
with an exponent $-2.17$, as shown by the dash line. Data points are averages over the logarithmic bins of the reciprocal degree $k_{r}$.}
\label{kr-of-Epinions}
\end{figure}

\begin{figure}[!ht]
\begin{center}
\includegraphics[width=4in]{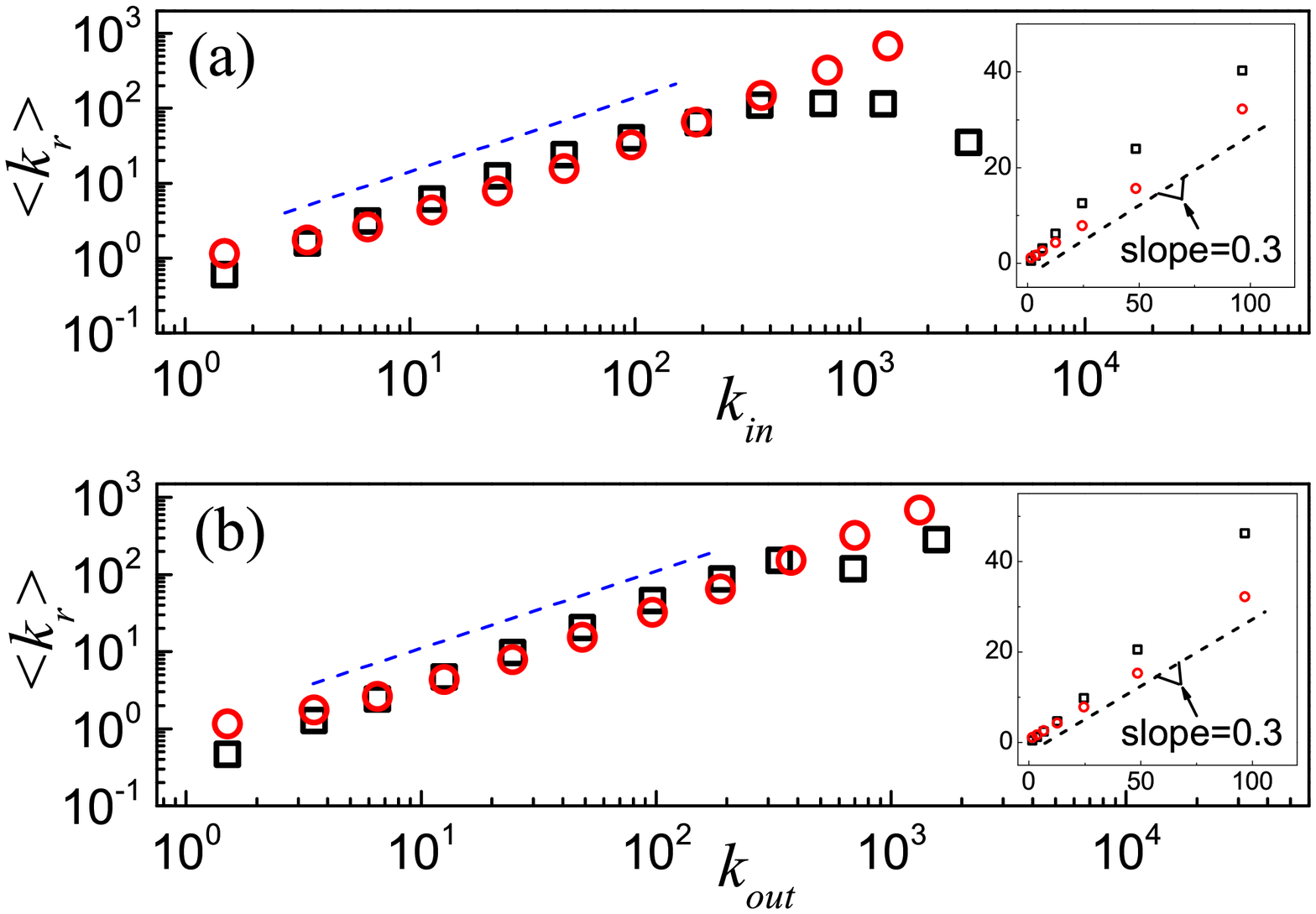}
\end{center}
\caption{\textbf{Mean reciprocal degree of nodes with (a) the same
indegree and (b) the same outdegree in
the Epinions social network and in the model.}  Results of the
Epinions social network (black squares) and simulation results
(red circles) based on the model are shown in a log-log scale in
the main panels.  Analytic treatment suggests that $\langle k_{r}
\rangle$ is linearly dependent on $k_{in}$ and $k_{out}$, and the
blue dash lines of slope 1 show its dependence.
The inset in each panel shows the
results in a linear scale and the dash line has a slope of $0.3$,
as given by Eqs.~\eqref{kr-scaling} and~\eqref{m-p}. Data points are averages over the logarithmic bins of the indegree $k_{in}$ and outdegree $k_{out}$, respectively.}
\label{kinavebiedge-koutavebiedge-Epinions}
\end{figure}

\begin{figure}[!ht]
\begin{center}
\includegraphics[width=4in]{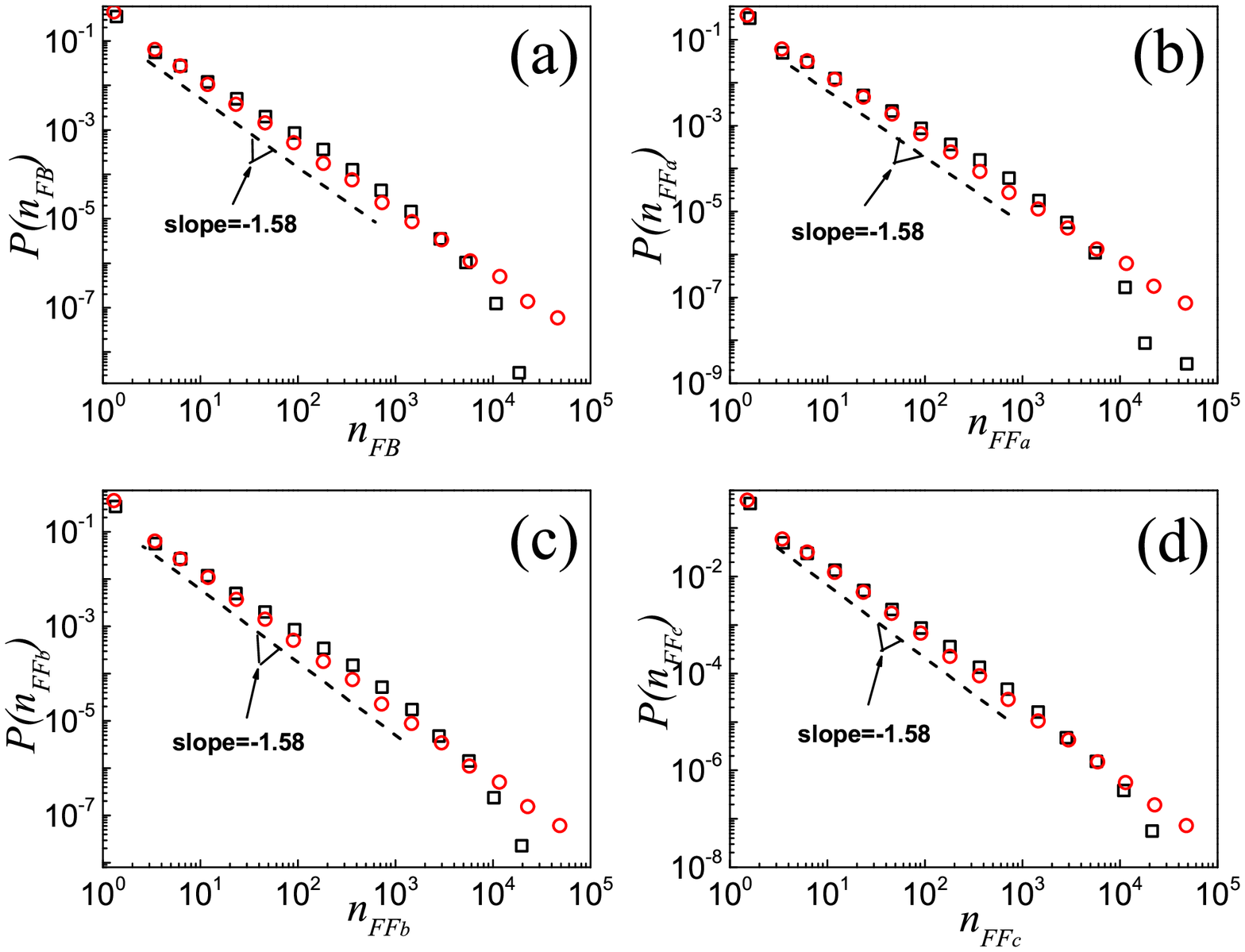}
\end{center}
\caption{\textbf{Distributions of four basic closed triples in the
Epinions social network and the model.} Distributions of closed
triples corresponding to (a) $FB$, (b) $FF_{a}$, (c) $FF_{b}$,
and (d) $FF_{c}$ loops in the Epinions social network (black
squares) and in the simulated network based on the model (red
circles). Analytic treatment (see Eqs.~\eqref{gamma-Delta}
and~\eqref{m-p}) suggests a scaling behavior with an exponent
$-1.58$, as shown by the dash lines. Data points are averages over the logarithmic bins of the $n_{FB}$, $n_{FFa}$, $n_{FFb}$ and $n_{FFc}$, respectively.}
\label{four-triangle-of-Epinions}
\end{figure}

\begin{figure}[!ht]
\begin{center}
\includegraphics[width=4in,height=3in]{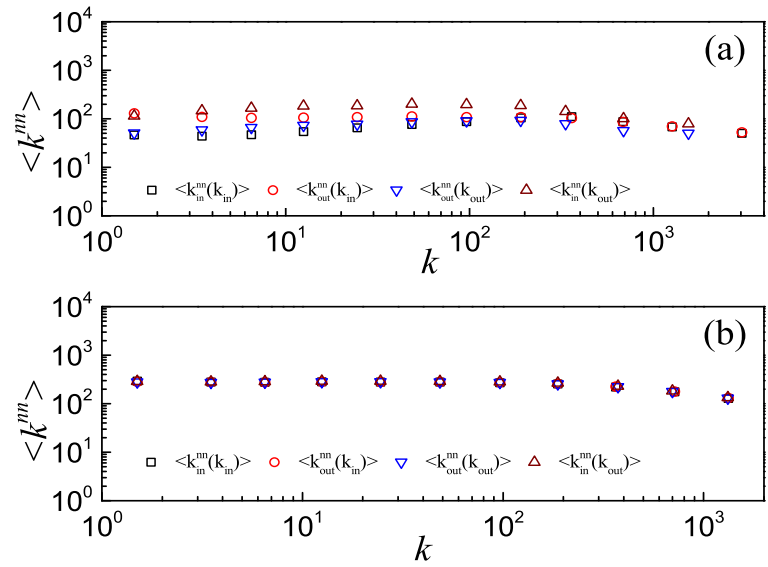}
\end{center}
\caption{\textbf{Degree correlations in the Epinions social network
and the model.}  Results of degree correlations as measured by
four quantities corresponding to the average nearest neighbor
degree $<k_{in}^{nn}(k_{in})>$ (squares),
$<k_{out}^{nn}(k_{in})>$ (circles), $<k_{out}^{nn}(k_{out})>$
(triangles), and $<k_{in}^{nn}(k_{out})>$ (inverted triangles)
for (a) the Epinions social network and (b) simulated network
based on the model. Data points are averages over the logarithmic bins of the indegree $k_{in}$ or outdegree $k_{out}$.}
\label{knn-of-Epinions}
\end{figure}

\begin{figure}[!ht]
\begin{center}
\includegraphics[width=4in]{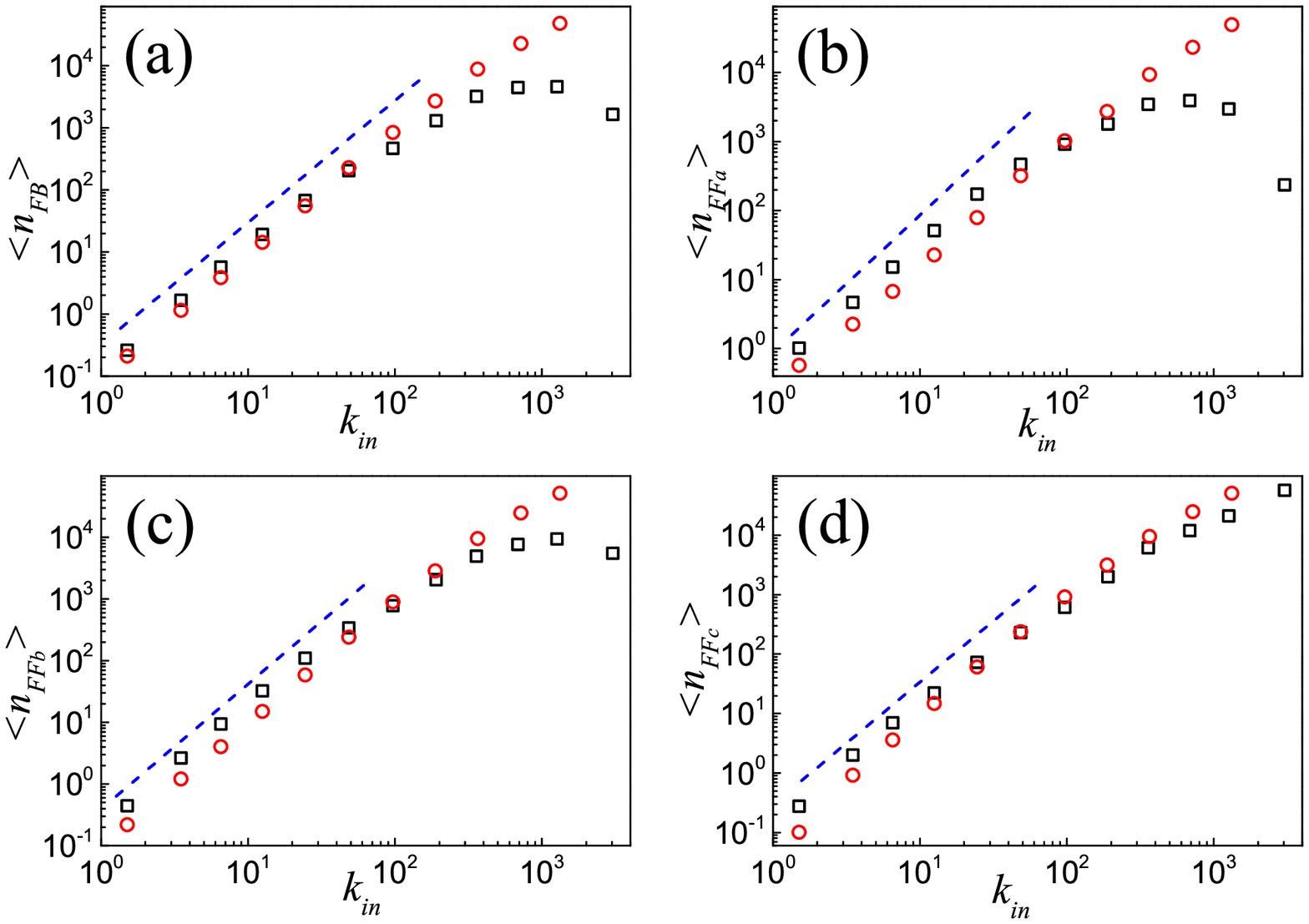}
\end{center}
\caption{\textbf{Mean number of the four closed triples for nodes
with the same indegree in the Epinions social network
and the model.}  Results for the mean number of closed
triples corresponding to (a) $FB$, (b) $FF_{a}$, (c) $FF_{b}$,
and (d) $FF_{c}$ loops for nodes with the same indegree are shown
for the Epinions social network (black squares) and simulated
network (red circles) based on the model.  Analytic treatment (see
Eq.~\eqref{nFB-scaling}) gives a scaling behavior with an exponent
$2$, as indicated by the dash lines. Data points are averages over the logarithmic bins of the indegree $k_{in}$.}
\label{kinAveTriangle-of-Epinions}
\end{figure}

\begin{figure}[!ht]
\begin{center}
\includegraphics[width=4in]{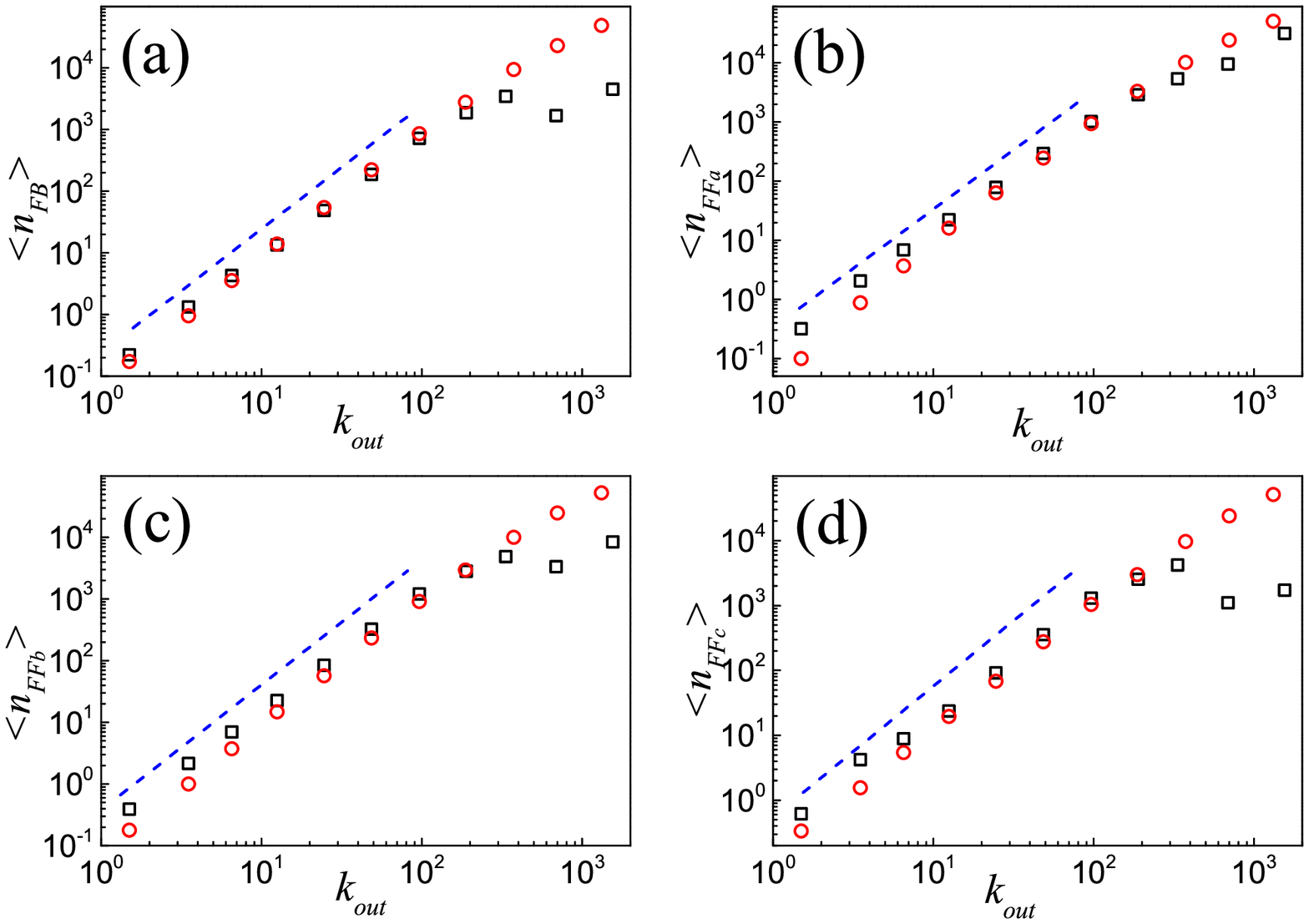}
\end{center}
\caption{\textbf{Mean number of the four closed triples for nodes
with the same outdegree in the Epinions social network
and the model.}  Results for the mean number of closed
triples corresponding to (a) $FB$, (b) $FF_{a}$, (c) $FF_{b}$,
and (d) $FF_{c}$ loops for nodes with the same outdegree are shown
for the Epinions social network (black squares) and simulated
network (red circles) based on the model.  Analytic treatment (see
Eq.~\eqref{nFB-scaling}) gives a scaling behavior with an exponent
$2$, as indicated by the dash lines. Data points are averages over the logarithmic bins of the outdegree $k_{out}$.}
\label{koutAveTriangle-of-Epinions}
\end{figure}

\begin{figure}[!ht]
\begin{center}
\includegraphics[width=4in]{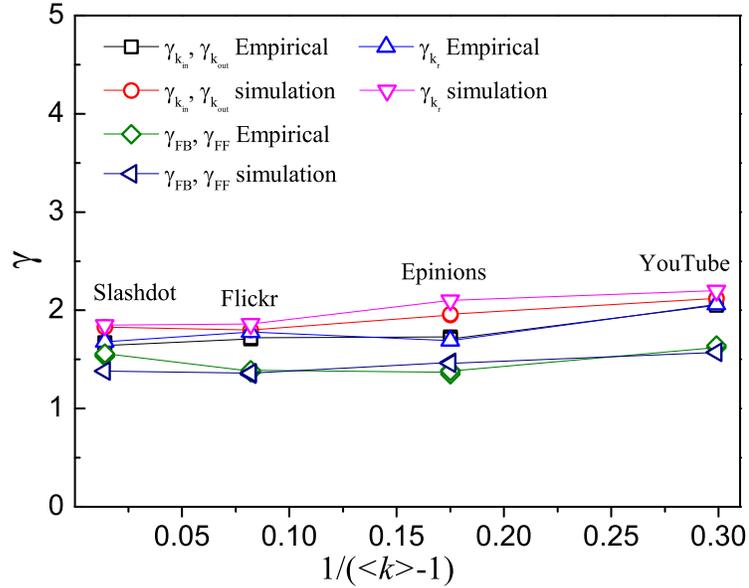}
\end{center}
\caption{\textbf{Values of the $\gamma$-exponents for various
distributions.}  Values of the $\gamma$-exponents for the various
distributions as determined by the maximum likelihood estimation
against $1/(\langle k \rangle -1)$ for each of the four
large-scale online social networks and the corresponding
simulated networks based on the model.  The lines are only guides
to the eye.} \label{exponents}
\end{figure}

\textbf{Figure \ref{four-triangles}. Three possible unclosed
triples and four basic closed triples for a
focal node (red).}\\
\textbf{Figure \ref{kin-and-kout-of-Epinions}. Indegree (a) and
outdegree (b) distributions of the Epinions social network (black squares) and
simulation results (red circles) based on the model.}\\
\textbf{Figure \ref{kintokout-of-Epinions}. Relationship between
the indegree and the outdegree of nodes in the Epinions social network and the model.}\\
\textbf{Figure \ref{kr-of-Epinions}. Reciprocal degree
distributions of the Epinions social network and the model.}\\
\textbf{Figure \ref{kinavebiedge-koutavebiedge-Epinions}. Mean
reciprocal degree of nodes with (a) the same indegree and (b) the
same outdegree in the Epinions social
network and in the model.}\\
\textbf{Figure \ref{four-triangle-of-Epinions}. Distributions of
four basic closed triples in the Epinions social network and the
model.}\\
\textbf{Figure \ref{knn-of-Epinions}. Degree correlations in the
Epinions social network and the model.}\\
\textbf{Figure \ref{kinAveTriangle-of-Epinions}. Mean number of
the four closed triples for nodes with the same indegree in the
Epinions social network and the model.}\\
\textbf{Figure \ref{koutAveTriangle-of-Epinions}. Mean number of
the four closed triples for nodes with the same outdegree in the
Epinions social network and the model.}\\
\textbf{Figure \ref{exponents}.  Values of the $\gamma$-exponents
for various distributions.}\\

\newpage

\begin{appendix}
\renewcommand
\thefigure{S\arabic{figure}}
\renewcommand
\thetable{S\arabic{table}}
\setcounter{figure}{0}
\setcounter{table}{0}
\newpage
\clearpage
\begin{flushleft}
{\Large \textbf{Appendix SI: Emergence of
scale-free close-knit friendship structure in online social
networks}}
\\
\end{flushleft}

This Appendix is divided into three sections. In
Sec. 1, we give the exponents of power-law fits carried out by the
maximum likelihood estimation. In Sec. 2, we present the degree
correlation by the Pearson correlation coefficient. In Sec. 3,
statistical properties of $Slashdot$, $Flickr$ and $YouTube$ are
presented.

\newpage

\begin{table}[!ht]
\caption{ \bf{The exponents of various distributions obtained by
power-law fits of real online social networks and the simulated
network based on the model using the maximum likelihood
estimation.} $x_{min}$ is the lower bound of the range for
fitting a power-law distribution, $\gamma$ is the corresponding
exponent and $KS$ is the goodness-of-fit value based on the
Kolmogorov-Smirnov statistic.} \centering
\begin{tabular}{lcccccccc}
\hline
\multicolumn{1}{c}{} & \multicolumn{3}{c}{Epinions} & & \multicolumn{3}{c}{model}\\
\cline{2-4}
\cline{6-8}
Distribution & $\gamma$ & $x_{min}$ & \textit{KS} & & $\gamma$ & $x_{min}$ & \textit{KS}\\
\hline
$P(k_{in})$ & 1.73 & 2 & 0.018 & & 1.95 & 3 & 0.009\\
$P(k_{out})$ & 1.71 & 2 & 0.022 & & 1.96 & 3 & 0.014\\
$P(k_{r})$ & 1.69 & 1 & 0.01 & & 2.1 & 4 & 0.025\\
$P(n_{FB})$ & 1.37 & 2 & 0.034 & & 1.47 & 3 & 0.006 \\
$P(n_{FFa})$ & 1.39 & 2 & 0.04 & & 1.46 & 5 & 0.009\\
$P(n_{FFb})$ & 1.35 & 2 &0.029 & & 1.46 & 2 & 0.008 \\
$P(n_{FFc})$ & 1.38 & 3 & 0.038 & & 1.46 & 5 & 0.009\\
\hline
\multicolumn{1}{c}{} & \multicolumn{3}{c}{Slashdot} & & \multicolumn{3}{c}{model}\\
\cline{2-4}
\cline{6-8}
Distribution & $\gamma$ & $x_{min}$ & \textit{KS} & & $\gamma$ & $x_{min}$ & \textit{KS}\\
\hline
$P(k_{in})$ & 1.67 & 2 & 0.045 & & 1.83 & 3 & 0.004\\
$P(k_{out})$ & 1.64 & 2 & 0.047 & & 1.83 & 3 & 0.003\\
$P(k_{r})$ & 1.68 & 2 & 0.047 & & 1.85 & 2 & 0.004\\
$P(n_{FB})$ & 1.55 & 6 & 0.027 & & 1.38 & 4 & 0.009 \\
$P(n_{FFa})$ & 1.53 & 6 & 0.029 & & 1.38 & 4 & 0.015\\
$P(n_{FFb})$ & 1.54 & 6 &0.03 & & 1.38 & 4 & 0.008 \\
$P(n_{FFc})$ & 1.56 & 6 & 0.034 & & 1.38 & 4 & 0.017\\
\hline
\multicolumn{1}{c}{} & \multicolumn{3}{c}{Flickr} & & \multicolumn{3}{c}{model}\\
\cline{2-4}
\cline{6-8}
Distribution & $\gamma$ & $x_{min}$ & \textit{KS} & & $\gamma$ & $x_{min}$ & \textit{KS}\\
\hline
$P(k_{in})$ & 1.71 & 2 & 0.015 & & 1.8 & 4 & 0.004\\
$P(k_{out})$ & 1.72 & 5 & 0.03 & & 1.8 & 4 & 0.005\\
$P(k_{r})$ & 1.78 & 5 & 0.025 & & 1.86 & 1 & 0.006\\
$P(n_{FB})$ & 1.38 & 2 & 0.029 & & 1.36 & 4 & 0.007\\
$P(n_{FFa})$ & 1.38 & 5 & 0.029 & & 1.35 & 4 & 0.01\\
$P(n_{FFb})$ & 1.37 & 2 &0.029 & & 1.36 & 4 & 0.007 \\
$P(n_{FFc})$ & 1.39 & 4 & 0.033 & & 1.36 & 5 & 0.009\\
\hline
\multicolumn{1}{c}{} & \multicolumn{3}{c}{YouTube} & & \multicolumn{3}{c}{model}\\
\cline{2-4}
\cline{6-8}
Distribution & $\gamma$ & $x_{min}$ & \textit{KS} & & $\gamma$ & $x_{min}$ & \textit{KS}\\
\hline
$P(k_{in})$ & 2.05 & 3 & 0.015 & & 2.12 & 3 & 0.013\\
$P(k_{out})$ & 2.08 & 6 & 0.018 & & 2.12 & 3 & 0.015\\
$P(k_{r})$ & 2.06 & 6 & 0.019 & & 2.2 & 3 & 0.006\\
$P(n_{FB})$ & 1.62 & 4 & 0.031 & & 1.57 & 2 & 0.01 \\
$P(n_{FFa})$ & 1.62 & 4 & 0.044 & & 1.57 & 4 & 0.01\\
$P(n_{FFb})$ & 1.62 & 4 & 0.029 & & 1.57 & 4 & 0.005 \\
$P(n_{FFc})$ & 1.64 & 4 & 0.042 & & 1.57 & 4 & 0.01\\
\hline
\end{tabular}
\label{tab:exponentsforEpinions}
\end{table}

\newpage
\section*{2. Degree correlation measured by
Pearson correlation coefficient}
\begin{table}[!ht]
\caption{\bf{Pearson correlation coefficient.} $r(in,in)$
quantifies the tendency of nodes with a high indegree to be
connected to another node with a high indegree.  The other
quantities carry a similar interpretation.} \centering
\begin{tabular}{lrrrrr}
\hline
Datasets & $r(in,in)$ & $r(in,out)$ & $r(out,out)$ & $r(out, in)$\\
\hline
Epinions & -0.009 & 0.073 & -0.016 & -0.053\\
Slashdot & -0.068 & -0.059 & -0.064 & -0.071\\
Flickr & 0.06 & 0.055 & -0.0025 & -0.001\\
YouTube & -0.03 & -0.032 & -0.036 & -0.035\\
\hline
\end{tabular}
\end{table}

\section*{3. Statistical properties of
$Slashdot$, $Flickr$ and $YouTube$ social networks}

\begin{figure}[!ht]
\begin{center}
\includegraphics[width=4in]{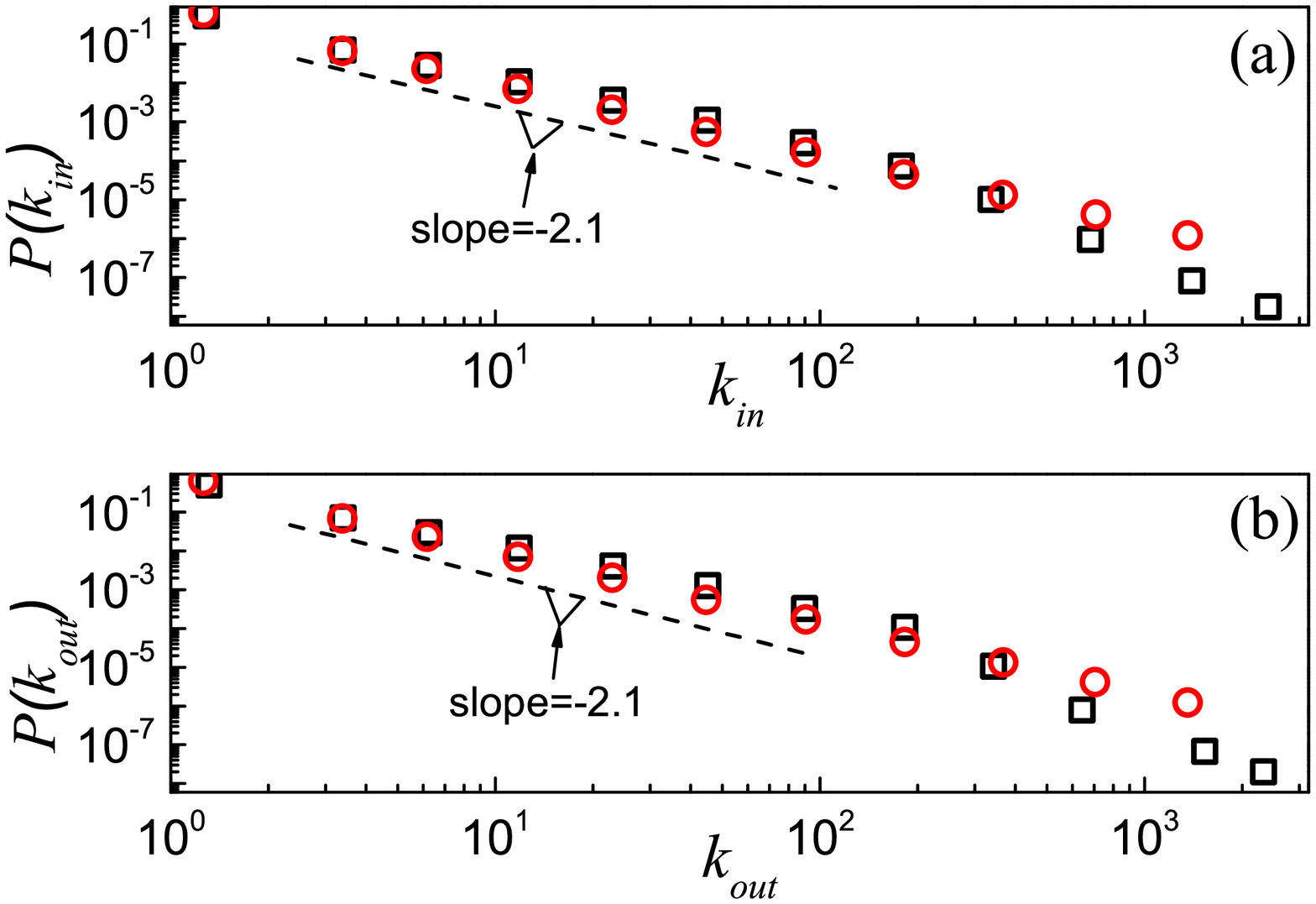}
\end{center}
\caption{\textbf{Indegree (a) and
outdegree (b) distributions of the Slashdot social network (black squares) and
simulation results (red circles) based on the model.}
The dashed lines in both panels have a
slope $-2.1$ as the analytic results in Eqs. (17) and (31)
suggested. The simulated network is generated by the model with
the parameters $N=82168$, $m\approx5.14$ and $p\approx0.67$ as
determined by the mean degree $\langle k \rangle$ and reciprocity
of the Slashdot social network. Data points are averages over the logarithmic bins of the indegree $k_{in}$ and outdegree $k_{out}$, respectively.}
\end{figure}

\begin{figure}[!ht]
\begin{center}
\includegraphics[width=4in]{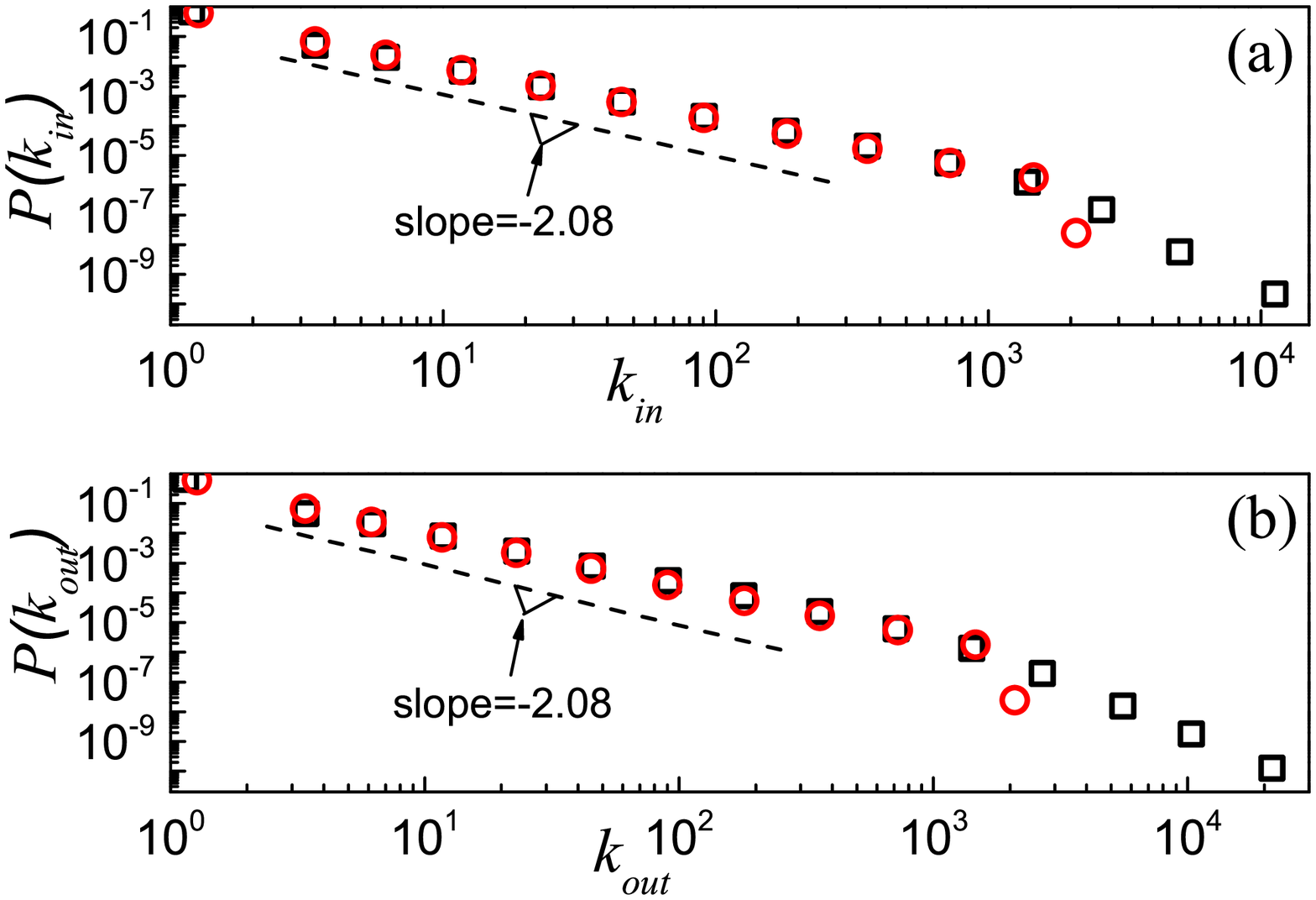}
\end{center}
\caption{\textbf{Indegree (a) and
outdegree (b) distributions of the Flickr social network (black squares) and
simulation results (red circles) based on the model.}
The dashed lines in both panels have a
slope $-2.08$ as the analytic results in Eqs. (17) and (31)
suggested. The simulated network is generated by the model with
the parameters $N=100000$, $m\approx8.07$ and $p\approx0.39$ as
determined by the mean degree $\langle k \rangle$ and reciprocity
of the Flickr social network. Data points are averages over the logarithmic bins of the indegree $k_{in}$ and outdegree $k_{out}$, respectively.}
\end{figure}

\begin{figure}[!ht]
\begin{center}
\includegraphics[width=4in]{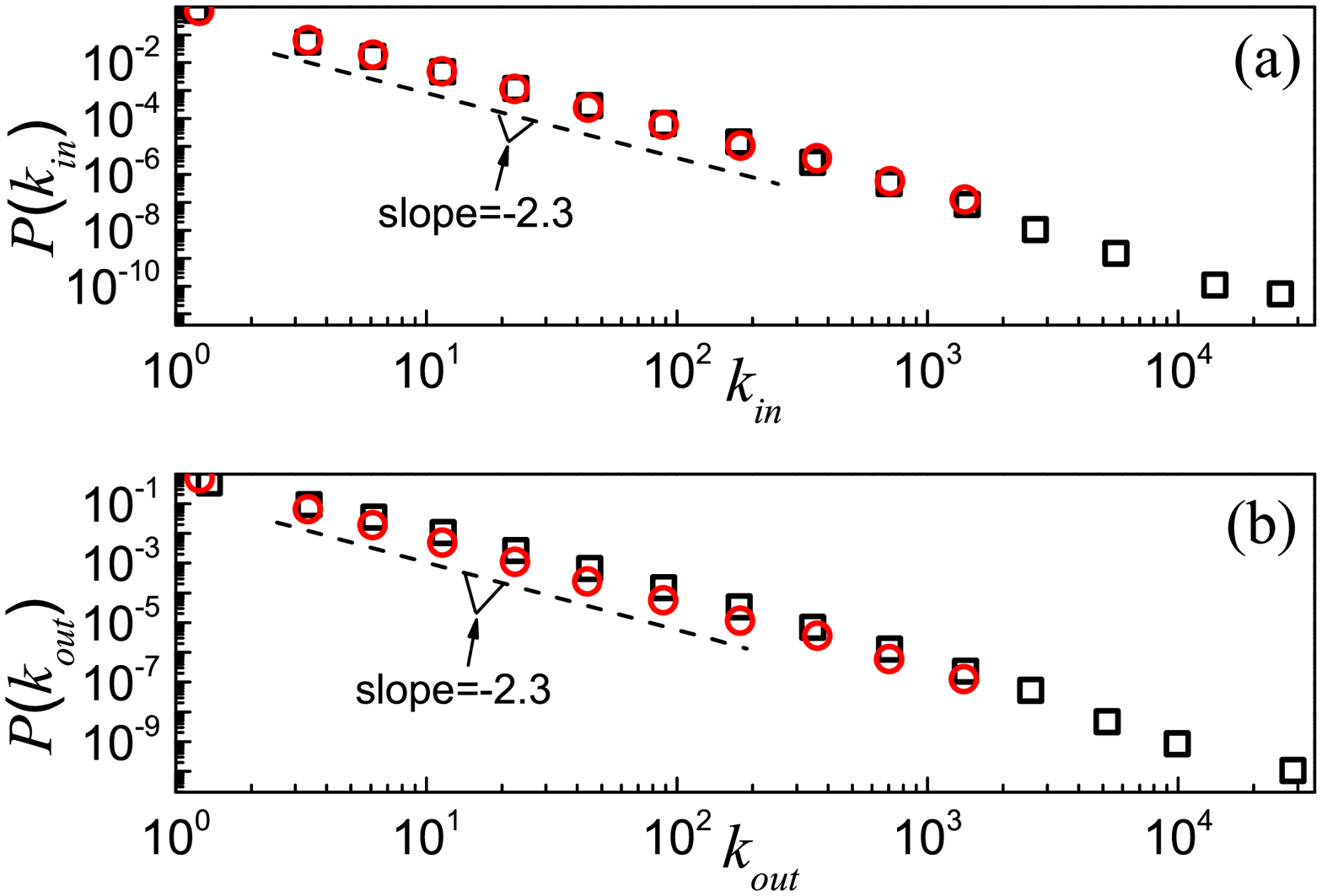}
\end{center}
\caption{\textbf{Indegree (a) and
outdegree (b) distributions of the YouTube social network (black squares) and
simulation results (red circles) based on the model.}
The dashed lines in both panels have a
slope $-2.3$ as the analytic results in Eqs. (17) and (31)
suggested. The simulated network is generated by the model with
the parameters $N=100000$, $m\approx4.34$ and $p\approx0.08$ as
determined by the mean degree $\langle k \rangle$ and reciprocity
of the YouTube social network. Data points are averages over the logarithmic bins of the indegree $k_{in}$ and outdegree $k_{out}$, respectively.}
\end{figure}

\begin{figure}[!ht]
\begin{center}
\includegraphics[width=3in]{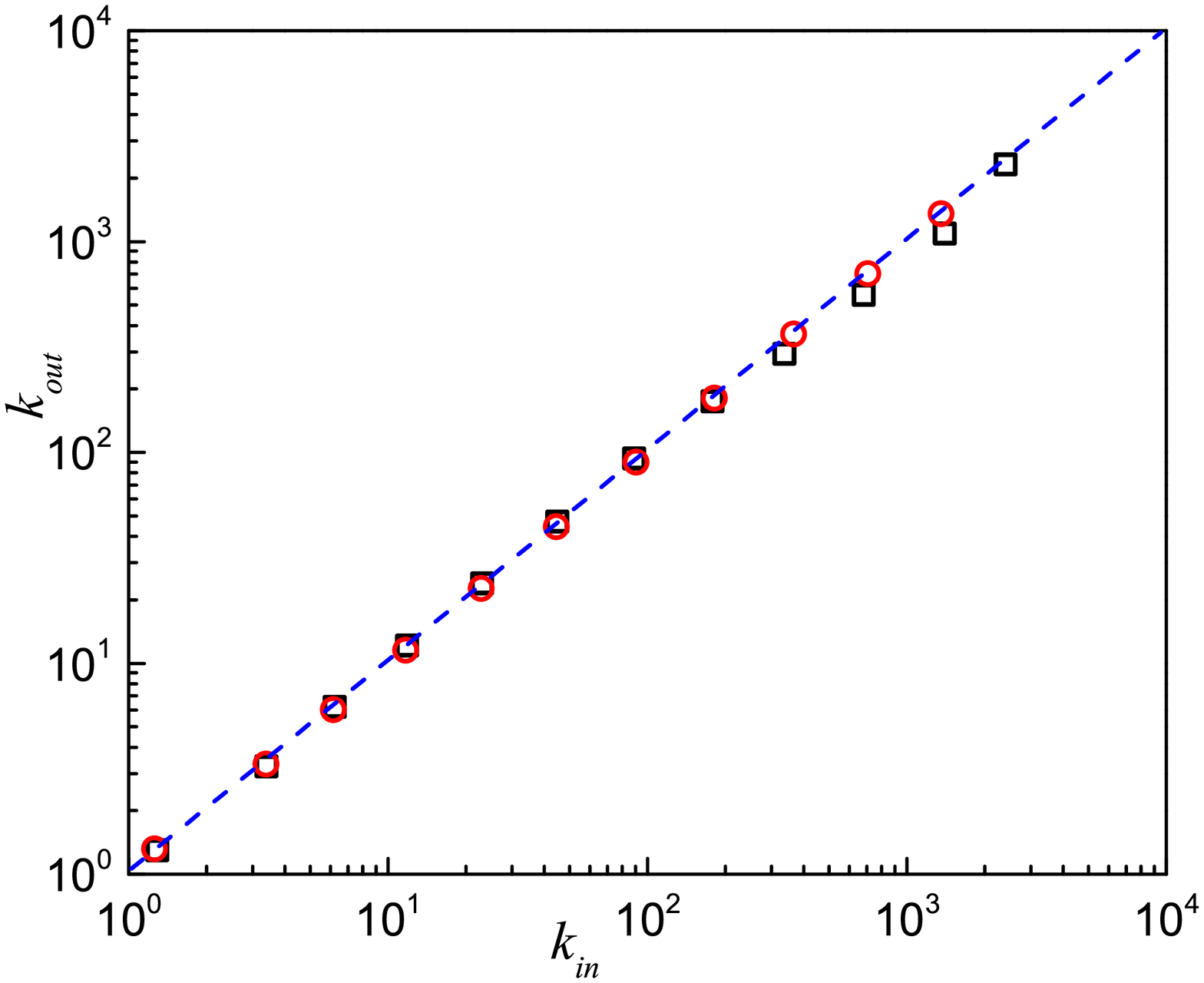}
\end{center}
\caption{\textbf{Relationship between the
indegree and the outdegree of nodes in the Slashdot social
network and the model.}
Results of the Slashdot social network (black
squares) and simulation results (red circles) based on the model
are shown. The blue dash line represents the relation function
$k_{in}=k_{out}$. Data points are averages over the logarithmic bins of the indegree $k_{in}$.}
\end{figure}

\begin{figure}[!ht]
\begin{center}
\includegraphics[width=3in]{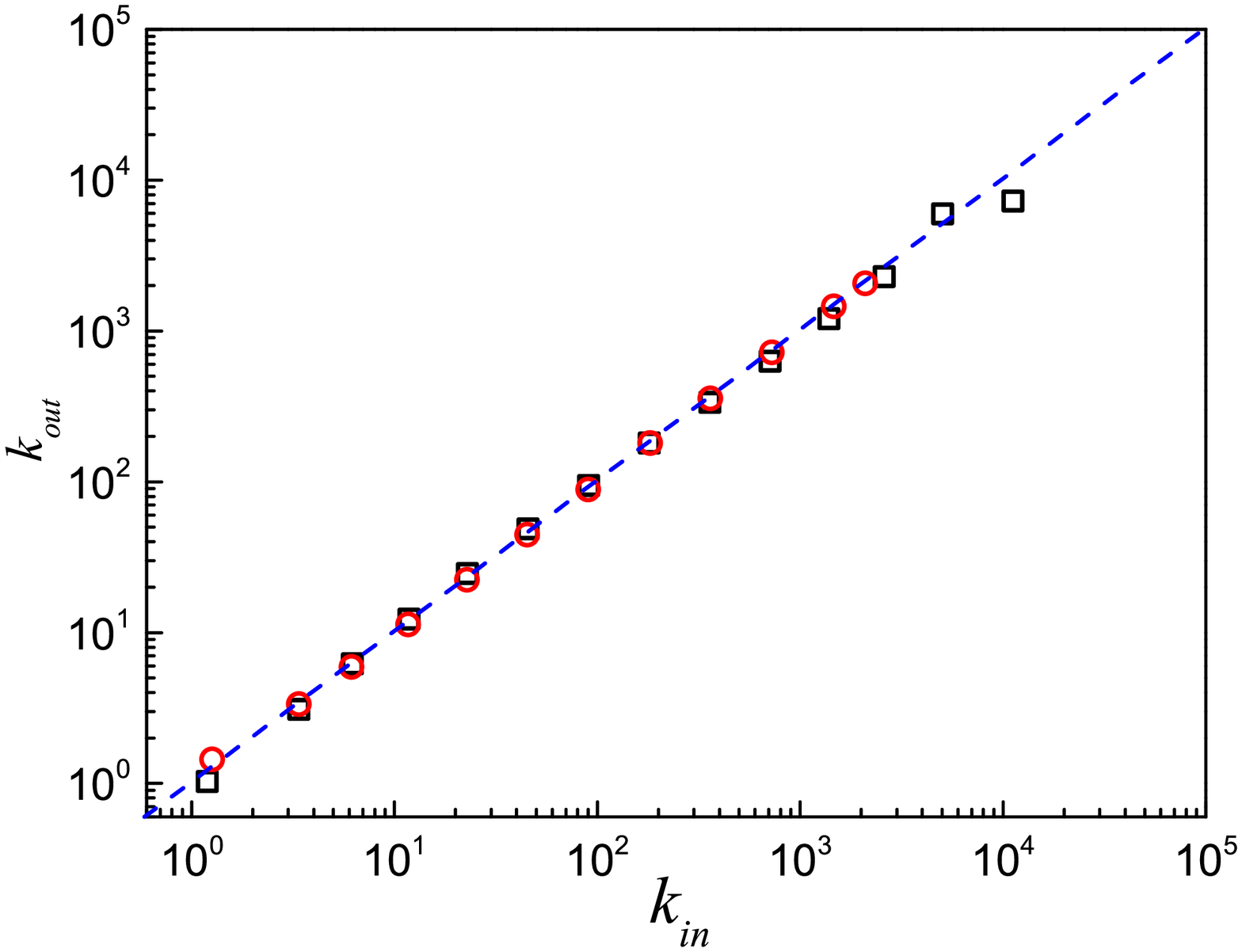}
\end{center}
\caption{\textbf{Relationship between the
indegree and the outdegree of nodes in the Flickr social network
and the model.}
Results of the Flickr social network (black squares) and simulation
results (red circles) based on the model are shown. The blue dash
line represents the relation function $k_{in}=k_{out}$. Data points are averages over the logarithmic bins of the indegree $k_{in}$.}
\end{figure}

\begin{figure}[!ht]
\begin{center}
\includegraphics[width=3in]{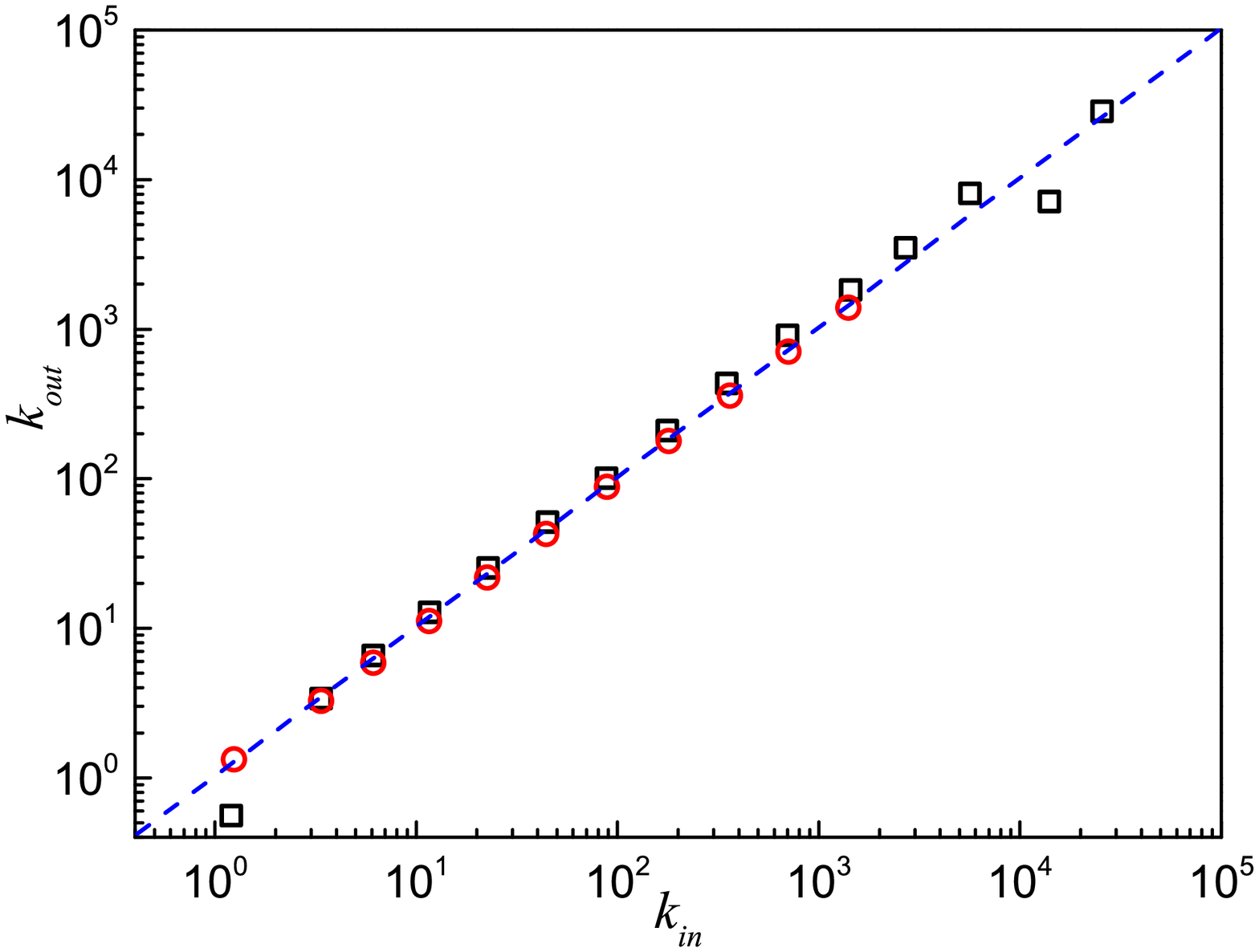}
\end{center}
\caption{\textbf{Relationship between the
indegree and the outdegree of nodes in the YouTube social network
and the model.}
Results of the YouTube social network (black
squares) and simulation results (red circles) based on the model
are shown. The blue dash line represents the relation function
$k_{in}=k_{out}$. Data points are averages over the logarithmic bins of the indegree $k_{in}$.}
\end{figure}

\clearpage

\begin{figure}
\begin{center}
\includegraphics[width=4in]{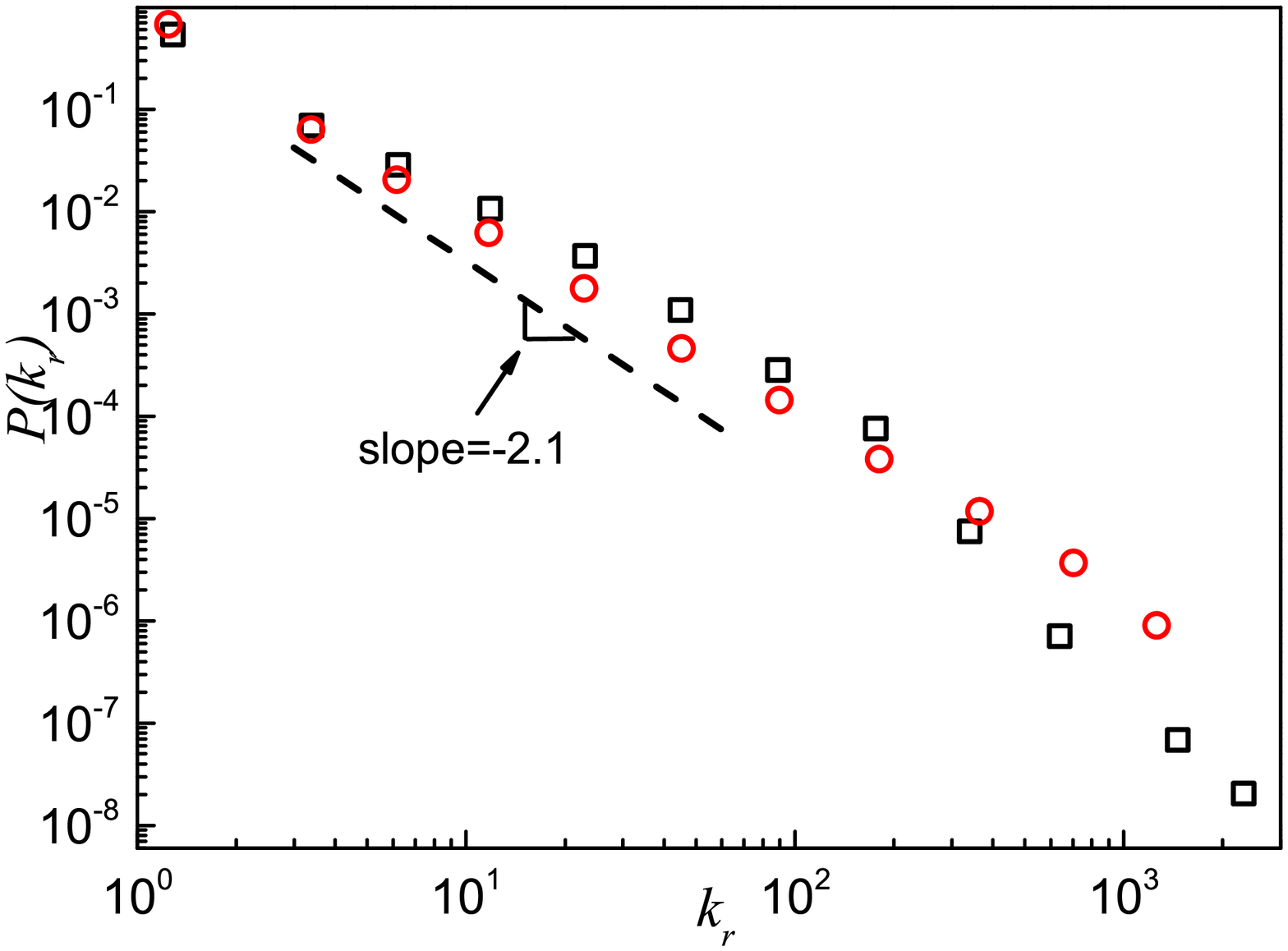}
\end{center}
\caption{\textbf{Reciprocal degree distributions of
the Slashdot social network and the model.}
Results of the Slashdot social network (black
squares) and simulation results (red circles) based on the model
are shown. Analytic treatment (see Eqs. (17) and (31)) suggests a
scaling behavior with an exponent $-2.1$, as shown by the dash
line. Data points are averages over the logarithmic bins of the reciprocal degree $k_{r}$.}
\label{kr-of-Slashdot}
\end{figure}

\begin{figure}[!ht]
\begin{center}
\includegraphics[width=4in]{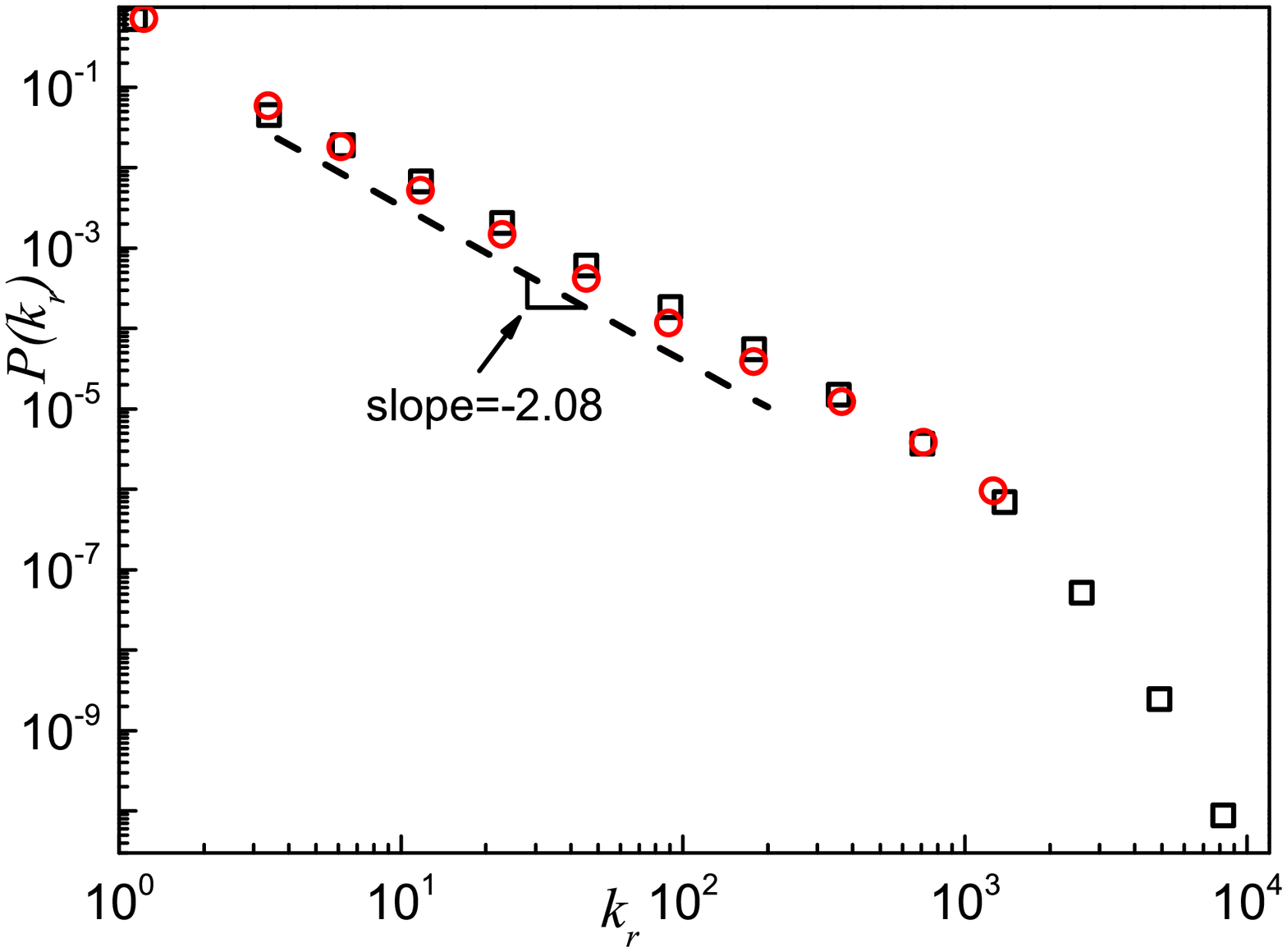}
\end{center}
\caption{\textbf{Reciprocal degree distributions of
the Flickr social network and the model.}
Results of the Flickr social network (black
squares) and simulation results (red circles) based on the model
are shown. Analytic treatment (see Eqs. (17) and (31)) suggests a
scaling behavior with an exponent $-2.08$, as shown by the dash
line. Data points are averages over the logarithmic bins of the reciprocal degree $k_{r}$.}
\label{kr-of-Flickr}
\end{figure}

\begin{figure}[!ht]
\begin{center}
\includegraphics[width=4in]{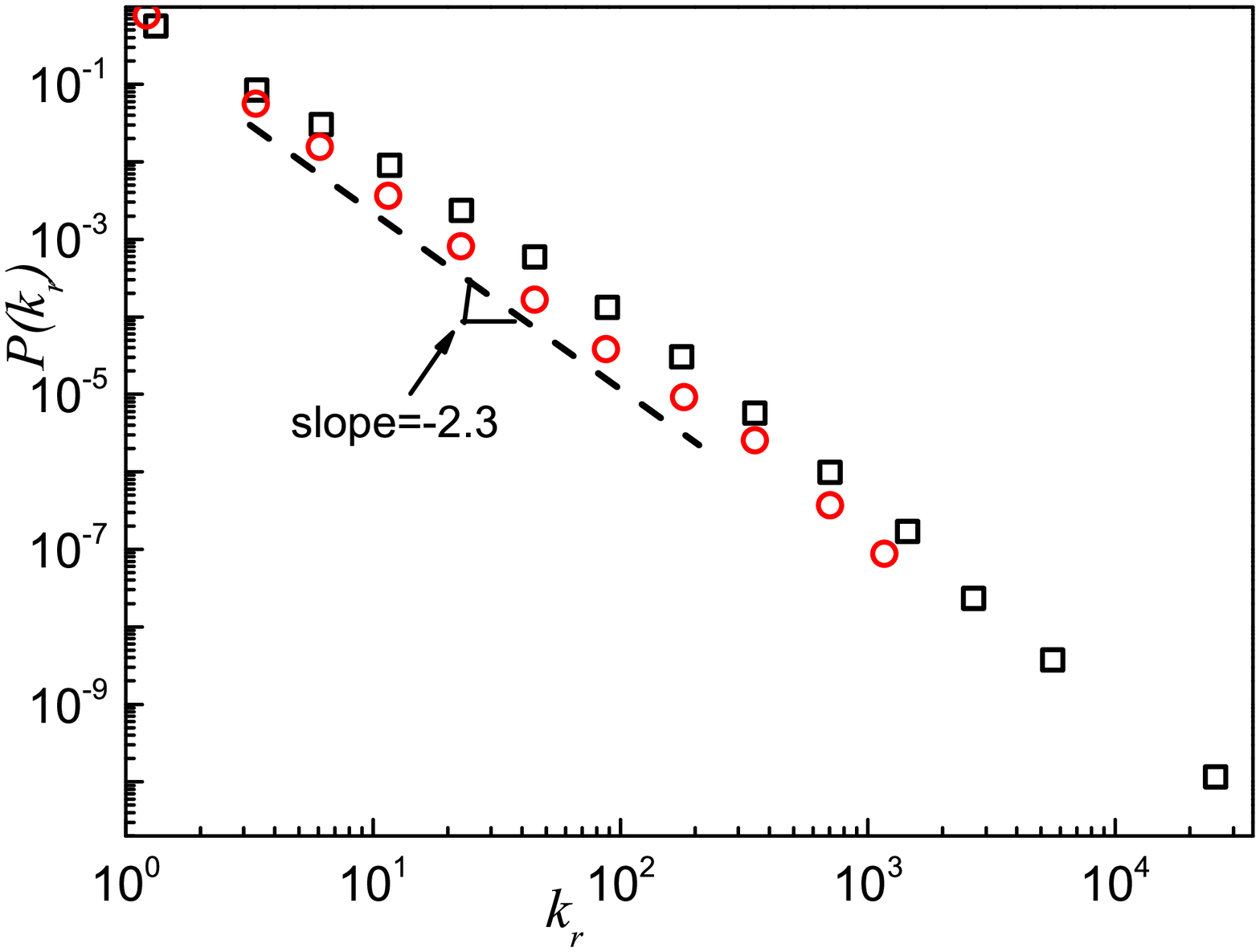}
\end{center}
\caption{\textbf{Reciprocal degree distributions of
the YouTube social network and the model.}
Results of the YouTube social network (black
squares) and simulation results (red circles) based on the model
are shown. Analytic treatment (see Eqs. (17) and (31)) suggests a
scaling behavior with an exponent $-2.3$, as shown by the dash
line. Data points are averages over the logarithmic bins of the reciprocal degree $k_{r}$.}
\label{kr-of-YouTube}
\end{figure}

\begin{figure}[!ht]
\begin{center}
\includegraphics[width=4in]{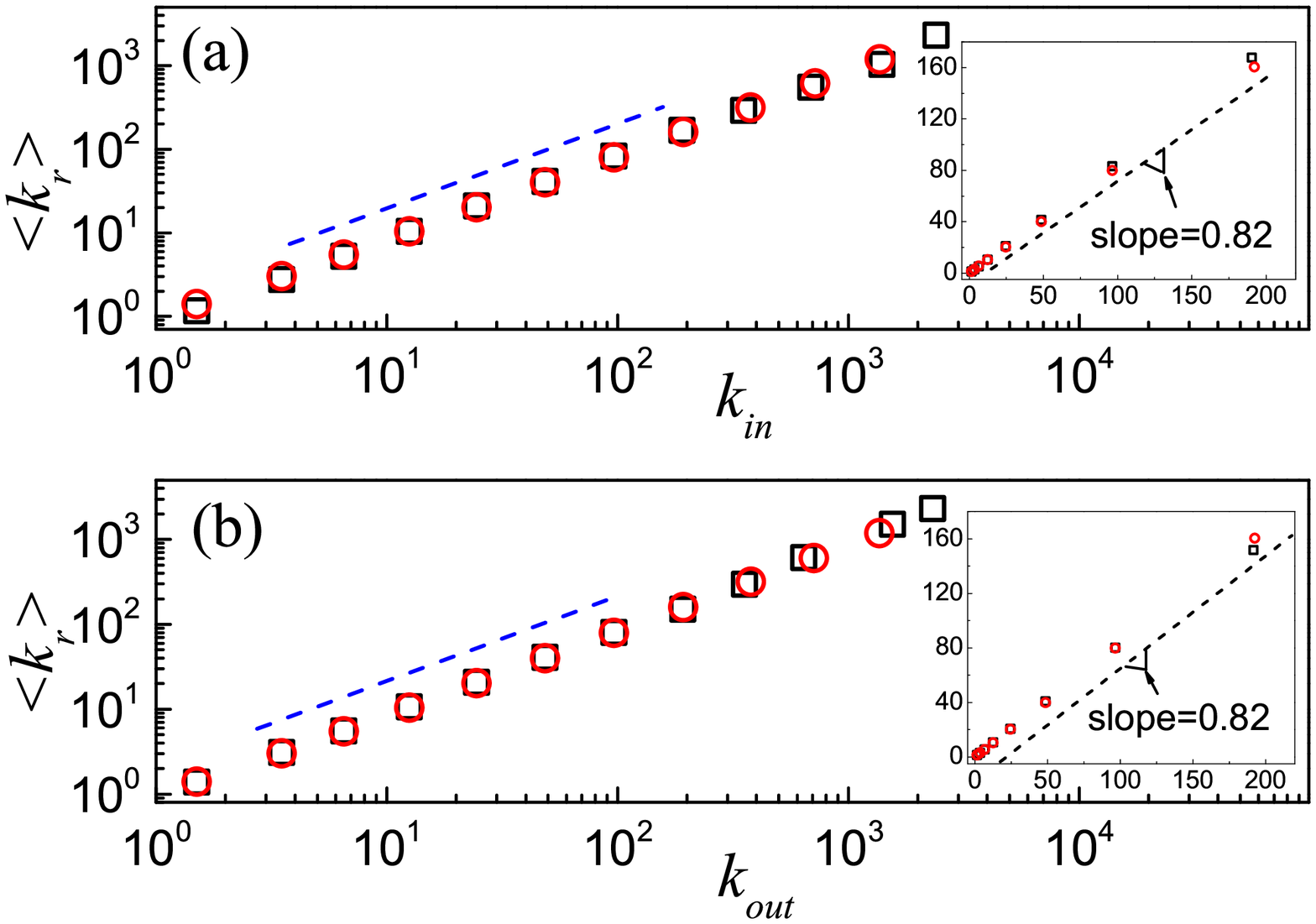}
\end{center}
\caption{\textbf{Mean reciprocal degree of nodes
with (a) the same indegree and (b) the same outdegree in the
Slashdot social network and in the model.}
Results of the Slashdot social network (black
squares) and simulation results (red circles) based on the model
are shown in a log-log scale in the main panels. Analytic
treatment suggests that $\langle k_{r}\rangle$ is linearly
dependent on $k_{in}$ and $k_{out}$, and the blue dash lines of
slope $1$ show its dependence. The inset in each panel shows the
results in a linear scale and the dash line has a slope of
$0.82$, as given by Eqs. $(20)$ and $(31)$.
Data points are averages over the logarithmic bins of the indegree $k_{in}$ and outdegree $k_{out}$, respectively.}
\label{kinavebiedge-koutavebiedge-Slashdot}
\end{figure}

\begin{figure}[!ht]
\begin{center}
\includegraphics[width=4in]{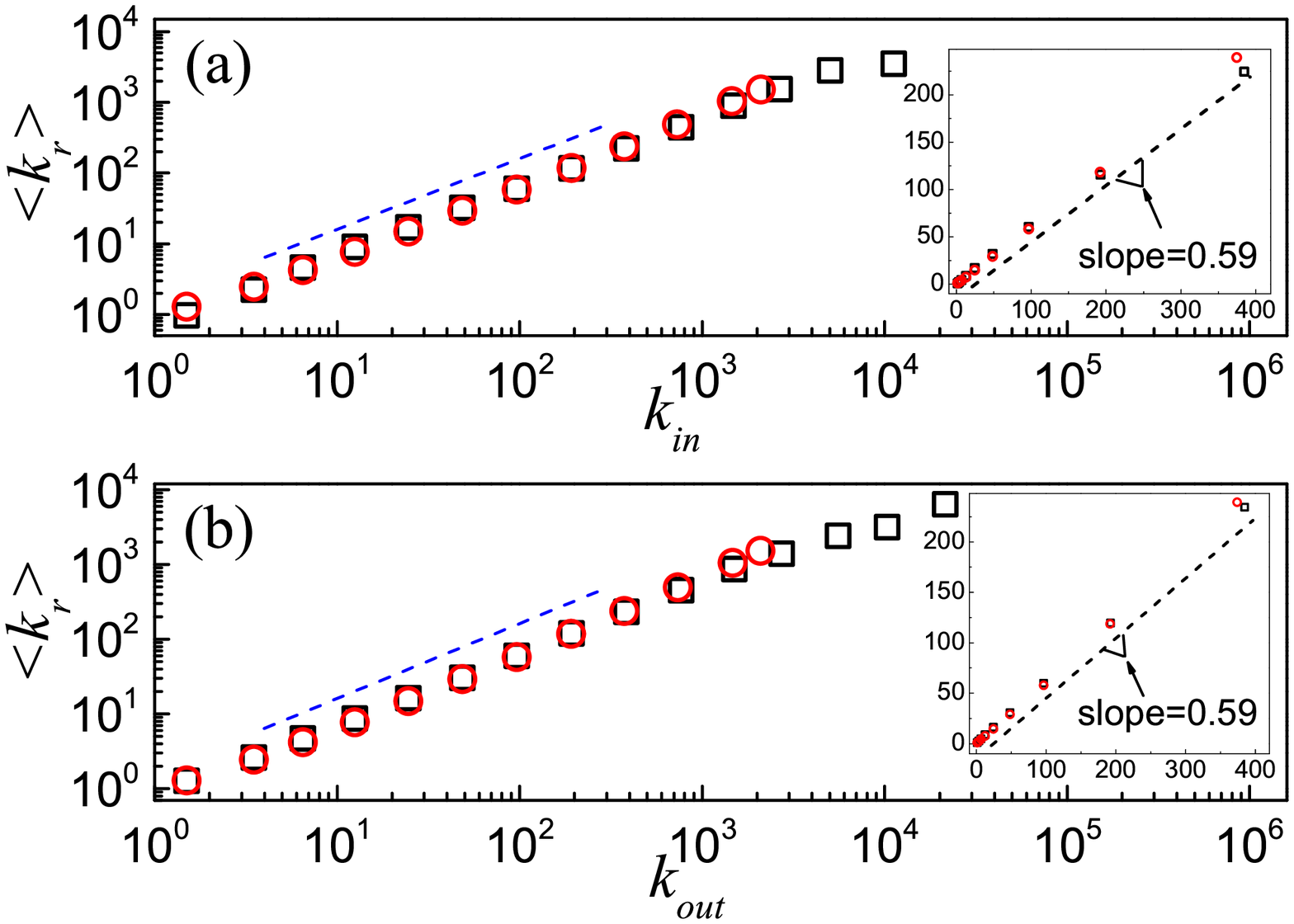}
\end{center}
\caption{\textbf{Mean reciprocal degree of nodes
with (a) the same indegree and (b) the same outdegree in the
Flickr social network and in the model.}
Results of the Flickr social network (black
squares) and simulation results (red circles) based on the model
are shown in a log-log scale in the main panels. Analytic
treatment suggests that $\langle k_{r}\rangle$ is linearly
dependent on $k_{in}$ and $k_{out}$, and the blue dash lines of
slope $1$ show its dependence. The inset in each panel shows the
results in a linear scale and the dash line has a slope of
$0.59$, as given by Eqs. $(20)$ and $(31)$.
Data points are averages over the logarithmic bins of the indegree $k_{in}$ and outdegree $k_{out}$, respectively.}
\label{kinavebiedge-koutavebiedge-Flickr}
\end{figure}

\begin{figure}[!ht]
\begin{center}
\includegraphics[width=4in]{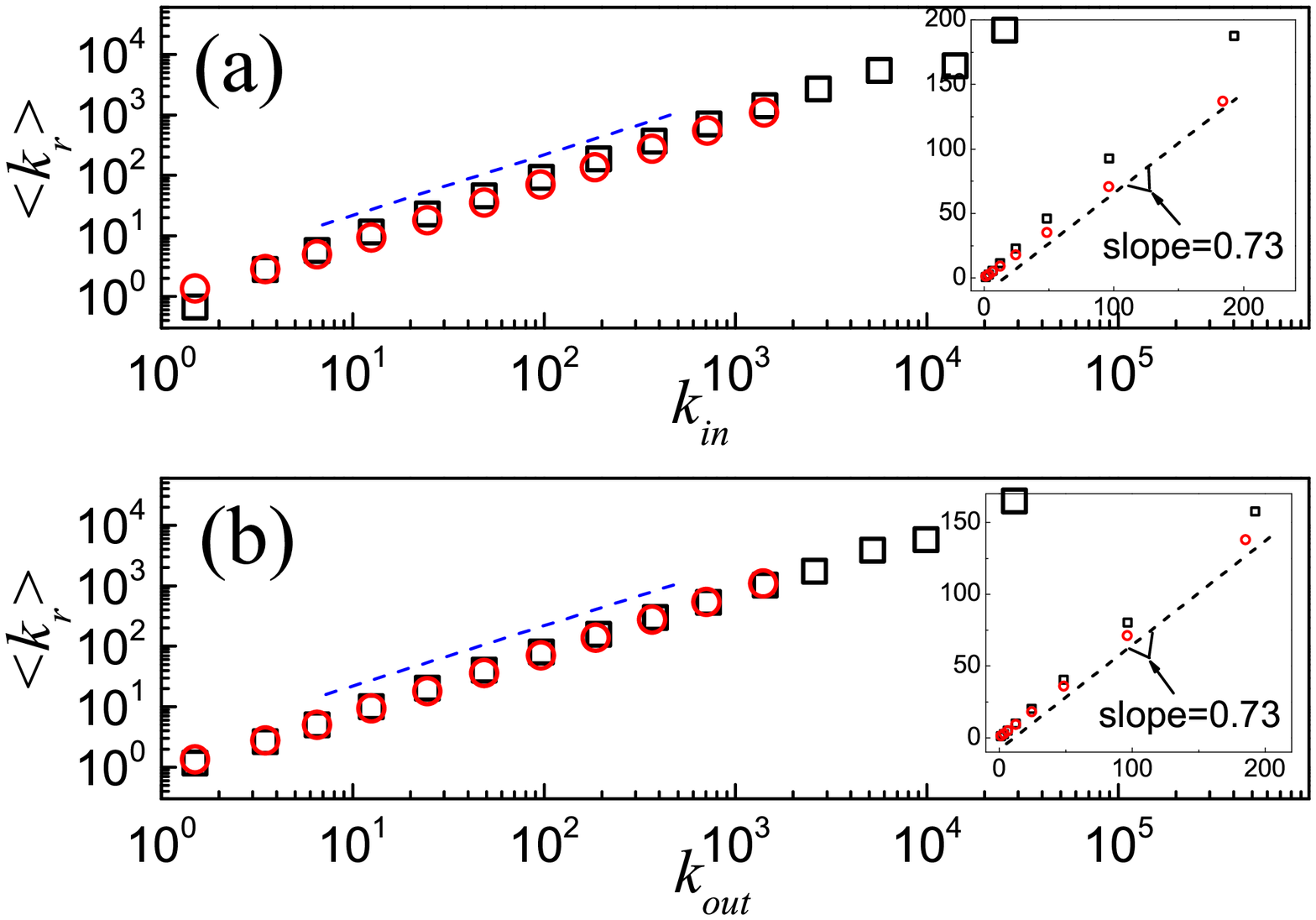}
\end{center}
\caption{\textbf{Mean reciprocal degree of nodes
with (a) the same indegree and (b) the same outdegree in the
YouTube social network and in the model.}
Results of the YouTube social network (black
squares) and simulation results (red circles) based on the model
are shown in a log-log scale in the main panels. Analytic
treatment suggests that $\langle k_{r}\rangle$ is linearly
dependent on $k_{in}$ and $k_{out}$, and the blue dash lines of
slope $1$ show its dependence. The inset in each panel shows the
results in a linear scale and the dash line has a slope of
$0.73$, as given by Eqs. $(20)$ and $(31)$.
Data points are averages over the logarithmic bins of the indegree $k_{in}$ and outdegree $k_{out}$, respectively.}
\label{kinavebiedge-koutavebiedge-YouTube}
\end{figure}

\begin{figure}[!ht]
\begin{center}
\includegraphics[width=4in]{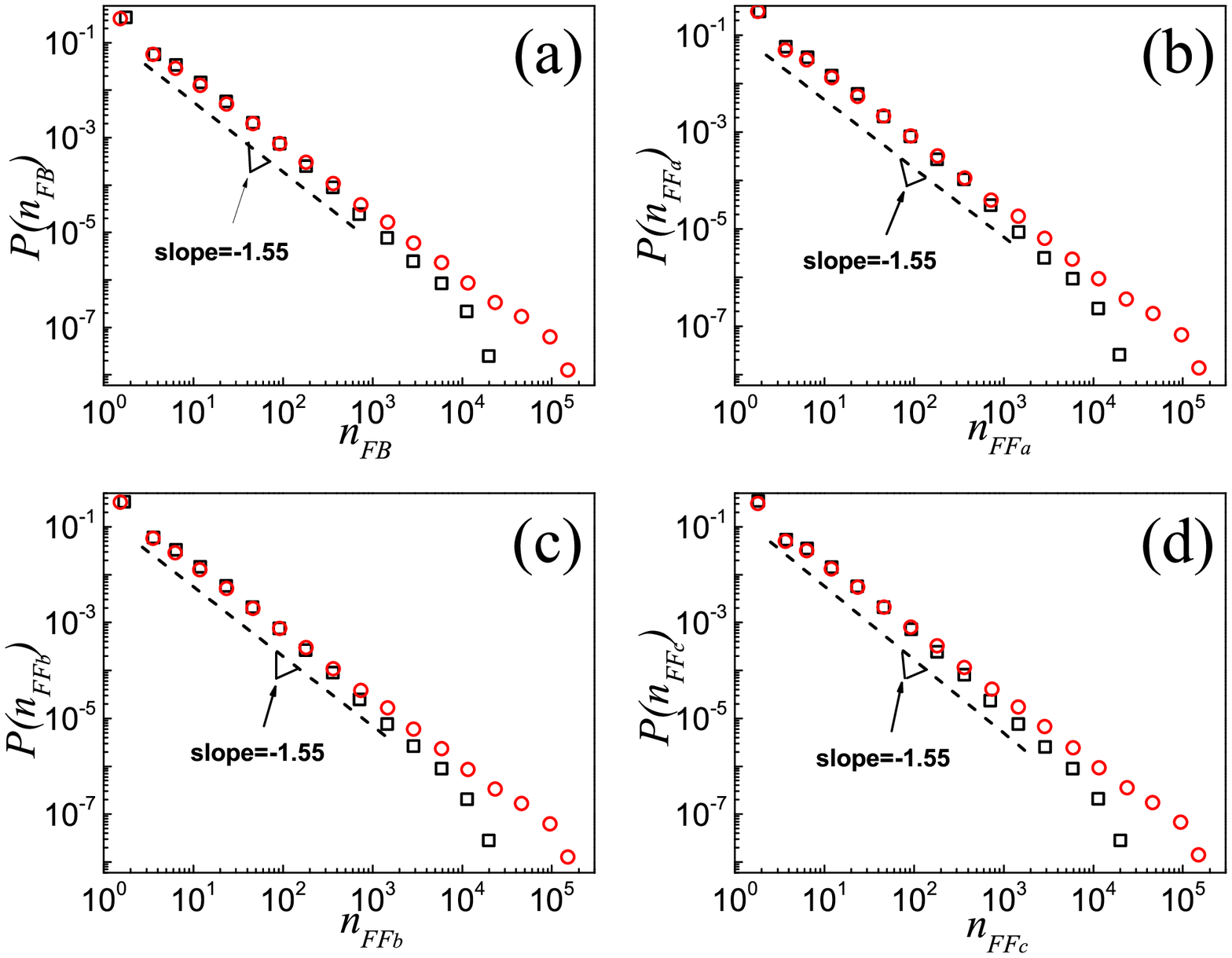}
\end{center}
\caption{\textbf{Distributions of four basic
closed triples in the slashdot social network and the model.}
Distributions of closed triples
corresponding to (a) $FB$, (b) $FF_{a}$, (c) $FF_{b}$, and (d)
$FF_{c}$ loops in the Slashdot social network (black squares) and
in the simulated network based on the model (red circles).
Analytic treatment (see Eqs. $(30)$ and $(31)$) suggests a
scaling behavior with an exponent $-1.55$, as shown by the dash
lines. Data points are averages over the logarithmic bins of the $n_{FB}$, $n_{FFa}$, $n_{FFb}$ and $n_{FFc}$, respectively.}
\end{figure}

\begin{figure}[!ht]
\begin{center}
\includegraphics[width=4in]{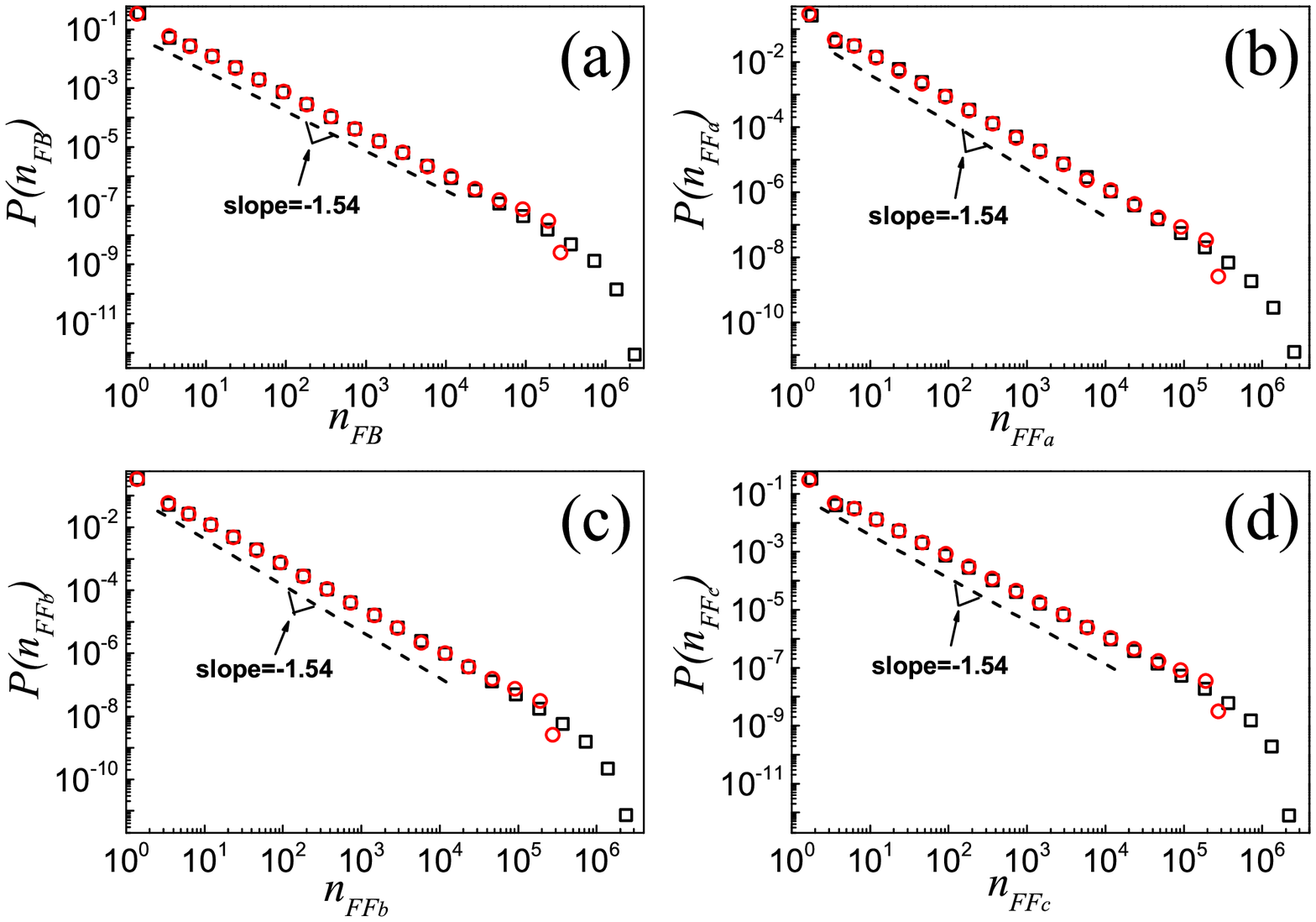}
\end{center}
\caption{\textbf{Distributions of four basic
closed triples in the Flickr social network and the model.}
Distributions of closed triples
corresponding to (a) $FB$, (b) $FF_{a}$, (c) $FF_{b}$, and (d)
$FF_{c}$ loops in the Flickr social network (black squares) and
in the simulated network based on the model (red circles).
Analytic treatment (see Eqs. $(30)$ and $(31)$) suggests a
scaling behavior with an exponent $-1.54$, as shown by the dash
lines. Data points are averages over the logarithmic bins of the $n_{FB}$, $n_{FFa}$, $n_{FFb}$ and $n_{FFc}$, respectively.}
\end{figure}

\begin{figure}[!ht]
\begin{center}
\includegraphics[width=4in]{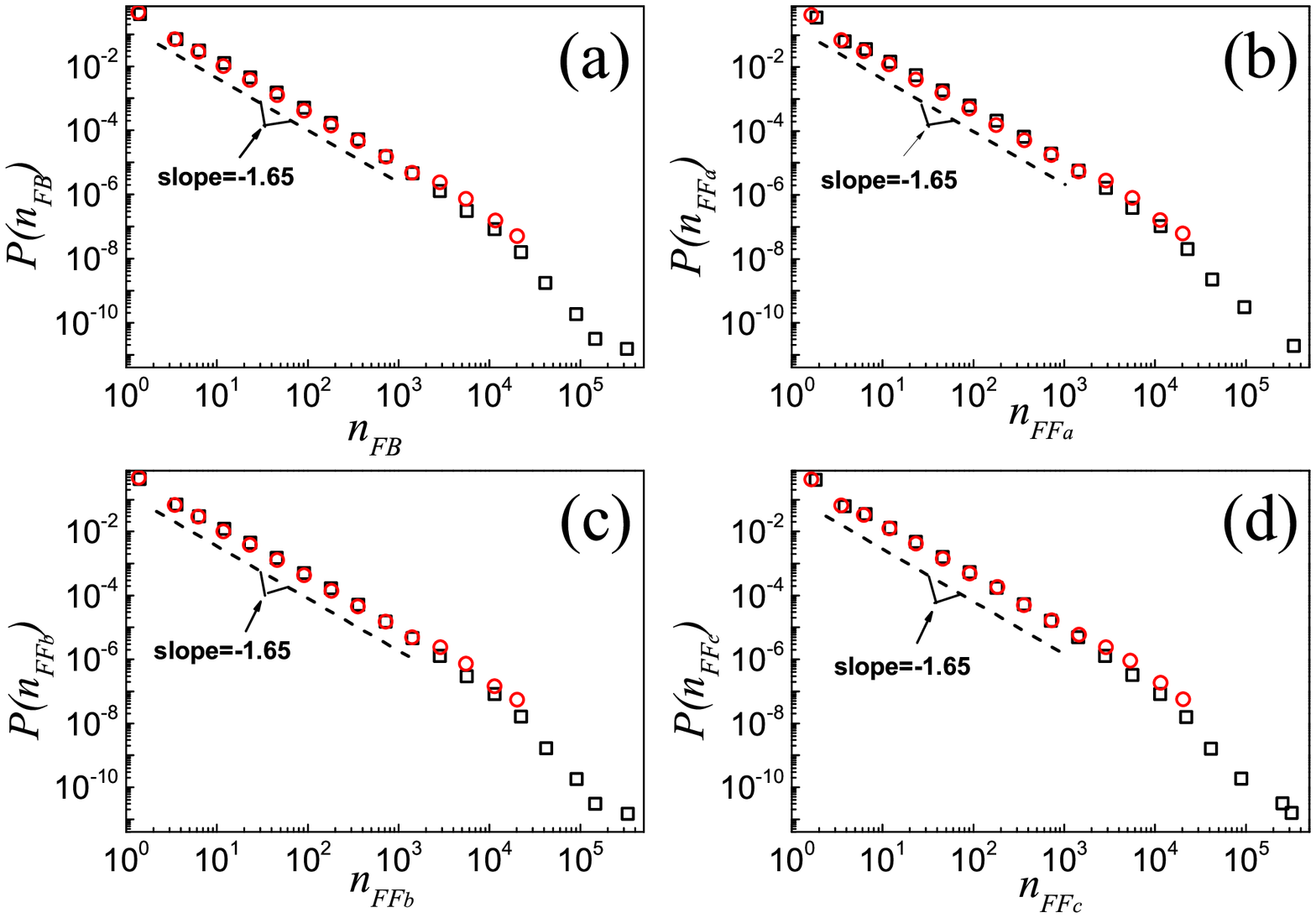}
\end{center}
\caption{\textbf{Distributions of four basic
closed triples in the YouTube social network and the model.}
Distributions of closed triples
corresponding to (a) $FB$, (b) $FF_{a}$, (c) $FF_{b}$, and (d)
$FF_{c}$ loops in the YouTube social network (black squares) and
in the simulated network based on the model (red circles).
Analytic treatment (see Eqs. $(30)$ and $(31)$) suggests a
scaling behavior with an exponent $-1.65$, as shown by the dash
lines. Data points are averages over the logarithmic bins of the $n_{FB}$, $n_{FFa}$, $n_{FFb}$ and $n_{FFc}$, respectively.}
\end{figure}

\begin{figure}[!ht]
\begin{center}
\includegraphics[width=4in,height=3in]{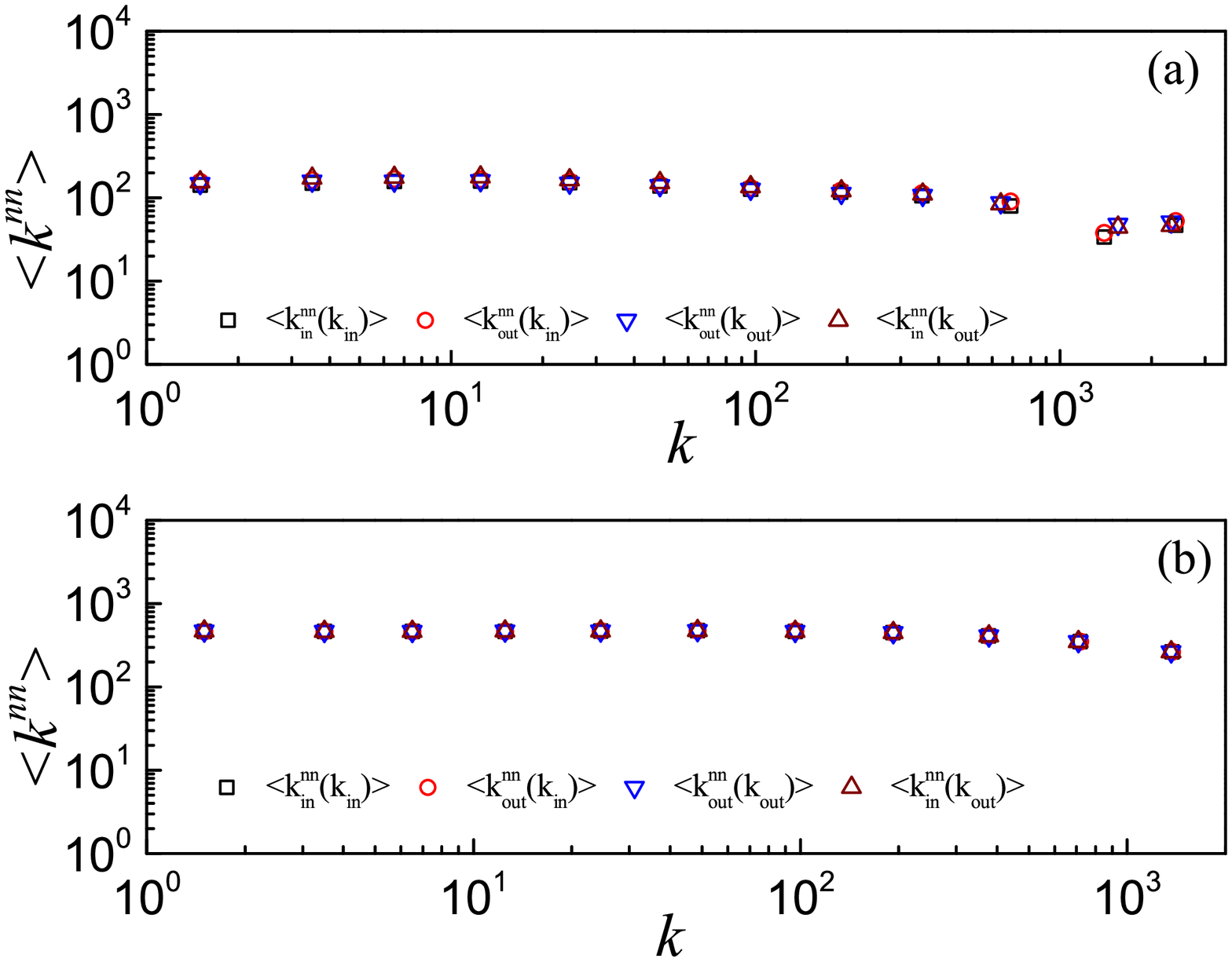}
\end{center}
\caption{\textbf{Degree correlations in the
Slashdot social network and the model.} Results
of degree correlations as measured by four quantities
corresponding to the average nearest neighbor degree
$<k_{in}^{nn}(k_{in})>$ (squares), $<k_{out}^{nn}(k_{in})>$
(circles), $<k_{out}^{nn}(k_{out})>$ (triangles), and
$<k_{in}^{nn}(k_{out})>$ (inverted triangles) for (a) Slashdot
social network and (b) simulated network based on the model.
Data points are averages over the logarithmic bins of the indegree $k_{in}$ or outdegree $k_{out}$.}
\end{figure}

\begin{figure}[!ht]
\begin{center}
\includegraphics[width=4in,height=3in]{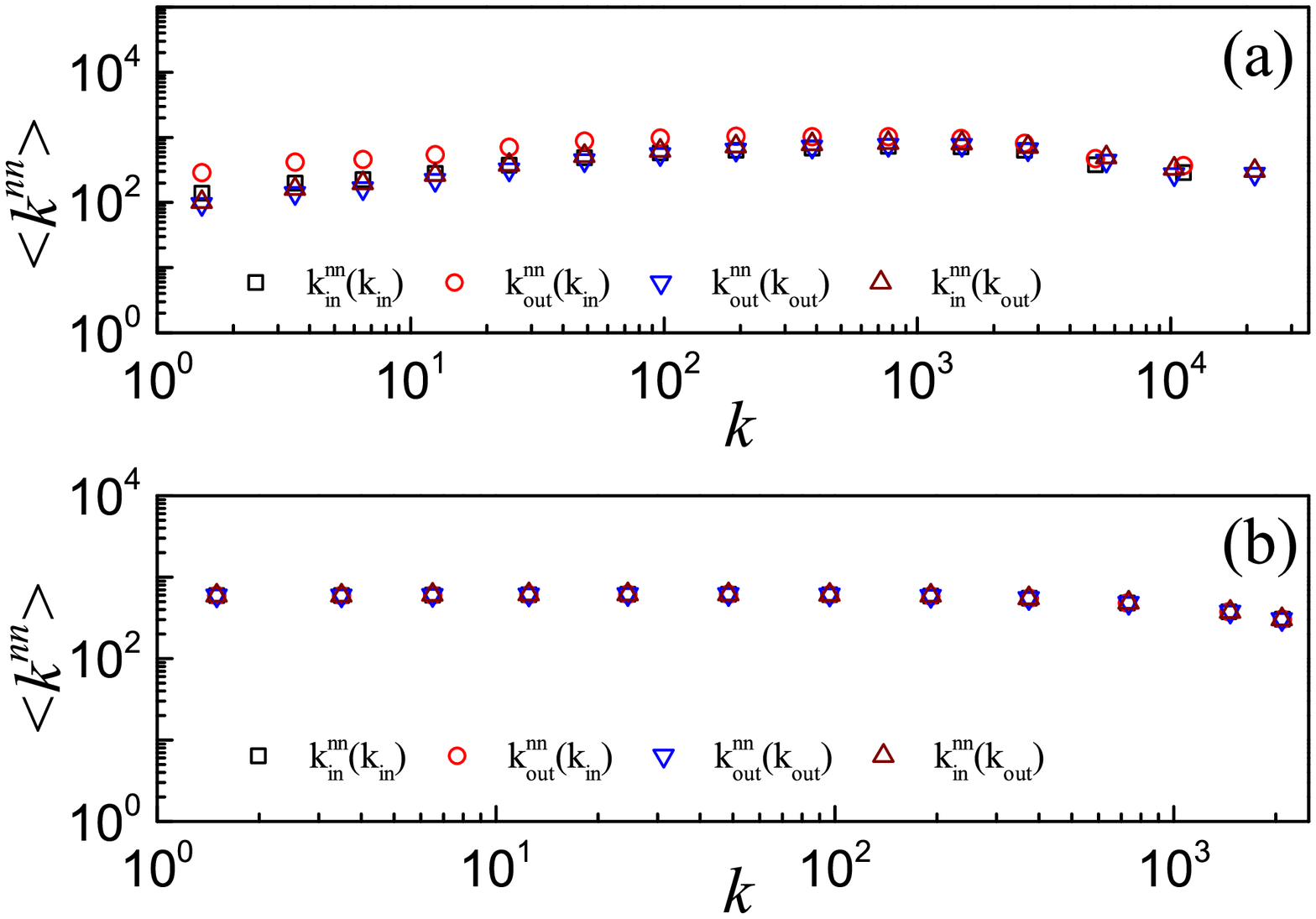}
\end{center}
\caption{\textbf{Degree correlations in the
Flickr social network and the model.} Results of
degree correlations as measured by four quantities corresponding
to the average nearest neighbor degree $<k_{in}^{nn}(k_{in})>$
(squares), $<k_{out}^{nn}(k_{in})>$ (circles),
$<k_{out}^{nn}(k_{out})>$ (triangles), and
$<k_{in}^{nn}(k_{out})>$ (inverted triangles) for (a) Flickr
social network and (b) simulated network based on the model.
Data points are averages over the logarithmic bins of the indegree $k_{in}$ or outdegree $k_{out}$.}
\end{figure}

\begin{figure}[!ht]
\begin{center}
\includegraphics[width=4in,height=3in]{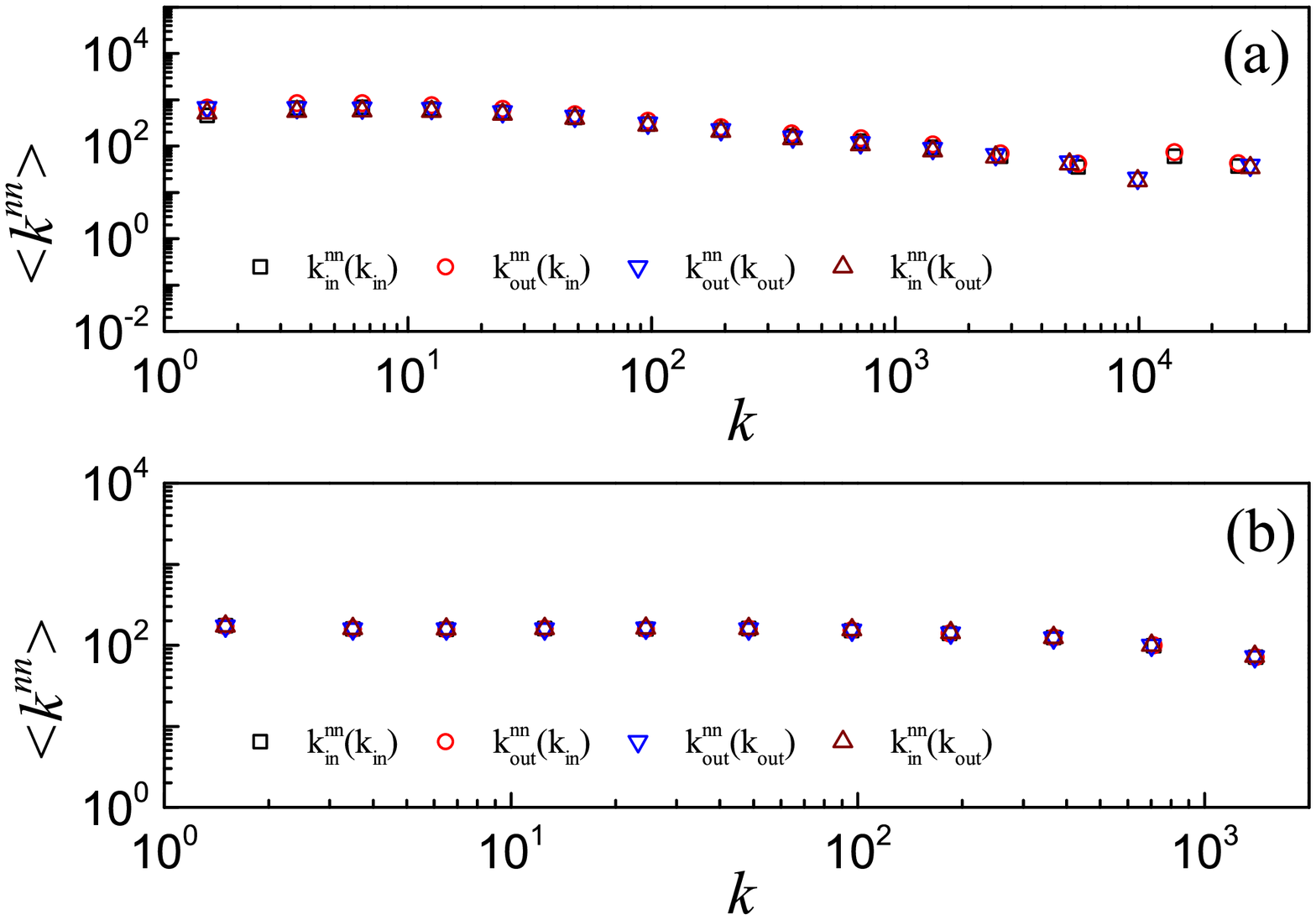}
\end{center}
\caption{\textbf{Degree correlations in the
YouTube social netowrk and the model.} Results
of degree correlations as measured by four quantities
corresponding to the average nearest neighbor degree
$<k_{in}^{nn}(k_{in})>$ (squares), $<k_{out}^{nn}(k_{in})>$
(circles), $<k_{out}^{nn}(k_{out})>$ (triangles), and
$<k_{in}^{nn}(k_{out})>$ (inverted triangles) for (a) YouTube
social network and (b) simulated network based on the model.
Data points are averages over the logarithmic bins of the indegree $k_{in}$ or outdegree $k_{out}$.}
\end{figure}

\begin{figure}[!ht]
\begin{center}
\includegraphics[width=4in]{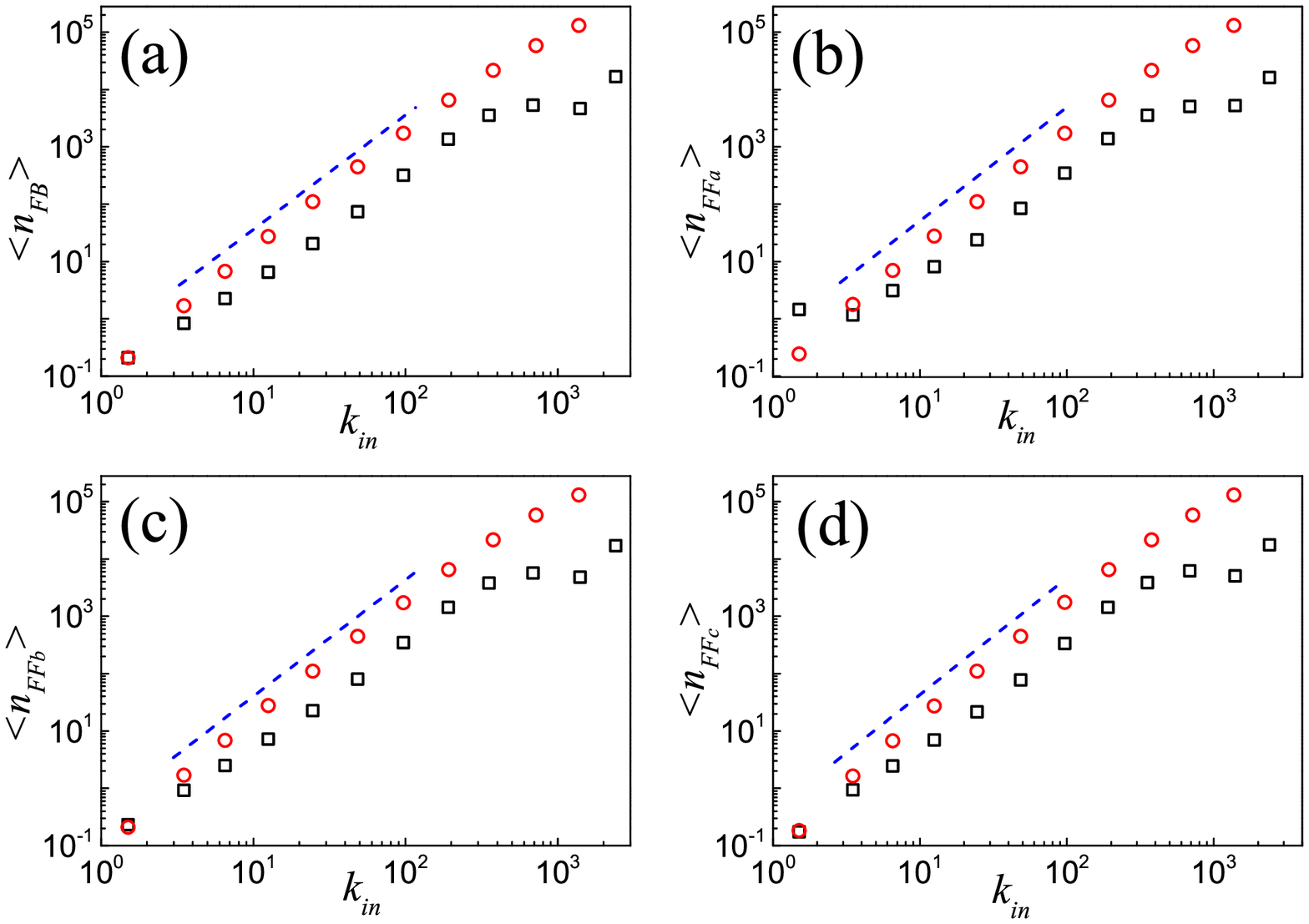}
\end{center}
\caption{\textbf{Mean number of the four closed
triples for nodes with the same indegree in the Slashdot social
network and the model.} Results for the mean number of closed triples corresponding to (a) $FB$, (b)
$FF_{a}$, (c) $FF_{b}$, and (d) $FF_{c}$ loops for nodes with the
same indegree are shown for the Slashdot social network (black
squares) and simulated network (red circles) based on the model.
Analytic treatment (see Eq. $(28)$) gives a scaling behavior with
an exponent $2$, as indicated by the dash line.
Data points are averages over the logarithmic bins of the indegree $k_{in}$.}
\label{kinAveTriangle-of-Slashdot}
\end{figure}

\begin{figure}[!ht]
\begin{center}
\includegraphics[width=4in]{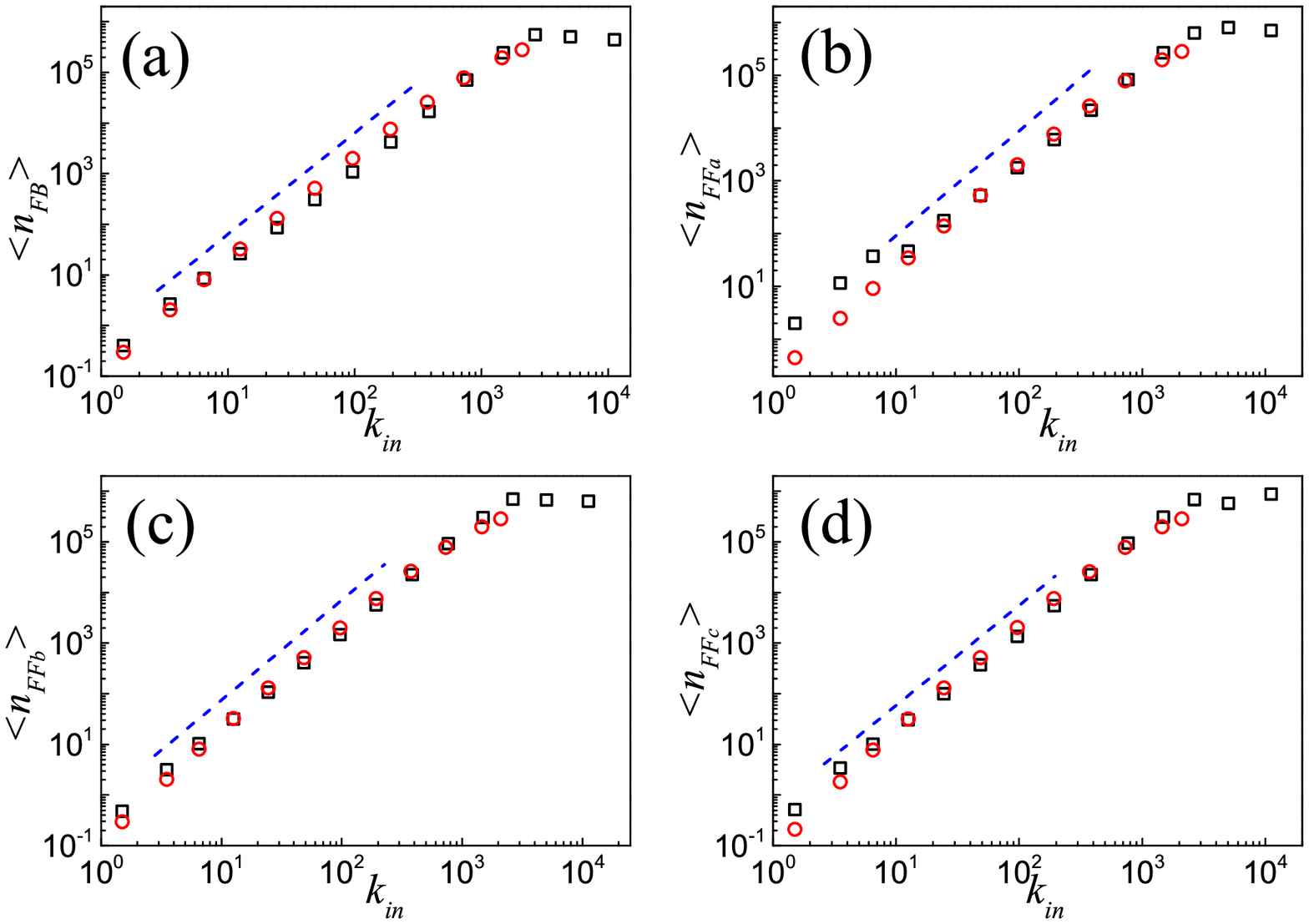}
\end{center}
\caption{\textbf{Mean number of the four closed
triples for nodes with the same indegree in the Flickr social
network and the model.} Results for the mean
number of closed triples corresponding to (a) $FB$, (b) $FF_{a}$,
(c) $FF_{b}$, and (d) $FF_{c}$ loops for nodes with the same
indegree are shown for the Flickr social network (black squares)
and simulated network (red circles) based on the model. Analytic
treatment (see Eq. $(28)$) gives a scaling behavior with an
exponent $2$, as indicated by the dash line.
Data points are averages over the logarithmic bins of the indegree $k_{in}$.}
\label{kinAveTriangle-of-Flickr}
\end{figure}

\begin{figure}[!ht]
\begin{center}
\includegraphics[width=4in]{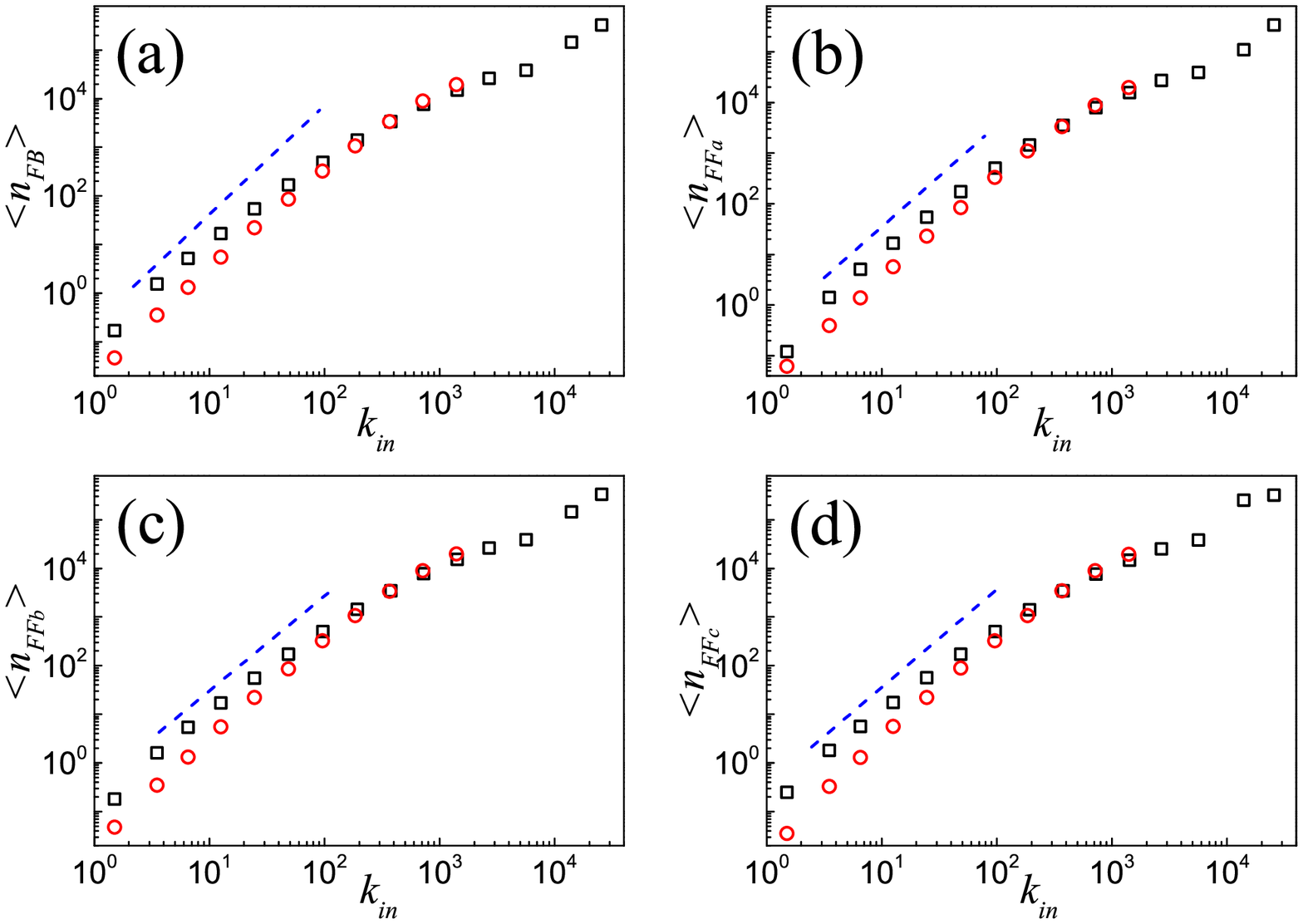}
\end{center}
\caption{\textbf{Mean number of the four closed
triples for nodes with the same indegree in the YouTube social
network and the model.} Results for the
mean number of closed triples corresponding to (a) $FB$, (b)
$FF_{a}$, (c) $FF_{b}$, and (d) $FF_{c}$ loops for nodes with the
same indegree are shown for the YouTube social network (black
squares) and simulated network (red circles) based on the model.
Analytic treatment (see Eq. $(28)$) gives a scaling behavior with
an exponent $2$, as indicated by the dash line.
Data points are averages over the logarithmic bins of the indegree $k_{in}$.}
\label{kinAveTriangle-of-YouTube}
\end{figure}

\clearpage

\begin{figure}[!ht]
\begin{center}
\includegraphics[width=4in]{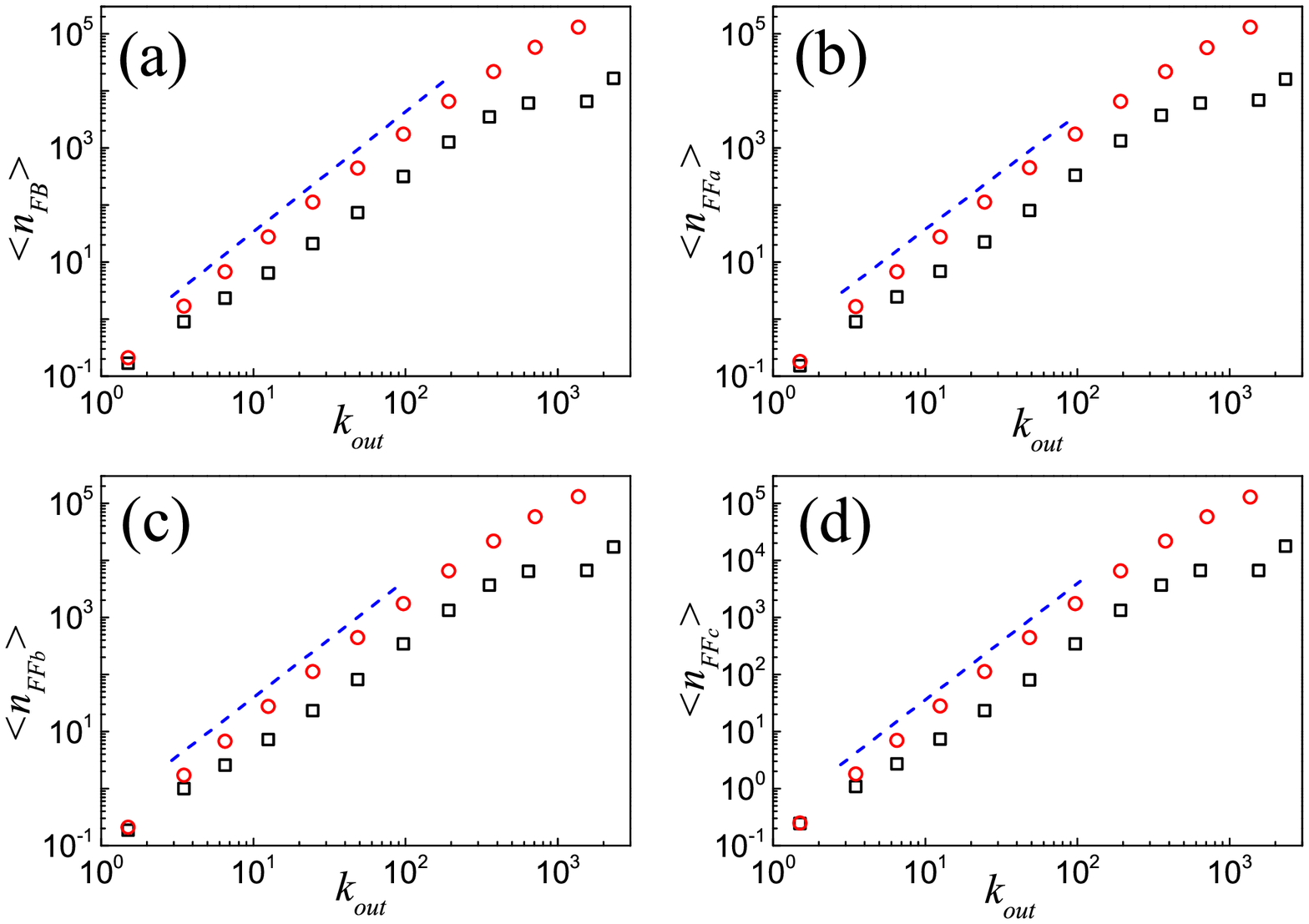}
\end{center}
\caption{\textbf{Mean number of the four closed
triples for nodes with the same outdegree in the Slashdot social
network and the model.} Results for the mean
number of closed triples corresponding to (a) $FB$, (b) $FF_{a}$,
(c) $FF_{b}$, and (d) $FF_{c}$ loops for nodes with the same
outdegree are shown for the Slashdot social network (black
squares) and simulated network (red circles) based on the model.
Analytic treatment (see Eq. $(28)$) gives a scaling behavior with
an exponent $2$, as indicated by the dash line.
Data points are averages over the logarithmic bins of the outdegree $k_{out}$.}
\label{koutAveTriangle-of-Slashdot}
\end{figure}

\begin{figure}[!ht]
\begin{center}
\includegraphics[width=4in]{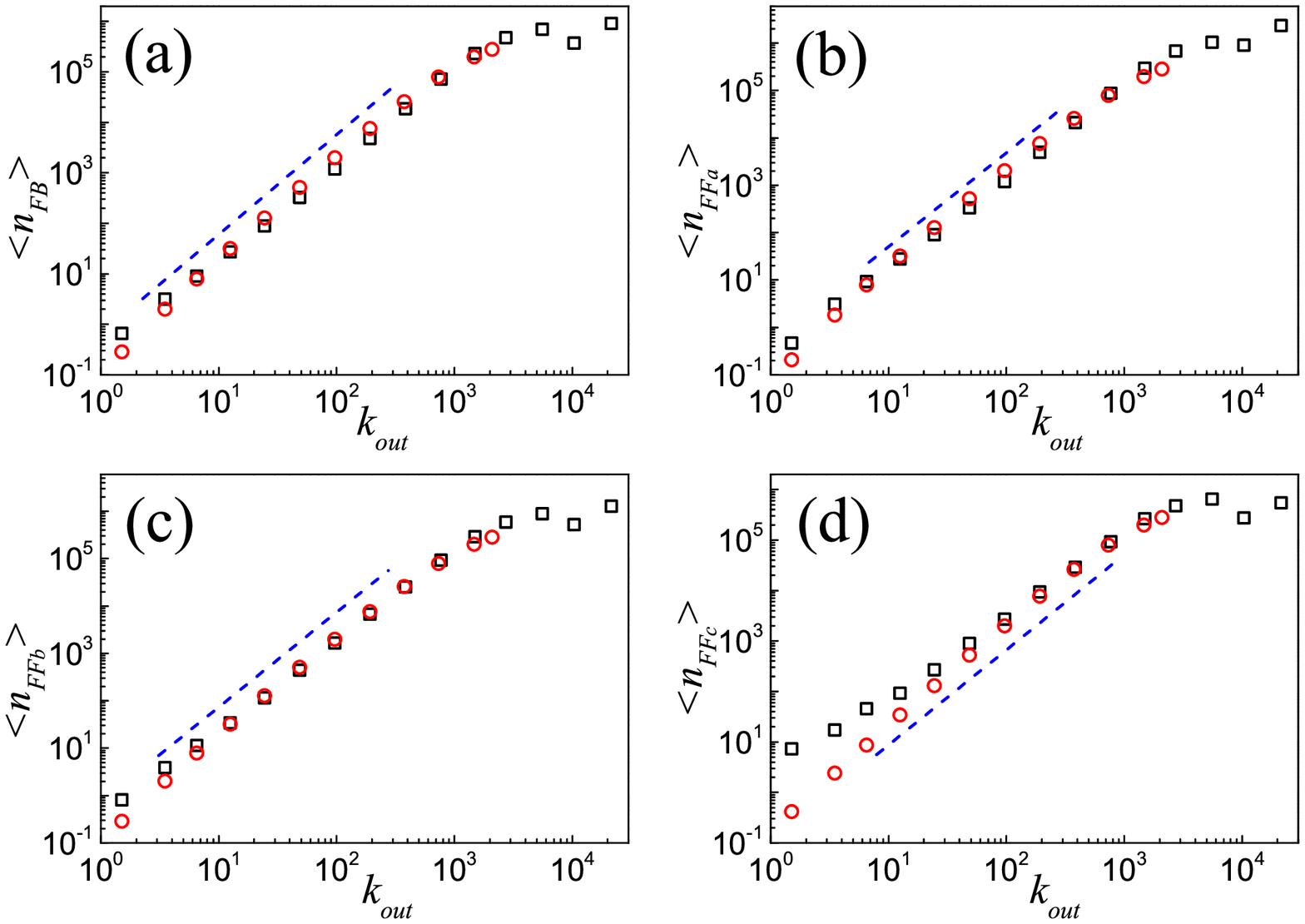}
\end{center}
\caption{\textbf{Mean number of the four closed
triples for nodes with the same outdegree in the Flickr social
network and the model.} Results for the mean
number of closed triples corresponding to (a) $FB$, (b) $FF_{a}$,
(c) $FF_{b}$, and (d) $FF_{c}$ loops for nodes with the same
outdegree are shown for the Flickr social network (black squares)
and simulated network (red circles) based on the model. Analytic
treatment (see Eq. $(28)$) gives a scaling behavior with an
exponent $2$, as indicated by the dash line.
Data points are averages over the logarithmic bins of the outdegree $k_{out}$.}
\label{koutAveTriangle-of-Flickr}
\end{figure}

\begin{figure}[!ht]
\begin{center}
\includegraphics[width=4in]{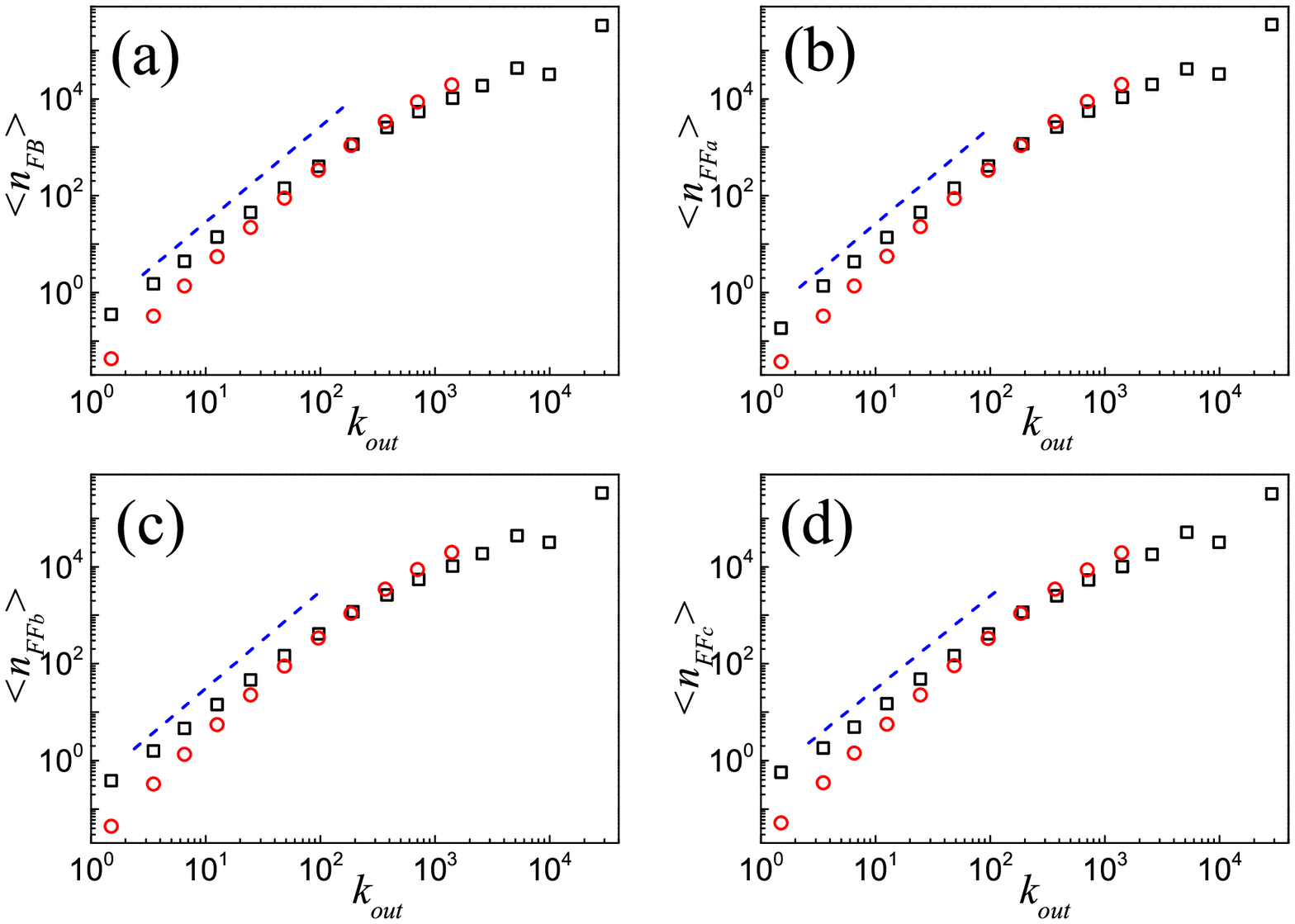}
\end{center}
\caption{\textbf{Mean number of the four closed
triples for nodes with the same outdegree in the YouTube social
network and the model.} Results for the mean
number of closed triples corresponding to (a) $FB$, (b) $FF_{a}$,
(c) $FF_{b}$, and (d) $FF_{c}$ loops for nodes with the same
outdegree are shown for the YouTube social network (black
squares) and simulated network (red circles) based on the model.
Analytic treatment (see Eq. $(28)$) gives a scaling behavior with
an exponent $2$, as indicated by the dash line.
Data points are averages over the logarithmic bins of the outdegree $k_{out}$.}
\label{koutAveTriangle-of-YouTube}
\end{figure}

\end{appendix}

\end{document}